\documentclass[preprint,10pt]{elsarticle}

\usepackage{hyperref}

\usepackage{subfig}
\usepackage{mathtools}
\usepackage{comment}
\usepackage{booktabs}
\usepackage{amssymb}
\usepackage{url}
\usepackage{tikz} 
\usetikzlibrary{shapes,arrows,positioning,chains,fit,arrows.meta}

\usepackage{float}
\usepackage{tabularx}
\usepackage{enumerate}
\usepackage{array}
\usepackage{xspace}

\usepackage{amsthm}
\usepackage{amsmath}
\usepackage{algorithm}
\usepackage[noend]{algpseudocode}
\usepackage{booktabs}

\usepackage{enumitem}
\usepackage{ulem}
\usepackage{framed}
\usepackage[a4paper,left=3cm,right=3cm,top=2.5cm,bottom=2.5cm]{geometry}

\newcommand{\fig}{\text{Fig.~}}
\newcommand{\eq}{\text{Eq.~}}
\newcommand{\sez}{\text{Sec.~}}
\newcommand{\alg}{\text{Algorithm~}}

\DeclareMathOperator*{\argmin}{arg\,min}

\usepackage{nomencl}
\usepackage{multicol}
\makenomenclature
\renewcommand*\nompreamble{\begin{multicols}{2}}
\renewcommand*\nompostamble{\end{multicols}}

\tikzstyle{P_node} = [circle,draw=darkgray,thick,align=center,minimum size=0.95cm,fill=cyan!50]
\tikzstyle{O_node} = [circle,draw=darkgray,thick,align=center,minimum size=0.95cm,fill=cyan!20]
  \tikzstyle{O_node_exp} = [circle,draw=darkgray,line width=.6mm,align=center,minimum size=0.97cm,fill=cyan!20]
\tikzstyle{D_node} = [circle,draw=darkgray,thick,align=center,minimum size=0.95cm,fill=green!70!cyan!30]
\tikzstyle{U_node_prob} = [circle, draw=darkgray, thick, minimum size=0.95cm,fill=red!40]
\tikzstyle{U_node_act} = [rectangle, draw=darkgray, line width=.6mm, minimum width=0.9cm, minimum height=0.9cm,fill=red!40]
\tikzstyle{R_node} = [rectangle,draw=darkgray,thick,align=center,minimum width=0.7cm, minimum height=0.7cm,fill=yellow!30,rotate=45]
\tikzstyle{op_node} = [rectangle,line width=0.1mm,draw=darkgray,minimum width=0.9cm, minimum height=0.9cm,fill=gray!25]
\tikzstyle{rectvia} = [rectangle,line width=0mm,draw=white,minimum width=0.5cm, minimum height=0.5cm,fill=white]

\makeatletter
\def\ps@pprintTitle{%
  \let\@oddhead\@empty
  \let\@evenhead\@empty
  \def\@oddfoot{\reset@font\hfil\thepage\hfil}
  \let\@evenfoot\@oddfoot
}
\makeatother

\begin{document}

\begin{frontmatter}
\title{Active Digital Twins via Active Inference}

\author[1]{Matteo~Torzoni\corref{cor1}}

\author[2]{Domenico~Maisto}

\author[3]{Andrea~Manzoni}

\author[2]{Francesco~Donnarumma}

\author[2]{Giovanni~Pezzulo}

\author[1]{Alberto~Corigliano}

\affiliation[1]{organization={Department of Civil and Environmental Engineering, Politecnico di Milano},
city={Milan},
postcode={20133},
country={Italy}}
\affiliation[2]{organization={Institute of Cognitive Sciences and Technologies, National Research Council},
city={Rome},
postcode={00185},
country={Italy}}
\affiliation[3]{organization={MOX -- Department of
Mathematics, Politecnico di Milano},
city={Milan},
postcode={20133},
country={Italy}}
\cortext[cor1]{Corresponding author: \texttt{matteo.torzoni@polimi.it}}

\begin{abstract}
Digital twins are transforming engineering and applied sciences by enabling real-time monitoring, simulation, and predictive analysis of physical systems and processes. However, conventional digital twins rely primarily on passive data assimilation, which limits their adaptability in uncertain and dynamic environments. This paper introduces the \textit{active digital twin} paradigm, based on active inference. Active inference is a neuroscience-inspired Bayesian framework for probabilistic reasoning and predictive modeling that unifies inference, decision-making, and learning under a single free energy minimization objective. By modeling the dynamics of the coupled physical--digital system as a partially observable Markov decision process, active digital twins autonomously balance pragmatic exploitation (maximizing goal-directed utility) and epistemic exploration (actively resolving uncertainty). As action becomes an integral part of the inference process, active digital twins actively seek information to maintain synchronization with, and learn from their physical counterparts. The proposed framework is assessed through virtual experiments of structural health monitoring and predictive maintenance of a railway bridge. The application showcases the step-by-step construction of a generative model enabling bidirectional perception--action interaction. The results demonstrate that active digital twins exhibit superior exploration capabilities compared to traditional reactive approaches, enabling enhanced autonomy and resilience.
\end{abstract}


\end{frontmatter}

\section{Introduction}
Over the past decade, the digital twin (DT) paradigm has emerged as a transformative approach for monitoring, control, and decision support, enabling diagnostic and predictive capabilities that surpass those of traditional computational models. As outlined in the 2024 report by the National Academies of Engineering, Science, and Medicine~\cite{Foundational}, DTs differ from both forward digital models and digital shadows~\cite{kritzinger20181016}. The former are designed to simulate how input parameters and internal states influence system behavior to generate observable outputs, while the latter focus on data assimilation and model updating. A DT is a tailored virtual representation that captures key attributes of a physical system or process~\cite{Ferrari2024}. This digital representation dynamically synchronizes with its physical counterpart by continuously assimilating sensor data and providing predictive capabilities. Specifically, DTs enable the simulation of what-if scenarios, supporting predictive decision-making aimed at maximizing utility. This paper proposes active inference (AIF)~\cite{parr2022active} as a new paradigm for DTs. By modeling the twin's evolution as a partially observable Markov decision process (POMDP)~\cite{2010artificial}, the AIF agent achieves intelligent automation under the \textit{free-energy principle}~\cite{friston2010free,friston2017graphical}. This results in a unified mathematical framework for a new class of \textit{active digital twins} (ADTs), equipped with spontaneous exploration capabilities.

Emerging from aeronautical and aerospace engineering~\cite{nasa2012digital,aiaa2020digital}, DT applications nowadays expand across several domains. These include structural health monitoring and predictive maintenance~\cite{Torzoni_DT,Torzoni_DT2,li2017dynamic}, additive manufacturing~\cite{phua2022digital}, smart cities~\cite{review_6}, energy transition~\cite{REIS2025125478}, urban sustainability~\cite{tzachor2022potential}, geotechnical engineering~\cite{geotechnical_DT_Straub}, subduction zone modeling~\cite{Tsunami_DT}, railway infrastructure management~\cite{arcieri_pomdp2}, aerial vehicles monitoring and control~\cite{mcclellan2022physics, tezzele2024adaptive}, spacecraft operations in orbit~\cite{Sebastian_space}, personalized medicine~\cite{corral2020digital,chaudhuri2023predictive}, and climate science~\cite{bauer2021digital}. Despite the growing interest in DTs, their implementation remains highly customized, typically tailored on the specific application, and often hard to deploy. The need for a widely accepted framework for DTs is therefore increasingly recognized in both research and industry. In~\cite{pgm_wilcox_dt}, Kapteyn et al. proposed an application-agnostic formulation for describing coupled physical--digital systems that evolve dynamically over time and interact via observed data and control inputs. A key contribution of their work is the abstraction of the coupled dynamical system into a generalized representation, which serves as the foundation for a mathematical description of DTs. This abstraction is consistent with agent-based representations in POMDPs, typically formalized using probabilistic graphical models~\cite{koller2009probabilistic}.

We introduce ADTs with enhanced exploratory capabilities, employing AIF agents based on discrete generative models to leverage and significantly extend the abstraction of physical--digital systems by Kapteyn et al.~\cite{pgm_wilcox_dt}. Active inference is a theoretical framework integrating perception, decision-making, and learning within the unified objective of \textit{free energy minimization}~\cite{parr2022active}. An AIF agent maintains an internal generative model of its environment, continuously updating its beliefs in response to sensory inputs. By minimizing variational free energy, the agent simultaneously fulfills two objectives: reducing the divergence between predicted and preferred future observations, and resolving expected uncertainty about hidden states through action. This dual mechanism naturally balances \textit{exploitation} of existing knowledge to achieve specific goals with \textit{exploration}, i.e., the acquisition of new information. These two imperatives can be referred to using interchangeable terminology. The exploitative or pragmatic behavior is associated with terms such as goal-directed behavior or utility maximization, while the exploratory or epistemic behavior is described using terms such as information seeking, information gain, or uncertainty resolution. The AIF framework has been applied in diverse domains, from neuroscience~\cite{Friston02102015,fitzgerald2015dopamine,parr2017uncertainty,pezzulo2018hierarchical,van2024hierarchical} -- for modeling decision-making under uncertainty -- to reinforcement learning~\cite{fountas2020deep,mazzaglia2021contrastive}, collective behavior~\cite{maisto2023interactive,heins2024collective}, and robotics~\cite{buckley2017free,lanillos2021active,taniguchi2023world,vijayaraghavan2025development}, demonstrating its versatility in modeling dynamic systems.

The generative model of an AIF agent functions as a self-updating engine that unifies the key aspects underpinning ADTs -- namely, data assimilation, state estimation, prediction, planning, and learning -- under a Bayesian framework that generalizes across applications. Furthermore, as demonstrated in the following sections, AIF agents naturally provide a mechanism for active information seeking, thereby unlocking the full potential of ADTs. When combined with goal-directed (pragmatic) behavior and possibly enhanced with learning capabilities, this information-seeking (epistemic) drive enables ADTs to engage in spontaneous exploration in response to (potentially critical) uncertainty, ultimately maximizing pragmatic utility.

Compared to alternative approaches for developing DTs, AIF offers remarkable advantages. Unlike reinforcement learning, which relies on trial-and-error exploration (often infeasible in real-world applications) and typically requires extensive datasets, AIF enables ADTs to infer hidden states and optimize future behavior using a compact generative model. Moreover, the free energy minimization imperative of AIF balances information-seeking (epistemic) and goal-directed (pragmatic) behaviors, without the need for manually tuned reward functions or random exploration. These features make AIF particularly well-suited for adaptive and robust ADTs.
 
We present the ADT paradigm through an application in structural health monitoring and predictive maintenance of engineering structures. Given the potentially high life-cycle costs -- economic, social, and safety -- associated with such systems, adopting a DT perspective is crucial to enable condition-based or predictive maintenance practices, replacing traditionally employed time-based methods~\cite{Glaser,Achenbach}. To this end, non-destructive tests and in-situ inspections are inadequate for continuous and global monitoring. Conversely, by assimilating sensor data from permanent data collection systems, vibration-based structural health monitoring techniques enable automated damage identification and evolution tracking~\cite{Torzoni_DML,Springer_Ubertini}. This paradigm shift has the potential to unlock personalized monitoring, management, and maintenance programs~\cite{DT_review, 5_TMM}, offering numerous benefits throughout the system life-cycle -- including more informed structural safety assessments, better resource allocation, and increased system availability~\cite{tesi_Matteo}.

\begin{figure}[!t]
\centering
\includegraphics[width=1\textwidth]{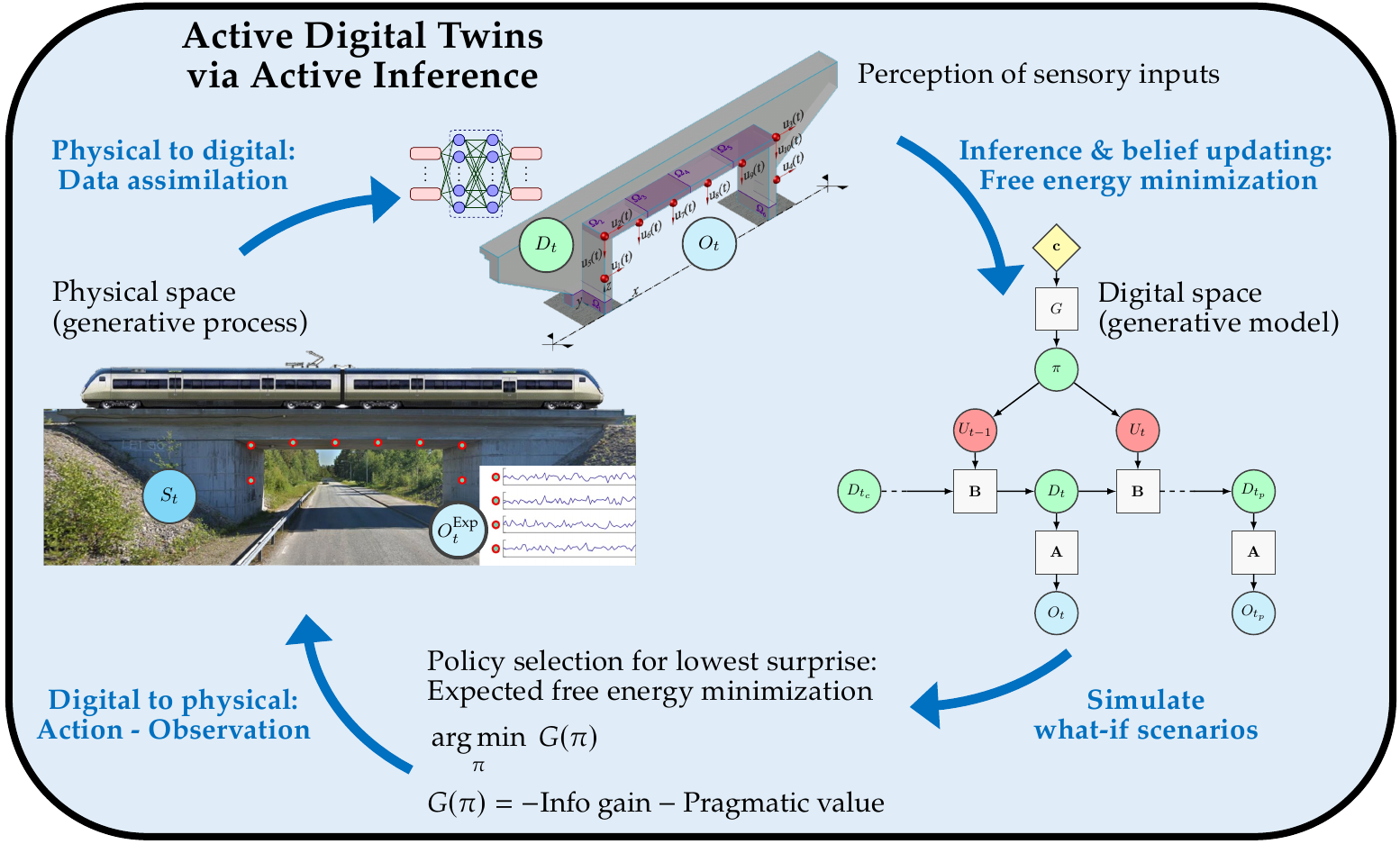}
\vspace{-0.15cm}
    \caption{Active digital twins via active inference -- Graphical abstraction of the end-to-end information flow. The dichotomy between the external physical process generating observational data (i.e., the \textit{generative process}) and the agent's internal model (i.e., the \textit{generative model}) is evident by the symmetry along the vertical axis. Meanwhile, the two forms of inference -- digital state estimation and policy selection -- exhibit a symmetry along the horizontal axis. A detailed schematic of generative models for both digital state and policy inference is presented in \fig\ref{fig:AIF_POMDP}.}
\label{fig:graph_abs}
\vspace{-0.25cm}
\end{figure}

A graphical abstraction of the computational flow is illustrated in \fig\ref{fig:graph_abs}. The end-to-end loop spans from the physical to the digital domain through data assimilation and inference, and then back to the asset through action and observation, while explicitly accounting for uncertainty quantification, propagation, and resolution. We refer to the monitored asset, whose physical state is hidden to the AIF agent and only indirectly accessible via the sensed structural response, as the \textit{external generative process}. The asset state evolves over time according to physical laws influenced by both its internal properties and external factors. These external factors might encompass long-term degradation mechanisms caused by chemical, physical, or mechanical aging, as well as sudden changes, such as discrete damage events or maintenance interventions~\cite{zakic1991classification}.

The digital counterpart (AIF agent) is defined by an \textit{internal generative model}, implemented as a probabilistic graphical model in the form of a dynamic Bayesian network (DBN)~\cite{koller2009probabilistic,2010artificial}. This factored representation provides a systematic way to maintain a posterior belief about latent variables that characterize the (hidden) structural health of the asset, such as damage presence, location, and severity, by continuously integrating new observations within a sequential Bayesian inference scheme. Belief updating is achieved by minimizing variational \textit{free energy}, which measures the discrepancy between the model's predicted observations and the actual sensor data.

In parallel, the internal generative model supports the forward simulation of future states. This enables the ADT to evaluate ``what-if'' trajectories for structural health evolution, conditioned on its current beliefs. This forecasting step involves modeling not only the asset's physical dynamics but also the agent's control, represented as latent variables encoding sequences of future actions, usually termed policies~\cite{agency}. Policy selection is then framed as an optimization problem, where the agent seeks to minimize the \textit{expected free energy} -- a quantity that balances ($i$) selecting policies that align future observations with (pragmatic) goal-directed prior preferences, and ($ii$) resolving uncertainty about hidden states through (epistemic) information-seeking. This formalism unifies inference and control: posterior beliefs are updated via free energy minimization, while action sequences are selected to minimize expected free energy, converting the problem of decision-making into a problem of inference under the generative model. Once an action is executed, the generative process evolves, and the bidirectional perception--action cycle restarts.

{\bf Open questions:} The limitations of conventional DTs, typically restricted to passive observation and open-loop simulation, have been recognized in~\cite{DT_IEEE}. However, the challenge of actively seeking information to enhance perception and learning remains largely unaddressed. Similar problems have long been studied in fields such as active vision, where perception is not limited to passively acquired images, but involves actively steering the sensing process to reduce uncertainty about the environment~\cite{Active_perception,Active_vision}. Likewise, robotic systems dynamically adjust sensing devices to enhance environmental exploration~\cite{probabilistic_robotics,planning_acting}, as in simultaneous localization and mapping (SLAM) tasks~\cite{slam}. Similar perception mechanisms are embedded in autonomous driving systems, where sensor attention is dynamically allocated according to environmental conditions and contextual priorities~\cite{auto_drive}. Active digital twins build on the same principle of closing the loop between perception and action, enabling these systems to autonomously improve situational awareness, refine their internal models through active exploration, and proactively manage the environment evolution.

{\bf Novelty:} While AIF has been widely applied in neuroscience, robotics, and decision theory, the novelty of this work lies in adopting it as a foundational framework for digital twins. By actively resolving uncertainty, ADTs foster adaptation and self-learning as means to maximize utility, in the spirit of intelligent automation~\cite{san2026evolution}. The balance between goal-directed (pragmatic) and information-seeking (epistemic) behaviors emerges naturally from the ADT's generative model through expected free energy minimization (see also \fig\ref{fig:graph_abs}). The framework is formalized as a POMDP encoded through a probabilistic graphical model, providing an application-agnostic platform that extends previous abstractions of physical--digital systems~\cite{pgm_wilcox_dt}. We also provide a reproducible procedure for constructing the ADT's generative model and demonstrate its advantages in structural health monitoring and predictive maintenance, highlighting behaviors unattainable by purely reactive digital twins.

The paper is organized as follows. Section~\ref{sez:PGM} describes the POMDP encoding the coupled dynamics of the physical--digital system. Section~\ref{sez:active_DT} illustrates how AIF agents are used to realize ADTs. Section~\ref{sez:results} assesses the proposed approach through simulated monitoring, management, and maintenance of a railway bridge, providing comparative results for different AIF agents featuring increasingly rich behavior. The outcomes, along with current limitations and policy recommendations, are discussed in \sez\ref{sez:discussion}. Conclusions and future developments are finally outlined in \sez\ref{sez:conclusion}.

\section{Partially observable Markov decision process for digital twins}
\label{sez:PGM}

\begin{figure}[!t]
\center
\begin{tikzpicture}[scale=.85, every node/.style={scale=1.}]

\node [P_node] (P_0) at (0,5) {};
\node [] () at (0,5) {$S_{0}$};
\node [D_node] (D_0) at (0,0) {};
\node [] at (0,0) {$D_{0}$};
\node [O_node_exp] (O_0) at (0,3) {};
\node [] at (0,3) {$O^\text{Exp}_{0}$};
\node [U_node_act] (U_A_0) at (2.5,3) {};
\node [] at (2.5,3) {$U_{0}$};
\node [R_node] (R_0) at (2.5,1) {};
\node [] at (2.5,1) {$R_{0}$};

\node [P_node] (P_1) at (5,5) {};
\node [] () at (5,5) {$S_{1}$};
\node [D_node] (D_1) at (5,0) {};
\node [] at (5,0) {$D_{1}$};
\node [O_node_exp] (O_1) at (5,3) {};
\node [] at (5,3) {$O^\text{Exp}_{1}$};
\node [U_node_act] (U_A_1) at (7.5,3) {};
\node [] at (7.5,3) {$U_{1}$};
\node [R_node] (R_1) at (7.5,1) {};
\node [] at (7.5,1) {$R_{1}$};

\node [P_node] (P_2) at (10,5) {};
\node [] () at (10,5) {$S_{2}$};
\node [D_node] (D_2) at (10,0) {};
\node [] at (10,0) {$D_{2}$};
\node [O_node_exp] (O_2) at (10,3) {};
\node [] at (10,3) {$O^\text{Exp}_{2}$};
\node [U_node_act] (U_A_2) at (12.5,3) {};
\node [] at (12.5,3) {$U_{2}$};
\node [R_node] (R_2) at (12.5,1) {};
\node [] at (12.5,1) {$R_{2}$};

\node [] (t0) at (-0.6,-2) {};
\node [] () at (0,-2) {$|$};
\node [] () at (0,-1.5) {$t=0$};
\node [] () at (5,-2) {$|$};
\node [] () at (5,-1.5) {$t=1$};
\node [] () at (10,-2) {$|$};
\node [] () at (10,-1.5) {$t=2$};
\node [] (t2) at (14,-2) {};

\node [] (P_E) at (14,5) {};
\node [] (D_E) at (14,0) {};
\node [] (U_E_P) at (14,4.2) {};
\node [] (U_E_D) at (14,1.2) {};

\draw[-latex,thick,black] (P_0) to (P_1);
\draw[-latex,thick,black] (P_1) to (P_2);

\draw[-latex,thick,black] (P_0) to (O_0);
\draw[-latex,thick,black] (D_0) to (O_0);
\draw[-latex,thick,black] (U_A_0) to (R_0);
\draw[-latex,thick,black] (D_0) to (R_0);

\draw[-latex,thick,black] (D_0) to (D_1);
\draw[-latex,thick,black] (U_A_0) to (P_1);
\draw[-latex,thick,black] (P_1) to (O_1);
\draw[-latex,thick,black] (D_1) to (O_1);
\draw[-latex,thick,black] (U_A_0) to (D_1);
\draw[-latex,thick,black] (U_A_1) to (R_1);
\draw[-latex,thick,black] (D_1) to (R_1);

\draw[-latex,thick,black] (D_1) to (D_2);
\draw[-latex,thick,black] (U_A_1) to (P_2);
\draw[-latex,thick,black] (P_2) to (O_2);
\draw[-latex,thick,black] (D_2) to (O_2);
\draw[-latex,thick,black] (U_A_1) to (D_2);
\draw[-latex,thick,black] (U_A_2) to (R_2);
\draw[-latex,thick,black] (D_2) to (R_2);

\draw[-latex,thick,black] (t0) to (t2);

\draw[-,thick,black] (D_2) to (D_E);
\draw[-,thick,black] (P_2) to (P_E);
\draw[-,thick,black] (U_A_2) to (U_E_P);
\draw[-,thick,black] (U_A_2) to (U_E_D);

\draw[darkgray,dotted] (-1.5,2) to (14,2);
\node [rotate=90] () at (-1.5,3.8) {\small Physical space};
\node [rotate=90] () at (-1.5,0.3) {\small Digital space};
\end{tikzpicture}
\vspace{-0.15cm}
\caption{Dynamic Bayesian network encoding the asset-twin dynamical system. Circular nodes represent random variables, square nodes denote taken actions, and diamond-shaped nodes symbolize the objective function. Nodes with bold outlines indicate observed quantities, while those with thin outlines represent latent variables to be inferred. Directed edges encode conditional dependencies between variables.}
\label{fig:unrolled_dbn}
\vspace{-0.25cm}
\end{figure}
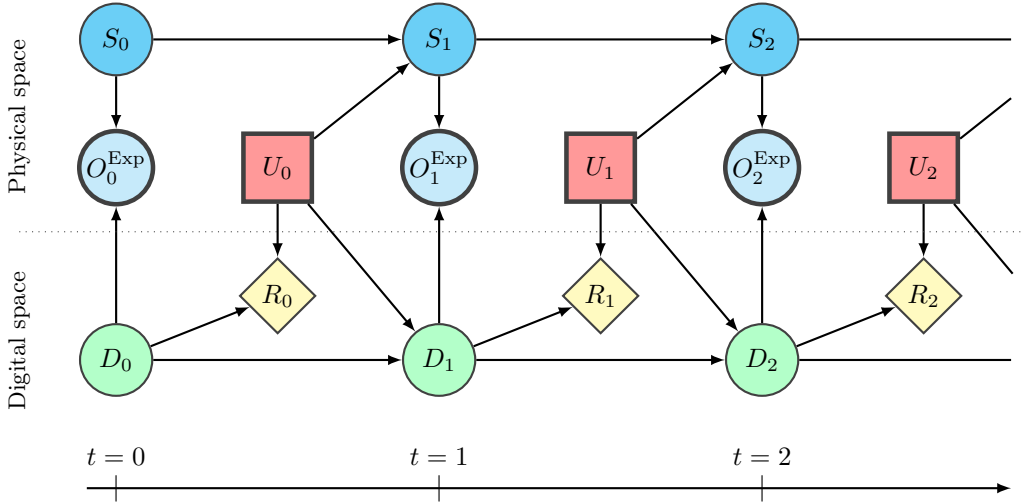

Figure~\ref{fig:unrolled_dbn} illustrates the probabilistic graphical model -- adapted from~\cite{pgm_wilcox_dt} -- that represents the dynamic interaction between the physical and virtual domains. This abstraction is inspired by classical POMDP formulations~\cite{2010artificial}. POMDPs are state-space models for decision-making in stochastic, partially observable environments, where system dynamics are typically described by Markov transition models. Unlike standard Markov decision processes, where a policy directly maps observable states to actions, POMDPs define the policy as a mapping from belief states -- probabilistic representations of hidden states inferred from observations -- to actions. 

The graph in \fig\ref{fig:unrolled_dbn} is a DBN, in which circular nodes represent random variables, square nodes denote taken actions, and diamond-shaped nodes symbolize the objective function. All variables are defined at discrete time steps. Each time the DT is updated through the assimilation of new observational data, the DBN advances by one time step, with $t\in\lbrace0,\ldots,T\rbrace$, where $t=0$ marks the moment the DT enters operation, and $t=T$ defines its lifetime horizon. Nodes with bold outlines indicate observed quantities, while those with thin outlines correspond to latent variables that must be inferred. The DBN is sparsely connected, with edges encoding conditional dependencies among the variables. For an overview of the fundamentals of DBNs, the reader is referred to~\cite{koller2009probabilistic,2010artificial}.

Capital letters denote random variables associated with the quantities in our abstraction, the corresponding lowercase letters refer to their specific realizations, and subscripts indicate their time index. Calligraphic letters denote the set of possible values each quantity can assume. For instance, the hidden physical state is denoted as $S_t\sim p(s_t)$, where $s_t$ represents a particular realization at time $t$, and $p(s_t)$ defines the probability that $S_t=s_t$ for any possible state $s_t\in \mathcal{S}$.

The digital state
$D_t \sim p(d_t)$ is designed to capture the essential features
of the (hidden) physical state that are relevant for diagnosis, prediction, and decision-making~\cite{Ferrari2024}. The digital state space $\mathcal{D}$ can represent a variety of information, including initial and/or boundary conditions, material properties, and other key characteristics to describe the asset under consideration.

The physical-to-digital information flow from $S_t$ to $D_t$ is mediated by the assimilation of observational data $O_t \sim p(o_t)$, enabling the inference of $D_t$. The observation space $\mathcal{O}$ may include sensor measurements, inspection results, or diagnostic reports. Since the physical state $S_t$ is only partially and indirectly observable, the digital state $D_t$ encodes posterior beliefs over possible system configurations at time $t$, reflecting the evidence provided by the available observations~\cite{Kamariotis_UQ,Cirak_STATFEM}. This perceptual process is realized through observation models -- one for each observation modality -- which relate digital states and observations in a probabilistic manner, such that $O_t$ is modeled as stochastically generated from $D_t$. Throughout the paper, we will use both $O_t$ and $O^\text{Exp}_t$ to represent observations: $O_t$ refers to predicted (expected) observations under the generative model, while $O^\text{Exp}_t$ denotes actual sensor data. Belief updates are driven by minimizing the discrepancy between predicted and actual observations.

The updated digital state $D_t$ informs the digital-to-physical information flow by guiding the selection of control actions to influence future physical states. In \fig\ref{fig:unrolled_dbn}, $U_t\sim p(u_t)$ denotes a decision variable representing the action taken. The action space $\mathcal{U}$ may include interventions that directly modify the physical state, adjustments to the operational conditions, or modifications to the observational process. Each action is associated with its own transition model -- one for each digital state factor -- across the digital state space, which serves as a control-dependent predictor that propagates the digital state beliefs forward in time. 

Finally, the reward node $R_t\sim p(r_t)$ quantifies the performance of the asset-twin system within a reward space $\mathcal{R}$. These rewards assess the expected ``quality'' of DBN trajectories to guide action selection toward optimal outcomes. In general, reward values may represent real costs associated with states and actions, or abstract metrics tuned to steer the system toward the desired behavior.

Formally, a POMDP can be defined as a seven-tuple $\langle\mathcal{D},\mathcal{O},\mathcal{U},\mathcal{R},\mathbf{A},\mathbf{B},\boldsymbol{\phi}\rangle$, where: $\mathcal{D}$ denotes the space of beliefs over hidden states; $\mathcal{O}$ is the space of possible observations; $\mathcal{U}$ is the space of available actions; $\mathcal{R}:\mathcal{D}\times\mathcal{U}\mapsto\mathbb{R}$ defines the reward function, which assigns a numerical value to beliefs-action pairs; $\mathbf{A}:\mathcal{O}\times\mathcal{D}\mapsto[0,1]$ is the observation model, encoding the conditional observation likelihood $p(O_t\mid D_t;\boldsymbol{\phi})$, which represents beliefs about how hidden states give rise to observations; $\mathbf{B}:\mathcal{D}\times\mathcal{D}\times\mathcal{U}\mapsto[0,1]$ is the transition model, encoding the conditional probability $p(D_t\mid D_{t-1},U_{t-1};\boldsymbol{\phi})$, which represents beliefs about the temporal evolution of hidden states conditioned on control actions; finally, $\boldsymbol{\phi}$ is a vector of parameters of the POMDP model.

In the following, we assume that digital states, observational data, and control actions are defined over discrete and finite spaces. This implies that these variables can only take value on a finite set of discrete levels. Consequently, categorical distributions give a natural choice for representing the corresponding probability distributions. These latter assign a probability value between $0$ and $1$ to each discrete outcome, under the constraint that probabilities across all levels must sum to one, as they represent a complete and mutually exclusive set of realizations. 

The joint probability distribution $p(O_t, D_t, U_t, R_t,\boldsymbol{\phi})$ over the POMDP factorizes -- according to the chain rule of probability -- into a product of categorical distributions (representing conditional likelihoods) and Dirichlet distributions (serving as priors). Numerically, these discrete distributions are organized as multidimensional arrays known as conditional probability tables (CPTs). The leading dimensions (rows) of a CPT correspond to the support of the random variable, while the lagging dimensions (columns) represent the conditioning variables. Each column specifies the probability distribution of a random variable given a particular configuration of its parent nodes, and the entries within each column sum to one, as they represent a complete set of mutually exclusive and exhaustive outcomes. If a node has no parents, its CPT reduces to a single column representing the prior probabilities of its possible values. The contents of these CPTs can be controlled through the model parameters included in $\boldsymbol{\phi}$.

The complete set of possible realizations of the unobserved variables -- conditioned on observational data $O^\text{Exp}_{0:t_c}=o^\text{Exp}_{0:t_c}$ and control actions $U_{0:t_c}=u_{0:t_c}$ -- from the initial time step $t = 0$ up to the current time $t_c$, with $t= 0,\ldots,t_c$, can be extracted by leveraging the conditional independence assumptions implied by the graph structure in \fig\ref{fig:unrolled_dbn}. The joint belief state can then be factorized according to the following sequential Bayesian inference formulation: 
\begin{equation}
\begin{split}
&p(D_{0:t_c}, R_{0:t_c}, \boldsymbol{\phi}\mid O^\text{Exp}_{0:t_c}=o^\text{Exp}_{0:t_c},U_{0:t_c}=u_{0:t_c}) =\\&\hspace{50pt} p(\boldsymbol{\phi})p(D_0;\boldsymbol{\phi})\prod_{t=1}^{t_c}p(D_t\mid D_{t-1},u_{t-1};\boldsymbol{\phi})\prod_{t=0}^{t_c}p(o_t^\text{Exp}\mid D_t;\boldsymbol{\phi})p(R_t \mid  D_t,u_t).
\end{split}
\label{eq:joint_factor}
\end{equation}
In \eq\eqref{eq:joint_factor}, the term $p(\boldsymbol{\phi})$ represents the prior distribution over the model parameters $\boldsymbol{\phi}$; inference over them typically evolves on a slower timescale than the inference of hidden states and control actions. $p(D_0;\boldsymbol{\phi})$ denotes the prior over the initial hidden states, representing the digital state belief at $t=0$, before any observation is incorporated. The term \mbox{$p(o_t^\text{Exp}\mid D_t;\boldsymbol{\phi})$} represents the sensory likelihood encoded in $\mathbf{A}$. Similarly, \mbox{$p(D_t\mid D_{t-1},u_{t-1};\boldsymbol{\phi})$} defines the transition likelihood encoded in $\mathbf{B}$. Finally, $p(R_t \mid  D_t,u_t)$ represents the likelihood of receiving a given reward, encapsulating the objective function evaluation. Note that selecting actions $U_t=u_t$ underpins solving the planning problem induced by the probabilistic graphical model. After forming a belief that measures the desirability of actions, such as \mbox{$p(U_t\mid D_t)$}, the actual action can be selected either as the best-point estimate or by sampling from this posterior, converting probabilistic control into a decision.

\section{Active inference for digital twins}
\label{sez:active_DT}

An attractive feature of AIF is that perception, learning, and action emerge as distinct manifestations of variational Bayesian inference~\cite{parr2022active}. Perception, or state estimation, is accomplished through inference over dynamically evolving hidden states, conditioned on assimilated observations and past actions. Learning corresponds to the gradual inference of model parameters that capture the statistical regularities of the environment. Action, in turn, is realized by inferring a posterior distribution over policies and sampling actions accordingly.

In the following, we describe the use of AIF agents to ``navigate'' the POMDP underlying the DT problem, enabling the full potential of ADTs. Section~\ref{sez:generative} introduces the AIF generative model, which encodes the probabilistic assumptions about the underlying environment. Section~\ref{sez:VFE} addresses digital state inference via variational free energy minimization. Section~\ref{sez:EFE} covers policy inference and action selection through expected free energy minimization. Section~\ref{sez:learning} describes the slow-scale learning of the model parameters that define the AIF generative model. Finally, \sez\ref{sez:active_behav} discusses the active information-seeking (epistemic) behavior that characterizes ADTs. 

\subsection{Active inference generative model}
\label{sez:generative}

\begin{figure}[!t]
\center
\begin{tikzpicture}[scale=.85, every node/.style={scale=1.}]

\node [D_node] (pi) at (5.5,9) {};
\node [] at (5.5,9) {$\pi$};
\node [op_node] (G) at (5.5,10.5) {};
\node [] at (5.5,10.5) {$G$};
\node [R_node] (C) at (5.5,12) {};
\node [] at (5.5,12) {$\mathbf{c}$};

\node [D_node] (D_0) at (0.65,6) {};
\node [] () at (0.65,6) {$D_{t_c}$};
\node [op_node] (B_1) at (3.5,6) {};
\node [] at (3.5,6) {$\mathbf{B}$};
\node [U_node_prob] (U_A_0) at (3.5,7.5) {};
\node [] at (3.5,7.5) {$U_{t-1}$};

\node [D_node] (D_1) at (5.5,6) {};
\node [] () at (5.5,6) {$D_{t}$};
\node [O_node] (O_1) at (5.5,3) {};
\node [] at (5.5,3) {${O}_{t}$};
\node [op_node] (A_1) at (5.5,4.5) {};
\node [] at (5.5,4.5) {$\mathbf{A}$};
\node [op_node] (B_2) at (7.5,6) {};
\node [] at (7.5,6) {$\mathbf{B}$};
\node [U_node_prob] (U_A_1) at (7.5,7.5) {};
\node [] at (7.5,7.5) {$U_{t}$};

\node [D_node] (D_2) at (10.35,6) {};
\node [] () at (10.35,6) {$D_{t_p}$};
\node [O_node] (O_2) at (10.35,3) {};
\node [] at (10.35,3) {${O}_{t_p}$};
\node [op_node] (A_2) at (10.35,4.5) {};
\node [] at (10.35,4.5) {$\mathbf{A}$};

\node [] (D_i1) at (2,6) {};
\node [] (D_i2) at (1.7,6) {};
\node [] (t_i1) at (2,1.5) {};
\node [] (t_i2) at (1.7,1.5) {};
\node [] (B_E) at (11.5,6) {};
\node [] (D_i3) at (9.05,6) {};
\node [] (D_i4) at (8.75,6) {};

\draw[-latex,thick,black] (C) to (G);
\draw[-latex,thick,black] (G) to (pi);
\draw[-latex,thick,black] (pi) to (U_A_0);
\draw[-latex,thick,black] (pi) to (U_A_1);
\draw[-latex,thick,black] (U_A_0) to (B_1);
\draw[-latex,thick,black] (U_A_1) to (B_2);
\draw[-latex,thick,black] (B_1) to (D_1);
\draw[-latex,thick,black] (D_1) to (B_2);
\draw[-latex,thick,black] (D_1) to (A_1);
\draw[-latex,thick,black] (D_2) to (A_2);
\draw[-latex,thick,black] (A_1) to (O_1);
\draw[-latex,thick,black] (A_2) to (O_2);

\draw[-,thick,dashed,black] (B_2) to (D_i3);
\draw[-latex,thick,black] (D_i4) to (D_2);
\draw[-,thick,dashed,black] (D_0) to (D_i1);
\draw[-latex,thick,black] (D_i2) to (B_1);

\end{tikzpicture}
\vspace{-0.15cm}
\caption{Dynamic Bayesian network encoding the active inference generative model used to predict future digital states and observations under each policy. Circular nodes represent random variables, while the diamond-shaped node denotes prior preferences that reflect a goal-directed (pragmatic) objective. Gray square nodes represent parametrized operators of the generative model. Directed edges encode conditional dependencies between variables.}
\label{fig:internal_dbn}
\vspace{-0.25cm}
\end{figure}
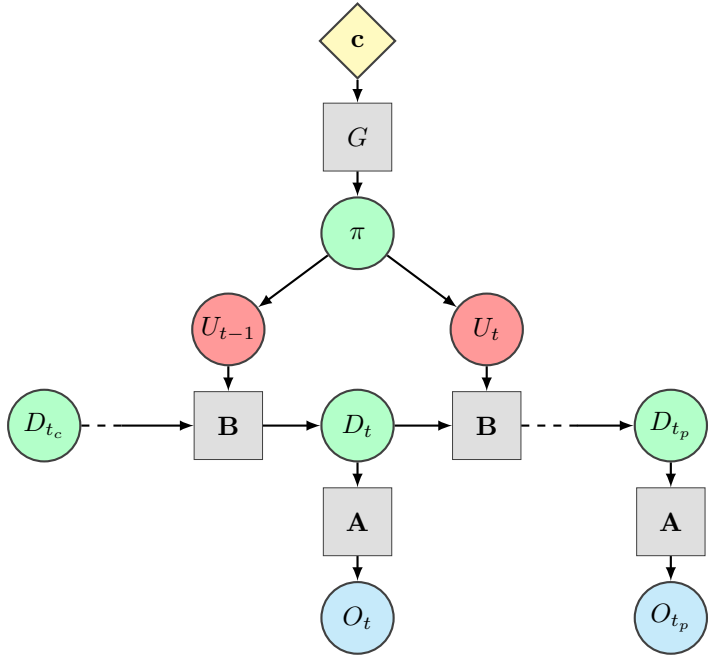

In AIF, the set of probabilistic assumptions about how the environment (or \textit{generative process}) produces observations (via the observation model $\mathbf{A}$) and how actions influence the environment evolution (via the transition model $\mathbf{B}$) is referred to as the POMDP generative model. This model is used to represent the joint distribution in \eq\eqref{eq:joint_factor}, from current time $t_c$ to a prediction horizon $t_p > t_c$. Specifically, for time- and space-discretized POMDPs, probabilistic estimates of future digital states and observations over the prediction time steps $t = t_c,\ldots,t_p$ are computed as:
\begin{equation}
p(O_{t_c:t_p},D_{t_c:t_p}, \boldsymbol{\phi} \mid  \pi)
=p(\boldsymbol{\phi})p(D_{t_c};\boldsymbol{\phi})\prod_{t=t_c+1}^{t_p}p(D_t\mid D_{t-1},\pi;\boldsymbol{\phi})\prod_{t=t_c}^{t_p}p(O_t\mid D_t;\boldsymbol{\phi}),
\label{eq:generative}
\end{equation}
which reflects unrolling the AIF generative model of \fig\ref{fig:internal_dbn} over $t = t_c,\ldots,t_p$. Compared to \eq\eqref{eq:joint_factor}, the factorization in \eq\eqref{eq:generative} introduces several modifications to align with the AIF framework. First, the control variable $U$ is replaced by a policy $\pi$, defined as a sequence of control states $\pi=\lbrace u_{t_c},\ldots,u_{t_p}\rbrace$. The generative model in \eq\eqref{eq:generative} is conditioned on a fixed policy $\pi$, which is how it is used for inference purposes. Policies are treated as latent variables to be inferred: the posterior over policies represents the agent beliefs about its intended actions, while single actions are realizations sampled from the posterior over control states. The policy-to-control mapping $p(U_t \mid \pi)$ assigns the control state at each time-step based on the selected policy. 

The second modification concerns the omission of the reward variable $R$. In AIF, utility-maximization goals are encoded as a prior distribution $\widetilde{p}(O_{t_c:t_p})$ over future observations. These preferences are specified through an unconditional CPT $\mathbf{c}:\mathcal{O}\mapsto[0,1]$. Such prior preferences guide policy selection toward goal-directed (pragmatic) behavior by favoring actions expected to produce preferred observations. This formal equivalence between rewards and priors eliminates the need for explicit cost functions. Further, it enables optimal control to be cast as an inference problem: the joint probability of observations, digital states, control states, and model parameters is maximized when the system samples from preferred observations. The square node $G$ in the graph represents the expected free energy, which quantifies the desirability of each policy by incorporating both pragmatic and information-seeking (epistemic) components, as explained in \sez\ref{sez:EFE}. Finally, the initial prior $p(D_{t_c};\boldsymbol{\phi})$ is typically represented by an unconditional CPT denoted as $\mathbf{d}:\mathcal{D}\mapsto[0,1]$. 

Graphically, the generative model illustrated in \fig\ref{fig:internal_dbn} shows several differences compared to the DBN in \fig\ref{fig:unrolled_dbn}. This formulation focuses on predicting future digital states $D_{t_c:t_p}$ and sampling (or generating) sequences of potential observations $O_{t_c:t_p}$ based on the probabilistic structure encoded in $\mathbf{A}$ and $\mathbf{B}$, conditioned on control actions $U_{t_c:t_p}$ that have not yet been executed. Accordingly, actions $U_t$ are modeled as (circular) random variables rather than (square) decision nodes, since they represent what-if scenarios beyond data assimilation. Moreover, the same color is used to represent both digital states and control policies in the graph, as both are latent variables of the generative model. Equipped with this generative model -- specified by the four-tuple $\langle\mathbf{A},\mathbf{B},\mathbf{c},\mathbf{d}\rangle$ -- AIF supports the inference over $D_t$, $\pi$, and $\boldsymbol{\phi}$, as described in the following sections.

\subsection{Digital state inference via variational free energy minimization}
\label{sez:VFE}

Given an observation $O^\text{Exp}_{t_c}=o^\text{Exp}_{t_c}$, the underlying digital state $D_{t_c}$ can be inferred by estimating a posterior distribution $p(D_{t_c}\mid O^\text{Exp}_{t_c}=o^\text{Exp}_{t_c})$, using Bayes' Rule:  
\begin{equation}
p(D_{t_c}\mid O^\text{Exp}_{t_c}=o^\text{Exp}_{t_c}) = \frac{p(o^\text{Exp}_{t_c}, D_{t_c})}{p(o^\text{Exp}_{t_c})}=\frac{p(o^\text{Exp}_{t_c}\mid D_{t_c})p( D_{t_c})}{
\sum_{d_{t_c}\in\mathcal{D}}p(o^\text{Exp}_{t_c},D_{t_c}=d_{t_c})
},
\label{eq:Bayes}
\end{equation}
where the (generative model) joint distribution $p(o^\text{Exp}_{t_c}, D_{t_c})$ is factorized into a likelihood term $p(o^\text{Exp}_{t_c}\mid D_{t_c})$ and a prior $p(D_{t_c})$. The denominator $p(o^\text{Exp}_{t_c})$ is the marginal likelihood or model evidence, which captures the probability of observing \mbox{$O^\text{Exp}_{t_c}=o^\text{Exp}_{t_c}$} under the generative model. 

Since Bayesian inversion to estimate hidden states from observations is generally intractable, AIF employs variational inference~\cite{murphy2023probabilistic} as an approximate Bayesian method, trading exactness for computational tractability. Specifically, we define a tractable variational distribution \linebreak\mbox{$Q(D_{t_c}; \boldsymbol{\theta}):\mathcal{D}\mapsto[0,1]$}, parametrized by $\boldsymbol{\theta}$, and optimize this surrogate distribution to make it as close as possible to the true posterior $p(D_{t_c}\mid O^\text{Exp}_{t_c}=o^\text{Exp}_{t_c})$. In our discrete POMDP setting, the variational parameters $\boldsymbol{\theta}$ correspond to the relative frequencies of each category in the support of a random variable. This leads to the following optimization problem:
\begin{equation}
\boldsymbol{\theta}^* = \argmin_{\boldsymbol{\theta}}\ \text{D}_\text{KL}\left[Q(D_{t_c}; \boldsymbol{\theta})\mid\mid p(D_{t_c}\mid o^\text{Exp}_{t_c}) \right],
\label{eq:optimiz_1}
\end{equation}
where $\text{D}_\text{KL}\left[Q(X)\mid\mid P(X\mid Y) \right]=\mathbb{E}_Q\left[\ln{Q(X)}-\ln{P(X\mid Y)}\right]$ denotes the Kullback-Leibler (KL) divergence between the approximate posterior $Q(X)$ and the true posterior $P(X\mid Y)$, for two generic random variables $X$ and $Y$. Here, $\mathbb{E}_Q$ denotes the expectation with respect to the variational posterior. However, this objective remains intractable because it depends on the true posterior $p(D_{t_c}\mid o^\text{Exp}_{t_c})$ that we seek to approximate. To circumvent this, we reformulate the objective as the variational free energy (VFE):
\begin{equation}
\begin{split}
\mathcal{F}_{t_c}(\boldsymbol{\theta}) &= \text{D}_\text{KL}\left[Q(D_{t_c}; \boldsymbol{\theta})\mid\mid p(D_{t_c}\mid o^\text{Exp}_{t_c}) \right] - \ln{p(o^\text{Exp}_{t_c})}\\
&=\sum_\mathcal{D}Q(D_{t_c}; \boldsymbol{\theta})\left[\ln{\frac{Q(D_{t_c}; \boldsymbol{\theta})}{p(D_{t_c}\mid o^\text{Exp}_{t_c})}}-\ln{p(o^\text{Exp}_{t_c})}\right]\\
&=\mathbb{E}_Q\left[\ln{Q(D_{t_c}; \boldsymbol{\theta})}-\ln{p(o^\text{Exp}_{t_c}, D_{t_c})}\right],
\label{eq:VFE}
\end{split}
\end{equation}
which serves as an upper bound on the negative log marginal likelihood ($-\ln{p(o^\text{Exp}_{t_c})}$), also known as the Bayesian surprise. Minimizing VFE thus brings the variational posterior closer to the true posterior while simultaneously increasing the marginal likelihood of the observation. The VFE objective leads to the following final form of the optimization problem:
\begin{equation}
\boldsymbol{\theta}^* = \argmin_{\boldsymbol{\theta}}\ \mathcal{F}_{t_c}(\boldsymbol{\theta}).
\label{eq:optimiz_2}
\end{equation}
At convergence, if $Q^*(D_{t_c}; \boldsymbol{\theta}^*)$ exactly matches the true posterior, the KL divergence vanishes, i.e., $\text{D}_\text{KL}=0$, and the VFE equals surprise: $\mathcal{F}_{t_c} =- \ln{p(o^\text{Exp}_{t_c})}$. By further minimizing surprise, the VFE then provides a useful objective not only for inference but also for learning the parameters of the generative model. The underlying rationale is that AIF agents aim to avoid surprising observations, and minimizing surprise is equivalent to maximizing model evidence.

With reference to the generative model formulation \eqref{eq:generative}, instantaneous inference over digital states involves approximating the true posterior $p(D_{t_c}\mid O^\text{Exp}_{t_c}=o^\text{Exp}_{t_c},D_{t_c-1},U_{t_c-1}=u_{t_c-1})$. This inference is conditioned on the current observation $O^\text{Exp}_{t_c}=o^\text{Exp}_{t_c}$, the previous (posterior) distribution over digital states $D_{t_c-1}$, and the previously executed action $U_{t_c-1}=u_{t_c-1}$, as:
\begin{equation}
\begin{split}
\boldsymbol{\theta}^* &= \argmin_{\boldsymbol{\theta}}\ \mathcal{F}_{t_c}(\boldsymbol{\theta})\\
&=\argmin_{\boldsymbol{\theta}}\ \mathbb{E}_Q\left[\ln{Q(D_{t_c}; \boldsymbol{\theta})}-\ln{p(o^\text{Exp}_{t_c}, D_{t_c}\mid D_{t_c-1},u_{t_c-1};\boldsymbol{\phi})}\right]\\
&=\argmin_{\boldsymbol{\theta}}\ \mathbb{E}_Q\left[\ln{Q(D_{t_c}; \boldsymbol{\theta})}-\ln{\left(p(o^\text{Exp}_{t_c}\mid D_{t_c};\boldsymbol{\phi})p(D_{t_c}\mid D_{t_c-1},u_{t_c-1};\boldsymbol{\phi})\right)}\right].
\end{split}
\label{eq:inference}
\end{equation}

The optimization problem in \eq\eqref{eq:inference} is solved using fixed-point iteration~\cite{wainwright2008graphical}, under the assumption of temporal factorization, where variational posteriors at different time steps are conditionally independent. As a result, the full VFE across trajectories decomposes into a sum of single-time-step free energies, enabling independent optimization at each time point. 

The posterior $Q(D_{t}; \boldsymbol{\theta})$ at a given time step $t$ can be further factorized across $F$ independent hidden state factors $D=\lbrace D^1,\ldots,D^F\rbrace$, following the mean-field approximation~\cite{Bishop}:
\begin{equation}
Q(D_{t}; \boldsymbol{\theta})=\prod_{f=1}^FQ(D^f_{t}; \boldsymbol{\theta}),
\end{equation}
where $Q(D^f_{t}; \boldsymbol{\theta})$ denotes the posterior over the $f$th hidden state factor, $f=1,\ldots,F$. These factors may represent distinct aspects of the generative process, potentially varying in dimensionality, transition dynamics, and association with specific observation modalities. Similarly, observations can be structured into $M$ distinct modalities \mbox{$O=\lbrace O^1,\ldots,O^M\rbrace$}, where each $O^m$, $m=1,\ldots,M$, corresponds to a separate sensory channel used by the agent at each time step. For example, in a DT application for the human health, one hidden state factor may represent a patient’s metabolic state, while another factor could encode cardiovascular function. Correspondingly, observation modalities may include blood glucose readings and heart rate measurements, each providing information about different latent physiological processes.
 
In this multi-modal, multi-factor setup, the observation likelihood array $\mathbf{A}$ becomes a collection of $M$ sub-arrays $\mathbf{A}=\lbrace \mathbf{A}^1,\ldots,\mathbf{A}^M\rbrace$, with each $\mathbf{A}^m$, $m=1,\ldots,M$, representing the observation model for the $m$th modality. Each sub-array encodes the likelihood \mbox{$p(O^m\mid D^1,\ldots,D^F;\boldsymbol{\phi})$}, capturing the dependency of that observation modality on the hidden state factors. Similarly, the transition model $\mathbf{B}$ is represented as a collection of $F$ sub-arrays $\mathbf{B}=\lbrace \mathbf{B}^1,\ldots,\mathbf{B}^F\rbrace$, under the assumption that hidden state factors evolve independently without influencing each other. Each $\mathbf{B}^f$, $f=1,\ldots,F$, encodes the dynamics $p(D^f_t\mid D^f_{t-1},u^f_{t-1};\boldsymbol{\phi})$, conditioned on the previous state and action for that factor. Note that control states are factorized analogously to hidden states, such that $U=\lbrace U^1,\ldots,U^F\rbrace$. Each control factor $U^f$ governs the transitions of the corresponding digital state factor $D^f$, with a dimensionality matching the number of possible control actions applicable to that aspect of the system.

This factored structure enables the encoding of complex conditional dependencies while significantly reducing memory requirements. For instance, if the model employs two separate hidden state factors to represent the location and identity of a phenomenon, the memory requirements for the factored representation scale linearly with the dimensionality of the two factors. An additional advantage of this factorization lies in its interpretability: by explicitly designing digital state factors to reflect intuitive features of the environment, the resulting generative model becomes more transparent and modular. In contrast, explicitly enumerating all possible combinations of ``where'' and ``what'' would incur polynomial memory complexity.

The marginal variational posteriors for each hidden state factor at the current time $t_c$ are computed via mean-field fixed-point iteration~\cite{wainwright2008graphical}. The algorithm proceeds by setting the gradient of the VFE $\mathcal{F}_{t_c}(\boldsymbol{\theta})$ to zero, and iteratively solving for each factorized component $Q(D^f_{t_c}; \boldsymbol{\theta})$, for $f=1,\ldots,F$. A detailed derivation of this procedure can be found in~\cite{pymdp}.

Note that in the AIF framework, there is no need to introduce an explicit node for representing quantities of interest, unlike the abstraction of physical--digital systems proposed in~\cite{pgm_wilcox_dt}. In their probabilistic graphical model, these variables are represented by a dedicated node and predicted from the updated digital state via the computational models comprising the DT. In contrast, under the AIF framework, such a node is redundant, as quantities of interest are naturally embedded within the observational data node. When observational evidence is unavailable for a particular modality, inference simply remains uninformed in that dimension of the observation space. Nevertheless, the updated digital state can still be used to predict expected values across any observation channel -- whether observed or unobserved -- via the corresponding observation model. The models may, in principle, incorporate arbitrarily complex forward mappings, ranging from high-fidelity physics-based simulators to purely data-driven surrogates or hybrid combinations of the two.

\subsection{Policy inference-action selection via expected free energy minimization}
\label{sez:EFE}

Given the updated variational posterior over the digital state $Q^*(D_{t_c}; \boldsymbol{\theta}^*)$, policy inference involves evaluating the quality of each admissible policy comprising future actions over a prediction horizon $t = t_c,\ldots,t_p$. In AIF, the desirability of (or preference for) each policy is quantified through the expected free energy (EFE). The EFE is the central quantity driving the behavior of ADTs and is formulated to evaluate sequences of actions (or policies) both on goal-directed (pragmatic) and information-seeking (epistemic) behaviors. Like the VFE, the EFE is a function of observations, hidden states, and policies. However, different from the VFE, it pertains to sequences of \textit{future} actions, where no actual observations are yet available, and it includes expectations over future digital states and future observations generated by the generative model. 

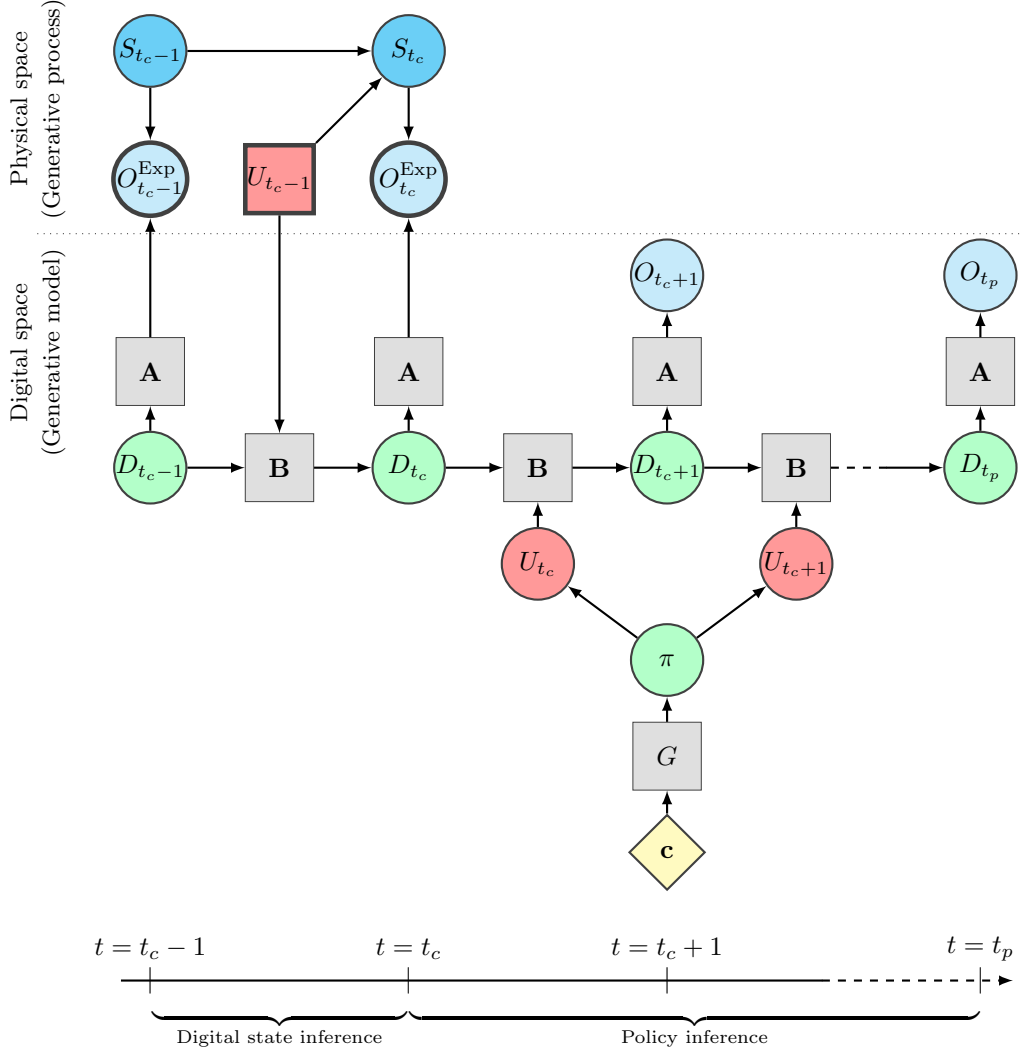
\begin{figure}[!t]
\center
\begin{tikzpicture}[scale=.85, every node/.style={scale=1.}]

\node [D_node] (pi) at (9.5,3) {};
\node [] at (9.5,3) {$\pi$};
\node [op_node] (G) at (9.5,1.5) {};
\node [] at (9.5,1.5) {$G$};
\node [R_node] (C) at (9.5,0) {};
\node [] at (9.5,0) {$\mathbf{c}$};

\node [P_node] (P_0) at (1.5,12.5) {};
\node [] () at (1.5,12.5) {$S_{t_c-1}$};
\node [D_node] (D_0) at (1.5,6) {};
\node [] () at (1.5,6) {$D_{t_c-1}$};
\node [O_node_exp] (O_0) at (1.5,10.5) {};
\node [] at (1.5,10.5) {${O}^\text{Exp}_{t_c-1}$};
\node [op_node] (A_0) at (1.5,7.5) {};
\node [] at (1.5,7.5) {$\mathbf{A}$};
\node [op_node] (B_0) at (3.5,6) {};
\node [] at (3.5,6) {$\mathbf{B}$};
\node [U_node_act] (U_0) at (3.5,10.5) {};
\node [] at (3.5,10.5) {$U_{t_c-1}$};

\node [P_node] (P_1) at (5.5,12.5) {};
\node [] () at (5.5,12.5) {$S_{t_c}$};
\node [D_node] (D_1) at (5.5,6) {};
\node [] () at (5.5,6) {$D_{t_c}$};
\node [O_node_exp] (O_1) at (5.5,10.5) {};
\node [] at (5.5,10.5) {${O}^\text{Exp}_{t_c}$};
\node [op_node] (A_1) at (5.5,7.5) {};
\node [] at (5.5,7.5) {$\mathbf{A}$};
\node [U_node_prob] (U_1) at (7.5,4.5) {};
\node [] at (7.5,4.5) {$U_{t_c}$};

\node [op_node] (B_1) at (7.5,6) {};
\node [] at (7.5,6) {$\mathbf{B}$};

\node [D_node] (D_2) at (9.5,6) {};
\node [] () at (9.5,6) {$D_{t_c+1}$};
\node [O_node] (O_2) at (9.5,9) {};
\node [] at (9.5,9) {${O}_{t_c+1}$};
\node [op_node] (A_2) at (9.5,7.5) {};
\node [] at (9.5,7.5) {$\mathbf{A}$};
\node [op_node] (B_2) at (11.5,6) {};
\node [] at (11.5,6) {$\mathbf{B}$};
\node [U_node_prob] (U_2) at (11.5,4.5) {};
\node [] at (11.5,4.5) {$U_{t_c+1}$};

\node [D_node] (D_3) at (14.35,6) {};
\node [] () at (14.35,6) {$D_{t_p}$};
\node [O_node] (O_3) at (14.35,9) {};
\node [] at (14.35,9) {${O}_{t_p}$};
\node [op_node] (A_3) at (14.35,7.5) {};
\node [] at (14.35,7.5) {$\mathbf{A}$};

\node [] (D_i1) at (2,6) {};
\node [] (D_i2) at (1.7,6) {};
\node [] (t_i1) at (2,1.5) {};
\node [] (t_i2) at (1.7,1.5) {};
\node [] (B_E) at (15.5,6) {};
\node [] (D_i3) at (13.05,6) {};
\node [] (D_i4) at (12.75,6) {};

\draw[-latex,thick,black] (C) to (G);
\draw[-latex,thick,black] (G) to (pi);
\draw[-latex,thick,black] (pi) to (U_1);
\draw[-latex,thick,black] (pi) to (U_2);

\draw[-latex,thick,black] (D_0) to (B_0);
\draw[-latex,thick,black] (D_0) to (A_0);
\draw[-latex,thick,black] (A_0) to (O_0);
\draw[-latex,thick,black] (P_0) to (O_0);
\draw[-latex,thick,black] (P_0) to (P_1);
\draw[-latex,thick,black] (U_0) to (B_0);
\draw[-latex,thick,black] (U_0) to (P_1);

\draw[-latex,thick,black] (B_0) to (D_1);
\draw[-latex,thick,black] (D_1) to (A_1);
\draw[-latex,thick,black] (A_1) to (O_1);
\draw[-latex,thick,black] (D_1) to (B_1);
\draw[-latex,thick,black] (P_1) to (O_1);
\draw[-latex,thick,black] (U_1) to (B_1);

\draw[-latex,thick,black] (B_1) to (D_2);
\draw[-latex,thick,black] (U_2) to (B_2);
\draw[-latex,thick,black] (D_2) to (A_2);
\draw[-latex,thick,black] (A_2) to (O_2);
\draw[-latex,thick,black] (D_2) to (B_2);

\draw[-,thick,dashed,black] (B_2) to (D_i3);
\draw[-latex,thick,black] (D_i4) to (D_3);
\draw[-latex,thick,black] (D_3) to (A_3);
\draw[-latex,thick,black] (A_3) to (O_3);

\node [] (t0) at (0.9,-2) {};
\node [] () at (1.5,-2) {$|$};
\node [] () at (1.5,-1.5) {$t=t_c-1$};
\node [] () at (5.5,-2) {$|$};
\node [] () at (5.5,-1.5) {$t=t_c$};
\node [] () at (9.5,-2) {$|$};
\node [] () at (9.5,-1.5) {$t=t_c+1$};
\node [] () at (14.35,-2) {$|$};
\node [] () at (14.35,-1.5) {$t=t_p$};
\node [] (ti3) at (12.05,-2) {};
\node [] (ti4) at (11.75,-2) {};
\node [] (t2) at (15,-2) {};
\draw[-,thick,black] (t0) to (ti3);
\draw[-latex,dashed,thick,black] (ti4) to (t2);

\node [] () at (3.5,-2.75) {$\underbrace{\hspace{3.35cm}}_{\text{Digital state inference}}$};
\node [] () at (9.925,-2.75) {$\underbrace{\hspace{7.5cm}}_{\text{Policy inference}}$};

\draw[darkgray,dotted] (-0.25,9.65) to (15,9.65);
\node [rotate=90] () at (-0.5,11.6) {\small Physical space};
\node [rotate=90] () at (-0.5,7.8) {\small Digital space};
\node [rotate=90] () at (0,11.6) {\small (Generative process)};
\node [rotate=90] () at (0,7.8) {\small (Generative model)};

\end{tikzpicture}
\vspace{-0.15cm}
\caption{A dynamic Bayesian network illustrating the use of active inference generative models to navigate the partially observable Markov decision process underlying the digital twin problem. Circular nodes represent random variables, red square nodes denote taken actions, gray square nodes represent parametrized operators of the generative model, and the diamond-shaped node symbolizes prior preferences that reflect a goal-directed (pragmatic) objective. Nodes with bold outlines indicate observed quantities, while those with thin outlines represent latent variables to be inferred. Directed edges encode conditional dependencies between variables. The upper left-to-right path represents the evolution of the physical space, while the lower path depicts the evolution of the digital space. Digital state inference is performed at the current time $t_c$, whereas policy inference involves propagating the updated digital state from $t_c$ to the prediction time $t_p$. \label{fig:AIF_POMDP}} 
\vspace{-0.25cm}
\end{figure}

The use of AIF generative models for digital state inference and policy inference is graphically summarized in \fig\ref{fig:AIF_POMDP}. Digital state inference integrates the prior belief at time $t_c-1$ with the observational data assimilated at $t_c$. In contrast, policy inference entails predictive modeling over the horizon $t = t_c,\ldots,t_p$, where the generative model operates without access to future sensory data or executed actions from the interfacing generative process.

The EFE associated to a generic policy $\pi$ is defined as:
\begin{equation}
G^\pi=\mathbb{E}_{Q(O_{t_c:t_p},D_{t_c:t_p}\mid\pi)}\left[\ln{Q(D_{t_c:t_p}\mid\pi)}-\ln{\widetilde{p}(O_{t_c:t_p},D_{t_c:t_p}\mid\pi)}\right],
\label{eq:EFE}
\end{equation}
where, for simplicity, we omit the explicit dependence of the variational posterior on the variational parameters $\boldsymbol{\theta}$, denoting it simply as $Q(D_t)$. Similarly, we omit the dependency of the generative model on the model parameters $\boldsymbol{\phi}$. In \eq\eqref{eq:EFE}, $\widetilde{p}(O_{t}, D_{t}\mid\pi)=p(D_{t}\mid O_{t},\pi)\widetilde{p}(O_{t})$ defines a generative model biased by the predictive prior over observations $\widetilde{p}(O_{t})$. This construction integrates the prior preferences encoded in $\mathbf{c}$ into the inference process (described below), enabling the AIF agent to act in ways that maximize the likelihood of preferred outcomes.

Given the assumed conditional independence of variational posteriors across time, the EFE at a generic time step $t\in\lbrace t_c,\ldots,t_p\rbrace$ for policy $\pi$ is given by:
\begin{equation}
\begin{split}
G_t^\pi&=\mathbb{E}_{Q(O_t,D_t\mid\pi)}\left[\ln{Q(D_{t}\mid\pi)}-\ln{\widetilde{p}(O_t,D_t\mid\pi)}\right]\\
&= -\underbrace{\mathbb{E}_{Q(O_t\mid\pi)}\left[\text{D}_\text{KL}\left[Q(D_{t}\mid O_{t},\pi)\mid\mid Q(D_{t}\mid\pi) \right] \right]}_{\text{Epistemic value (information gain)}} - \underbrace{\mathbb{E}_{Q(O_t\mid\pi)}\left[\ln{\widetilde{p}(O_{t})}\right]}_{\text{Pragmatic value (utility)}}\\
&\hspace{50pt}+\underbrace{\mathbb{E}_{Q(O_t\mid\pi)}\left[\text{D}_\text{KL}\left[Q(D_{t}\mid O_{t},\pi)\mid\mid p(D_{t}\mid O_{t},\pi) \right] \right]}_{\text{Expected variational approximation error ($\geq0$)}},
\end{split}
\label{eq:EFE_2}
\end{equation}
with the complete derivation provided in \ref{sez:efe_eq}, as adapted to the ADT framework from~\cite{pymdp}. In \eq\eqref{eq:EFE_2}, the first term denotes the epistemic value~\cite{Friston02102015}, which promotes information-seeking behavior. It favors policies under which the agent is expected to \textit{explore} states that yield high information gain about the digital state. This gain is quantified as the divergence between predicted digital states conditioned and unconditioned on observations under the same policy. The second term corresponds to the pragmatic value, which reflects goal-directed behavior. It favors policies that lead the agent to states expected to generate outcomes aligned with prior preferences $\widetilde{p}(O_{t})$. The final term captures the expected approximation error -- the divergence between the true digital state posterior and its variational approximation -- which is typically assumed to be negligible.

The epistemic drive in \eq\eqref{eq:EFE_2} is a crucial component that enables ADTs to exhibit spontaneous exploratory behavior. Epistemic actions in ADTs encompass decisions that gather information or improve the digital state observability. These may include, for instance, installing new sensors, scheduling targeted inspections, or testing model predictions. For example, in a manufacturing ADT, the agent might deliberately vary process parameters within safe limits to resolve uncertainty about machine wear dynamics. In a personalized medicine context, the ADT might recommend a low-risk diagnostic test to disambiguate between competing hypotheses about a patient’s physiological condition. In both cases, the primary objective of these actions is not immediate (pragmatic) utility maximization, but rather to refine the generative model and enhance the understanding of the environment. 

The EFE of temporally deep policies is given by the sum of time step-specific contributions:
\begin{equation}
G^\pi=\sum_{t=t_c}^{t_p}G_t^\pi,
\end{equation}
where each term is evaluated based on the agent’s predictive beliefs over future digital states and observations. The computation begins from the current posterior belief $Q^*(D_{t_c})$, which is then propagated over the prediction horizon $t=t_c,\ldots,t_p$ using the policy-specific transition and observation models. This process generates the posterior predictive densities $Q(O_{t_c:t_p}, D_{t_c:t_p}\mid\pi)$, which are subsequently used to evaluate the goal-directed (pragmatic) and information-seeking (epistemic) values at each time step.
 
Let $\Pi=\lbrace\pi_1,\ldots,\pi_P\rbrace$ denote the set of $P$ feasible policies, constructed through the combinatorial enumeration of sequences of actions from the action space $\mathcal{U}$ over the time horizon $t=t_c,\ldots,t_p$. The EFE vector $\mathbf{G}=(G^{\pi_1},\ldots,G^{\pi_P})^\top\in\mathbb{R}^P$, which assigns a scalar EFE to each policy, defines a prior over policies according to:
\begin{equation}
p(\pi) = \sigma(-\gamma\mathbf{G}),
\label{eq:EFE_prior}
\end{equation}
where $\sigma(x)=\frac{\exp(x)}{\sum_x\exp(x)}$ is the Softmax function, and $\gamma\in\mathbb{R}^+$ is an inverse temperature parameter that modulates the precision over policies. Higher $\gamma$ values yield more deterministic preferences.

Under the prior \eqref{eq:EFE_prior}, AIF agents perform policy inference by optimizing a variational posterior over policies $Q(\pi)$~\cite{pymdp}, to minimize the following VFE expansion over the prediction horizon $t=t_c,\ldots,t_p$:
\begin{equation}
\begin{split}
\mathcal{F}_{t_c:t_p}&=\mathbb{E}_{Q(D_{t_c:t_p},\pi)}\left[\ln{Q(D_{t_c:t_p},\pi)}-\ln{p(O_{t_c:t_p}, D_{t_c:t_p},\pi)}\right]\\
&=\mathbb{E}_{Q(D_{t_c:t_p},\pi)}\left[\ln{Q(D_{t_c:t_p}\mid\pi)}+\ln{Q(\pi)}-\ln{p(O_{t_c:t_p}, D_{t_c:t_p}\mid\pi)}-\ln{p(\pi)}\right]\\
&=\mathbb{E}_{Q(\pi)}\left[\ln{Q(\pi)}-\ln{p(\pi)}\right]\\
&\hspace{50pt}+\mathbb{E}_{Q(\pi)}\left[\mathbb{E}_{Q(D_{t_c:t_p}\mid\pi)}\left[\ln{Q(D_{t_c:t_p}\mid\pi)}-\ln{p(O_{t_c:t_p}, D_{t_c:t_p}\mid\pi)}\right]\right]\\
&=\text{D}_\text{KL}\left[Q(\pi)\mid\mid p(\pi) \right]+\mathbb{E}_{Q(\pi)}\left[\mathcal{F}^\pi_{t_c:t_p}\right],
\end{split}
\label{eq:VFE_2}
\end{equation}
which measures the KL divergence between the approximate posterior $Q(D_{t_c:t_p},\pi)$ and the generative model $p(O_{t_c:t_p},D_{t_c:t_p},\pi)$ as a sum of two contributions. The first is the KL divergence between the variational posterior over policies and the corresponding prior \eqref{eq:EFE_prior}, thereby incorporating the EFE into the inference process. The second term is a policy-weighted average of the free energy across all policies, where $\mathcal{F}^\pi_{t_c:t_p}$ denotes the free energy associated with a single policy $\pi$:
\begin{equation}
\begin{split}
\mathcal{F}^\pi_{t_c:t_p}&=-\mathbb{E}_{Q(D_{t_c:t_p}\mid\pi)}\left[\ln{p(O_{t_c:t_p}, D_{t_c:t_p}\mid\pi)}-\ln{Q(D_{t_c:t_p}\mid\pi)}\right]\\
&=-\mathbb{E}_{Q(D_{t_c:t_p}\mid\pi)}\left[\ln{p(O_{t_c:t_p}, D_{t_c:t_p}\mid\pi)}\right]-H\left[Q(D_{t_c:t_p}\mid\pi)\right],
\end{split}
\end{equation}
with $H\left[Q(D_{t_c:t_p}\mid\pi)\right] = \mathbb{E}_{Q(D_{t_c:t_p}\mid\pi)} \left[-\ln{Q(D_{t_c:t_p}\mid\pi)}\right]$ being the variational posterior entropy, which quantifies the uncertainty in the beliefs about future digital states under policy $\pi$.

By evaluating each policy independently and computing its associated free energy, the optimal posterior $Q^*(\pi)$ is obtained by minimizing the total VFE $\mathcal{F}_{t_c:t_p}$ with respect to $Q(\pi)$. This is achieved by enforcing the stationarity of $\mathcal{F}_{t_c:t_p}$ with respect to $Q(\pi)$, leading to a Softmax distribution through the following update rule:
\begin{equation}
Q^*(\pi)=\argmin_{Q(\pi)} \mathcal{F}_{t_c:t_p} = \sigma(\ln{p(\pi)}-\mathcal{F}^\pi_{t_c:t_p}),
\end{equation}
assigning higher probability to policies with lower free energy while remaining close to the prior. 

The posterior over control states $Q^*(U_{t})$ is formed by marginalizing over policies as follows:
\begin{equation}
Q^*(U_{t})=\sum_{\pi\in\Pi}p(U_t \mid \pi)Q^*(\pi),
\end{equation}
where $p(U_t \mid \pi)$ defines a deterministic mapping from policies to control states. The actual action $U_{t_c}=u_{t_c}$ to be executed on the system can eventually be selected either as the maximum a-posteriori estimate or by sampling from $Q^*(U_{t_c})$.

\subsection{Learning of the generative model via parameter inference}
\label{sez:learning}
In this section, we describe the learning of the parameters $\boldsymbol{\phi}$ that define the AIF generative model, based on the outcomes of inference. ``Learning'' $\boldsymbol{\phi}$ is a generative model's parameter updating occurring at a slower timescale than the faster inference processes for digital states and policies. Nevertheless, the update equations for $\boldsymbol{\phi}$ follow the same variational principles of digital state inference, where a variational posterior over $\boldsymbol{\phi}$ is optimized through VFE minimization.

In our discrete setting, posterior inference over $\boldsymbol{\phi}$ is performed by parametrizing the likelihood and prior distributions of the generative model with Dirichlet distributions, following an approach similar to~\cite{tezzele2024adaptive,Torzoni_DT2}. The choice to treat model parameters $\boldsymbol{\phi}$ as the parameters of Dirichlet distributions is motivated by their conjugacy to the categorical distribution. This formulation enables online learning via closed-form Bayesian updates, allowing evidence about the system response to actions to be incorporated efficiently, while ensuring that the posterior remains within the Dirichlet family. The approach is computationally scalable and supports continual refinement of ADTs, even when initialized with potentially inaccurate or uncertain priors and likelihoods.

By decomposing $\boldsymbol{\phi}$ into subsets corresponding to the categorical and Dirichlet parameters associated with the arrays $\mathbf{A}$, $\mathbf{B}$, and $\mathbf{d}$, we can highlight the stochastic parametrization underlying each likelihood and prior distribution. In this work, we focus on learning the parameters of the $\mathbf{B}$ array, although the same reasoning can be applied identically to $\mathbf{A}$ and $\mathbf{d}$. The stochastic parametrization of the transition model $\mathbf{B}\in\mathbb{R}^{\mid\mathcal{D}\mid\times\mid\mathcal{D}\mid\times\mid\mathcal{U}\mid}$, encoding the transition distribution $p(D_t\mid D_{t-1},u_{t-1};\mathfrak{B})$, is given by the tensor of categorical parameters \mbox{$\mathfrak{B}\in\mathbb{R}^{\mid\mathcal{D}\mid\times\mid\mathcal{D}\mid\times\mid\mathcal{U}\mid}$}, such that
\begin{align}
D_t\mid D_{t-1},u_{t-1};\mathfrak{B}&\sim\text{Cat}(\mathfrak{B}),\\
p(\mathfrak{B})&=\prod_{d\in\mathcal{D}}\prod_{u\in\mathcal{U}} p(\mathfrak{B}_{\bullet,d,u}),\quad \mathfrak{B}_{\bullet,d,u}\sim\text{Dir}(\mathfrak{b}_{\bullet,d,u}),
\end{align}
where $|\mathcal{X}|\in\mathbb{N}$ denotes the cardinality of a generic set $\mathcal{X}$; $\text{Cat}(\bullet)$ and $\text{Dir}(\bullet)$ denote categorical and Dirichlet distributions, respectively; the notation $\mathbf{X}_{\bullet,j,k}$ refers to the $j$th column of the $k$th slice of a generic tensor $\mathbf{X}$. The tensor $\mathfrak{b} \in\mathbb{R}^{\mid\mathcal{D}\mid\times\mid\mathcal{D}\mid\times\mid\mathcal{U}\mid}$ collects the (positive) concentration parameters defining the Dirichlet prior over $\mathfrak{B}$. These parameters can be interpreted as pseudo-counts representing prior beliefs about the frequency of each state transition given a digital state and an action. For notational simplicity, we assume the generative model is not factorized into multiple digital state factors or observation modalities. However, the formulation extends naturally to multi-modal, multi-factor settings by introducing additional parametrized dimensions.

By specializing $\boldsymbol{\phi}$ as the parametrization of $\mathbf{B}$, the generative model~\eqref{eq:generative} becomes:
\begin{equation}
p(O_{t_c:t_p},D_{t_c:t_p},\mathfrak{B},\pi)=p(\mathfrak{B})p(\pi)p(D_{t_c})\prod_{t=t_c+1}^{t_p}p(D_t\mid D_{t-1},\pi;\mathfrak{B})\prod_{t=t_c}^{t_p}p(O_t\mid D_t).
\label{eq:generative_2}
\end{equation}

Learning is formulated as the approximate inference of  $\mathfrak{B}$ by minimizing the VFE with respect to its approximate posterior $Q(\mathfrak{B})$. The full variational posterior is assumed to factorize as:
\begin{equation}
Q(D_{t_c:t_p},\mathfrak{B},\pi)=Q(\mathfrak{B})Q(\pi)\prod_{t=t_c}^{t_p}Q(D_{t}\mid\mathfrak{B},\pi),
\label{eq:full_posterior}
\end{equation}
where the variational distribution $Q(\mathfrak{B})$ is modeled as
\begin{align}
Q(\mathfrak{B})&=\prod_{d\in\mathcal{D}}\prod_{u\in\mathcal{U}} Q(\mathfrak{B}_{\bullet,d,u}), &Q(\mathfrak{B}_{\bullet,d,u})&=\text{Dir}
(\widehat{\mathfrak{b}}_{\bullet,d,u}),
\end{align}
and $\widehat{\mathfrak{b}}\in\mathbb{R}^{\mid\mathcal{D}\mid\times\mid\mathcal{D}\mid\times\mid\mathcal{U}\mid}$ plays the same role as $\mathfrak{b}$ in parametrizing the Dirichlet distribution, while being treated as variational parameters to be optimized. Accordingly, the full VFE objective for the generative model \eqref{eq:generative_2} is given by:
\begin{equation}
\mathcal{F}_{t_c:t_p}=\mathbb{E}_{Q(D_{t_c:t_p},\mathfrak{B},\pi)}\left[\ln{Q(D_{t_c:t_p},\mathfrak{B},\pi)}-\ln{p(O_{t_c:t_p}, D_{t_c:t_p},\mathfrak{B},\pi)}\right],
\end{equation}
which decomposes into sums of KL and expected log-likelihood terms following the factorization induced by the generative model and the variational posterior.

The update rule for the Dirichlet parameters $\mathfrak{b}$ directly follows from conjugacy. Specifically, given the Dirichlet prior parameters $\mathfrak{b}$ over the generative model, the digital state posterior $Q^*(D_{t_c})$ at the current time step $t_c$, the previous digital state posterior $Q^*(D_{t_c-1})$, and the action \mbox{$U_{t_c-1}=u_{t_c-1}$} taken at the previous time step, the  fixed-point update rule for the variational posterior Dirichlet parameters $\widehat{\mathfrak{b}}$ is:
\begin{equation}
\widehat{\mathfrak{b}}_{\bullet,\bullet,u_{t_c-1}}^*=\mathfrak{b}_{\bullet,\bullet,u_{t_c-1}}+\eta(Q^*(D_{t_c})\otimes Q^*(D_{t_c-1})),
\end{equation}
which corresponds to an update applied to the $u_{t_c-1}$th slice of $\widehat{\mathfrak{b}}$, where $\otimes$ denotes the outer product, and $\eta\in\mathbb{R}$, with $0\leq\eta\leq1$, is a learning rate parameter that scales the update step.

\subsection{Epistemic behavior of active digital twins}
\label{sez:active_behav}

If the agent also maintains a variational posterior over the model parameters $Q(\boldsymbol{\phi})$, as discussed in \sez\ref{sez:learning}, the EFE expression \eqref{eq:EFE_2} can be extended to capture the epistemic value associated with expected information gain not only over digital states but also over $\boldsymbol{\phi}$. The full derivation is provided in \ref{sez:efe_eq}, and it reveals an additional epistemic term that quantifies the expected information gain about the parameters governing the categorical distributions associated with the  $\mathbf{A}$, $\mathbf{B}$, and $\mathbf{d}$ arrays. When the AIF agent maintains and updates beliefs over these parameters, this term steers policy inference toward action-observation trajectories that are expected to produce informative updates to the generative model. We point out that the epistemic value over $\boldsymbol{\phi}$ is not exploited in the numerical demonstrations presented in \sez\ref{sez:results}. Nevertheless, it is retained in the formulation of the ADT framework, as this capability may enable essential functionalities depending on the context and specific application objectives.

Epistemic actions aimed at refining the generative model can be regarded as forms of autonomous calibration, wherein the ADT steers its operation into underexplored regimes or perturbs its environment to test and improve its generative model. For instance, a sensor might be temporarily activated solely to evaluate its reliability while updating a likelihood model deemed uncertain. This behavior highlights the distinction between passive and active learning: while passive learning consists in assimilating externally provided or randomly encountered data, active learning reflects the strategic initiation of data acquisition to accelerate model refinement and enhance future decision-making. The information-seeking (epistemic) behavior of ADTs thus emerges from their capacity for self-adaptive inference and learning, pursued alongside goal-directed (pragmatic) objectives. This dual optimization is embedded in policy inference through EFE minimization, which unifies goal-directed exploration and utility maximization within a single computational framework that moves beyond the passive replication of physical systems.

\subsection{Algorithmic description}
\label{sez:alg}

An algorithmic description of a single step of the AIF loop for ADTs is provided in \alg\ref{alg:algo}. Given the generative model, an observation sampled from the generative process, the posterior over digital states from the previous time step, and the action taken at the previous time step, one step of the loop involves: (1) performing inference over digital states based on the new observation; (2) using the posterior belief over digital states to perform policy inference and select the next action; (3) updating the generative model through learning informed by inference results.

\begin{algorithm}[t]
\hspace*{\algorithmicindent} \textbf{input}: 
generative model $\langle\mathbf{A},\mathbf{B},\mathbf{c},\mathbf{d}\rangle$\\
\hspace*{49pt} assimilated observation $O^\text{Exp}_{t_c}=o^\text{Exp}_{t_c}$\\
\hspace*{49pt} digital state posterior $Q^*(D_{t_c-1})$ at previous time step\\
\hspace*{49pt} action $U_{t_c-1}=u_{t_c-1}$ executed at previous time step
\begin{algorithmic}[1]
\Statex
\Statex$\triangleright$ digital state inference by minimizing variational free energy $\mathcal{F}_{t_c}$
\State infer digital state posterior $Q^*(D_{t_c})$
\Statex
\Statex$\triangleright$ policy inference and action selection by minimizing future variational free energy $\mathcal{F}_{t_c:t_p}$
\State compute posterior predictive distributions $Q(O_{t_c:t_p}, D_{t_c:t_p}\mid\pi)$
\State evaluate epistemic and pragmatic values over $t=t_c,\ldots t_p$ under each policy
\State infer control policies posterior $Q^*(\pi)$
\State select action $U_{t_c}=u_{t_c}$ by taking the best-point estimate or sampling from $Q^*(U_{t_c})$
\Statex
\Statex$\triangleright$ learning by minimizing variational free energy $\mathcal{F}_{t_c}$
\State update transition model $\mathbf{B}$ by updating the variational posterior Dirichlet parameters $\widehat{\mathfrak{b}}$
\Statex
\end{algorithmic}
\hspace*{\algorithmicindent} \Return updated generative model $\langle\mathbf{A},\mathbf{B},\mathbf{c},\mathbf{d}\rangle$\\ 
\hspace*{50.4pt} updated posterior distribution over control policies $Q^*(\pi)$\\
\hspace*{50.4pt} control action to be executed $U_{t_c}=u_{t_c}$\\
\hspace*{50.4pt} posterior predictive density over digital states $Q(D_{t_c:t_p})$\\
\hspace*{50.4pt} posterior predictive density over actions  $Q(U_{t_c:t_p})$
\caption{Active inference loop for active digital twins.}
\label{alg:algo}
\end{algorithm}

\section{Numerical demonstrations}
\label{sez:results}

This section demonstrates the proposed methodology through the simulated monitoring, management, and maintenance planning of the H{\"o}rnefors railway bridge~\cite{kth3}. Although this case study focuses specifically on structural health monitoring (SHM), the underlying framework broadly applies to a wide range of systems or domains. 

Section~\ref{sez:physical} introduces the monitored physical asset. Section~\ref{sez:assembly} describes the composition of the handled vibration data and the numerical models used to generate labeled examples under various damage scenarios. Section~\ref{sez:assimilation} outlines the assimilation of observational data for structural health identification using artificial neural networks. Section~\ref{sez:gen_model} details the step-by-step construction of the AIF generative model, namely the four-tuple $\langle\mathbf{A},\mathbf{B},\mathbf{c},\mathbf{d}\rangle$. Section~\ref{sez:results_1} presents ADT simulations results under purely goal-directed behavior, which serve as a baseline for comparison with simulations involving mixed pragmatic-epistemic behavior, discussed in \sez\ref{sez:results_2}. Finally, \sez\ref{sez:results_3} reports additional results assessing the robustness of the SHM framework to incomplete data streams and to uncertainty in the extent of the damaged region.

The AIF agents based on discrete, Markovian generative models have been simulated using the open-source \texttt{Python} library \texttt{pymdp}~\cite{pymdp}. Compared to other AIF libraries, such as the \texttt{MATLAB} toolbox \texttt{DEM} ~\cite{smith2022step} and the \texttt{C++} library \texttt{cpp-AIF}~\cite{gregoretti2024cpp}, \texttt{pymdp} offers notable advantages in terms of user-friendliness, flexibility, and customizability, although featuring lower process representation (\texttt{DEM}) and less computational efficiency (\texttt{cpp-AIF}). The simulations have been run on a PC featuring an \texttt{Intel\textregistered~Core\textsuperscript{TM} i9-14900KF} CPU @ 3.2 GHz and 64 GB RAM.

\subsection{Physical asset}
\label{sez:physical}
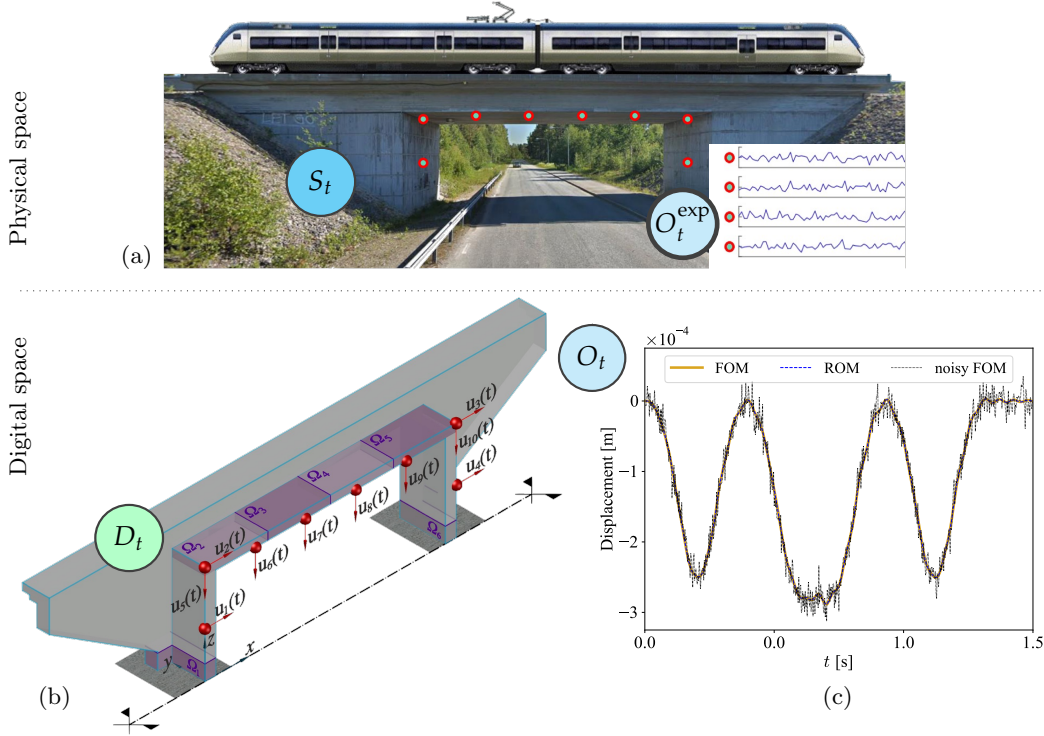
\begin{figure}[!t]
\hspace{-0.2cm}
\begin{tikzpicture}[scale=.9, every node/.style={scale=1.}]

\node[draw=none,fill=none] at (0,0){\includegraphics[width=.9\textwidth, angle=0]{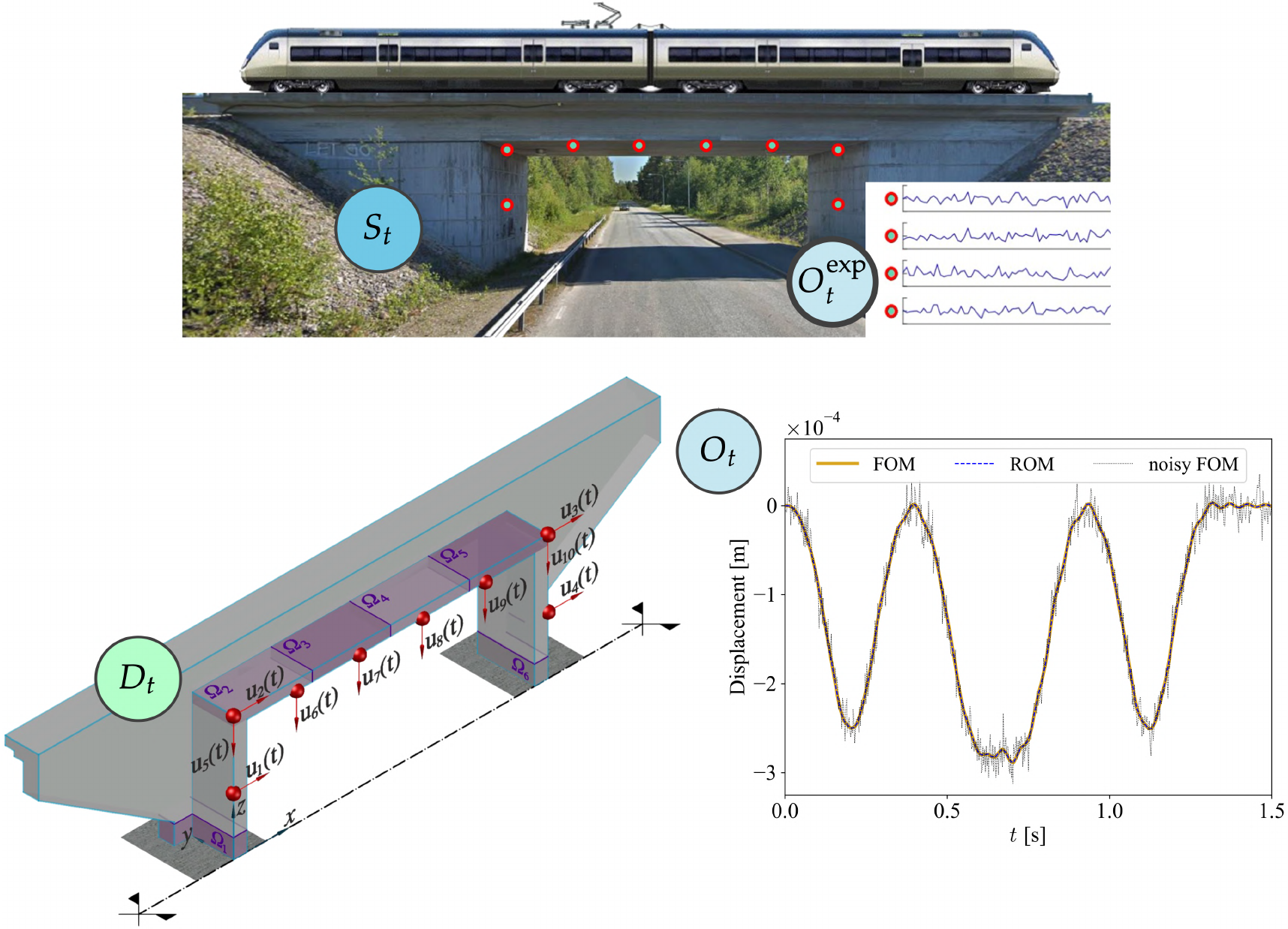}};

\draw[darkgray,dotted] (-7.5,1.15) to (7.5,1.15);
\node [rotate=90] () at (-7.5,2.95) {\small Physical space};
\node [rotate=90] () at (-7.5,-0.55) {\small Digital space};
\node [] () at (-5.8,1.65) {\small (a)};
\node [] () at (-7,-4.85) {\small (b)};
\node [] () at (4.45,-4.85) {\small (c)};

\end{tikzpicture}
\vspace{-0.15cm}
\caption{Physical asset and its digital twin. (a) The physical space corresponds to the H{\"o}rnefors bridge. (b) The digital space represents a structural health monitoring schematization, including details of synthetic recordings related to displacements $u_1(t),\ldots,u_{10}(t)$, and predefined damage regions $\Omega_1,\ldots,\Omega_6$. (c) Exemplary vertical displacement time history at midspan, comparing full-order model (FOM), reduced-order model (ROM), and noisy FOM approximations.} 
\label{fig:bridge_plus_digital}
\vspace{-0.25cm}
\end{figure}

The H{\"o}rnefors railway bridge, shown in \fig\ref{fig:bridge_plus_digital}(a),
is an integral reinforced concrete structure along the Swedish Bothnia line. It spans $15.7~\text{m}$, with a clearance height of $4.7~\text{m}$ and a width of $5.9~\text{m}$ (excluding edge beams). The main structural elements have a thickness of $0.5~\text{m}$ for the deck, $0.7~\text{m}$ for the frame walls, and $0.8~\text{m}$ for the wing walls. The foundation system comprises two slabs connected by stay beams, supported by pile groups. The concrete is of grade C35/45, characterized by the following material properties: Young’s modulus $E=34~\text{GPa}$, Poisson’s ratio $\nu= 0.2$, and density $\rho=2500~\text{kg/m}^3$. The bridge supports a single railway track with sleepers spaced at $0.65~\text{m}$ intervals, resting on a ballast layer that is $0.6~\text{m}$ deep and $4.3~\text{m}$ wide, with a density of $\rho_B=1800~\text{kg/m}^3$. The structure is subjected to dynamic loading from \textit{Gr{\"o}na T{\r a}get} trains operating at speeds between $v\in[160,215]~\text{km/h}$. We specifically consider configurations involving two-car trainsets, totaling eight axles, with each axle bearing a mass of $\psi\in[16,22]~\text{ton}$. The geometrical and mechanical parameters, as well as the moving load model, are adapted from~\cite{metodologico}. The physical state space $\mathcal{S}$ represents the ground-truth variability in the bridge structural health.

\subsection{Offline data assembly}
\label{sez:assembly}
The bridge monitoring system provides displacement data in the form of multivariate time series, denoted as $\mathbf{U}(\boldsymbol{\mu})=[\mathbf{u}_1(\boldsymbol{\mu}),\ldots,\mathbf{u}_{N_s}(\boldsymbol{\mu})]\in\mathbb{R}^{L\times N_s}$. These consist of $N_s=10$ individual time series corresponding to the degrees of freedom (dofs) indicated in \fig\ref{fig:bridge_plus_digital}(b). Each series contains $L$ samples equally spaced over the time interval $[0,1.5~\text{s}]$, acquired with a sampling frequency of $400~\text{Hz}$. The vector $\boldsymbol{\mu} \in\mathbb{R}^{N_\text{par}}$ collects $N_\text{par}$ control parameters, which are assumed to represent the operational and damage conditions. For the problem settings we consider, each observation spans a relatively short time interval, within which these conditions are regarded as constant.

We simulate the monitored asset using a physics-based computational model. Specifically, the structure is modeled as a linear-elastic continuum under the assumption of linearized kinematics, and the equations of elasto-dynamics describe its dynamic response to train transits. The model is spatially discretized using linear tetrahedral finite elements, and its solution is advanced in time to generate synthetic observational data, controlled by the parameter vector $\boldsymbol{\mu}$. 

The full-order model (FOM) is described in detail in~\cite{Torzoni_DT}; here, we summarize its key features. The finite element mesh consists of elements with a nominal size of $0.8~\textup{m}$, refined to $0.15~\textup{m}$ along the deck, resulting in a total of $17,292$ dofs. The ballast layer is accounted for by increasing the density of the deck and edge beams to represent an equivalent mass. Embankment effects are captured using distributed springs applied along the surfaces in contact with the ground, implemented via a Robin-type boundary condition with an elastic coefficient of $10^{8}~\textup{N/m}^3$. Structural damping is introduced using Rayleigh damping, calibrated to yield a $5\%$ damping ratio in the first two vibrational modes. The dynamic response is computed over the time interval $[0,1.5~\text{s}]$, uniformly partitioned into $L=600$ time steps, using an implicit Newmark time integration scheme~\cite{hughes2000finite}.

Damage-induced variations in the structural dynamic response are modeled as localized reductions in effective stiffness. Assuming that each observation spans a time window short enough compared to the timescale of damage progression, the structural behavior can be treated as linear within that interval. This enables a separation of timescales between the slow evolution of damage and the structural health assessment~\cite{Azam_Mariani}. While the precise damage mechanisms are typically confirmed through on-site inspections following early detection, the degradation patterns in integral bridges that can be described in this way include: cracking in concrete due to thermal gradients, freeze-thaw cycles, or overloading; progressive deterioration from alkali-silica reactions, which may lead to cracking and spalling; cracking from stress concentrations caused by differential settlements; and surface erosion from prolonged environmental exposure.

The digital state space $\mathcal{D}$ includes a set of predefined configurations of damage presence, location, and severity. These are modeled by parametrizing the stiffness matrix using two variables $y\in\mathbb{N}$ and $\delta\in\mathbb{R}$, both included in the parameter vector $\boldsymbol{\mu}$. The discrete variable \mbox{$y\in\lbrace0,\ldots,6\rbrace$} designates the damage region, with $y=0$ denoting the undamaged baseline. For the damage cases $y=1,\ldots,6$, we consider $N_\Omega=6$ predefined subdomains $\Omega_m$, for $m=1,\ldots,6$, each representing a potential damage location as shown in \fig\ref{fig:bridge_plus_digital}. Within each subdomain, the material stiffness may be reduced by a factor $\delta\in[30\%,80\%]$, which remains constant throughout the passage of a train.

To reduce the computational cost of solving the FOM for arbitrary values of $\boldsymbol{\mu}$, we employ a projection-based reduced-order model (ROM). The reduction is performed using a Galerkin reduced basis method~\cite{RB, chinestaenc2017}, relying on a low-dimensional set of basis functions computed through proper orthogonal decomposition. Following the method of snapshots~\cite{sirovich}, the ROM is constructed upon $400$ FOM solutions for different configurations of the input parameters $\boldsymbol{\mu} = (v,\psi,y,\delta)^\top$, which are taken as uniformly distributed and sampled via the Latin hypercube rule. The dimension of the reduced-order expansion is determined based on an energy retention criterion. By setting a tolerance of $10^{-3}$ for the fraction of discarded energy, the number of dofs is reduced to $133$. Both the FOM and ROM have been implemented in the \texttt{Matlab} environment, using the \texttt{redbKIT} library~\cite{Redbkit}. For a more detailed description, the reader is referred to~\cite{Torzoni_DT}.

A representative example of displacement time histories is reported in \fig\ref{fig:bridge_plus_digital}(c), showing the vertical displacement at midspan obtained from both the FOM and ROM. To emulate measurement noise and assess its potential impact on the handled structural response, signals are corrupted with additive Gaussian noise, yielding a signal-to-noise ratio of $120$.

\subsection{Data assimilation via artificial neural networks}
\label{sez:assimilation}
The vibration recordings are assimilated for structural health diagnostics by leveraging the flexibility of deep learning (DL) models for SHM applications, as demonstrated in~\cite{Torzoni_MF,temperatura,Torzoni_EWSHM}.

Data-driven approaches to SHM follow a pattern recognition paradigm~\cite{Bishop}, in which damage is assessed by comparing measurements with data previously collected under known structural conditions. This process relies on two key components: (i) feature selection and extraction, and (ii) statistical modeling to associate these features with specific damage patterns~\cite{Farrar01}. A major challenge lies in identifying damage-sensitive features that remain robust under varying operational and environmental conditions. DL offers an automated alternative for selecting and extracting optimized features by capturing temporal correlations within and across time series data~\cite{Avci_review,fink2020potential}.

In our framework, each time a train crosses the bridge, the vibration recordings $\mathbf{U}$ are initially processed by a DL classifier, which outputs confidence scores indicating the likelihood that $\mathbf{U}$ corresponds to each damage class defined by the $y$ parameter. The class with the highest confidence is selected as the best-point estimate for categorizing the measurements. Whenever damage is detected and localized within a region $\Omega_m$, $m=1,\ldots,6$, the vibration recordings $\mathbf{U}$ are further processed by a dedicated regression model -- one for each damageable region -- to estimate the severity of damage $\delta$. These initial estimates are then incorporated into the AIF framework as assimilated observation $O^\text{Exp}_{t_c}=o^\text{Exp}_{t_c}$, as detailed below. 
 
The DL architectures have been implemented through the \texttt{Tensorflow}-based \texttt{Keras} API~\cite{chollet2015keras}, and trained on a single \texttt{Nvidia GeForce RTX\textsuperscript{TM} 3080} GPU card. The training has been performed in a supervised fashion using $10,000$ noisy data instances generated from ROM simulations. For a comprehensive description the reader is referred to~\cite{Torzoni_DT}. 

\subsection{Active digital twin framework}
\label{sez:gen_model}

The outcomes from the DL models are integrated into our POMDP framework by discretizing the range of $\delta$ into $N_\delta=6$ intervals: $\lbrace[30\%,35\%],$ $[35\%,45\%],$ $[45\%,55\%],$ $[55\%,65\%],$ $[65\%,75\%],$ $[75\%,80\%]\rbrace$. This discretization results in a total of $N_\Omega N_\delta+1 =37$ possible damage scenarios, each specifying a combination of damage location and severity. By ordering them first by location and then by severity, this post-processed output constitutes the first observation modality $O^{\Omega\delta}$.

The observation space $\mathcal{O}$ is completed with a second observation modality $O^{u}$ corresponding to the action taken prior to data assimilation, such that $O=\lbrace O^{\Omega\delta},O^{u}\rbrace$. Including this additional perceptual channel provides two key benefits. First, it enables prior preferences (via the $\mathbf{c}$ array) to account not only for the costs associated with structural health states but also for those linked to actions. Second, as detailed below, it naturally supports the formulation of an action-conditioned observation model, introducing an inductive bias that facilitates digital state identification.

The digital state space $\mathcal{D}$ is structured into three factors $D=\lbrace D^\Omega,D^\delta,D^\text{Epi}\rbrace$, corresponding to: the damage location $D^\Omega=\lbrace\Omega_1,\ldots,\Omega_6\rbrace$; the discretized percentage reduction in material stiffness $D^\delta=\lbrace0\%,$ $[30\%,35\%],$ $[35\%,45\%],$ $[45\%,55\%],$ $[55\%,65\%],$ $[65\%,75\%],$ $[75\%,80\%]\rbrace$; and an epistemic switch $D^\text{Epi}=\lbrace\texttt{Epi},$ $\texttt{Non-Epi}\rbrace$, which indicates whether the AIF agent is likely to engage in information-seeking (epistemic) behavior. This third factor allows the agent to autonomously switch between acting as an active information seeker or as a utility maximizer that is confident in its beliefs. It is worth noting that this factorization is neither the only viable option nor necessarily the most appropriate. This reflects the inherent subjectivity involved in shaping the digital state space. For example, $D^\Omega$ and $D^\delta$ could have been merged into a single enumerated representation, similar to the one used for $O^{\Omega\delta}$, at the expense of increased computational complexity and reduced interpretability. Alternatively, a more expressive but computationally demanding option would involve defining six separate $D^\delta$ factors, one for each of the $N_\Omega$ damageable regions. Finally, note that including $D^\text{Epi}$ is essential to enable epistemic behavior, as this factor leads to distinct observation models associated with the $\texttt{Epi}$ and $\texttt{Non-Epi}$ states, as discussed further below.

The action space $\mathcal{U}$ comprises four control actions, each producing specific effects:
\begin{enumerate}[leftmargin=1.5em]
    \item \textit{Do nothing} (DN): the structural health state evolves according to a stochastic deterioration process, while regular revenue is maintained.
    \item \textit{Maintenance} (MA): a high-cost maintenance intervention is executed to mitigate existing damage. Although this action improves the structural condition, it may not fully restore the system to a pristine (damage-free) state.
    \item \textit{Restrict operations} (RO): traffic is limited to lightweight trains with axle load below $18~\text{ton}$, thereby reducing the rate of structural degradation. However, this also leads to a reduction in the revenue generated by the infrastructure.
    \item \textit{Read sensors} (RE): a moderate-cost, high-fidelity sensing action is performed to resolve uncertainty in the structural health state. This action provides high epistemic value by decreasing the entropy of the digital state posterior, thus increasing the mutual information between latent states and expected observations. This effect reflects the use of high-quality sensors, controlled forced vibration tests, or in-situ inspection. From the perspective of the generative model, performing an inspection is equivalent to reading vibration recordings from sensors, albeit with significantly higher information content and a corresponding higher cost.
\end{enumerate}

The observation likelihood array $\mathbf{A}=\lbrace\mathbf{A}^{\Omega\delta},\mathbf{A}^{u}\rbrace$ comprises two observations models: \linebreak\mbox{$\mathbf{A}^{\Omega\delta}\in\mathbb{R}^{\mid O^{\Omega\delta}\mid\times\mid D^\Omega\mid\times\mid D^\delta\mid\times\mid D^\text{Epi}\mid}$} and $\mathbf{A}^{u}\in\mathbb{R}^{\mid O^{u}\mid\times\mid D^\Omega\mid\times\mid D^\delta\mid\times\mid D^\text{Epi}\mid}$, respectively encoding the conditional sensory likelihoods $p(O^{\Omega\delta}\mid D^\Omega,D^\delta,D^\text{Epi})$ and $p(O^{u}\mid D^\Omega,D^\delta,D^\text{Epi})$ for the first and second observation modalities. Conceptually, these tensors are designed to answer two distinct questions: (i) what might the agent believe about the pre-classified signals? and (ii) what might the agent infer about its previous action?

\begin{figure}[!t]
\centering
\subfloat[\label{fig:A_1_epi}]{\includegraphics[width=.45\textwidth]{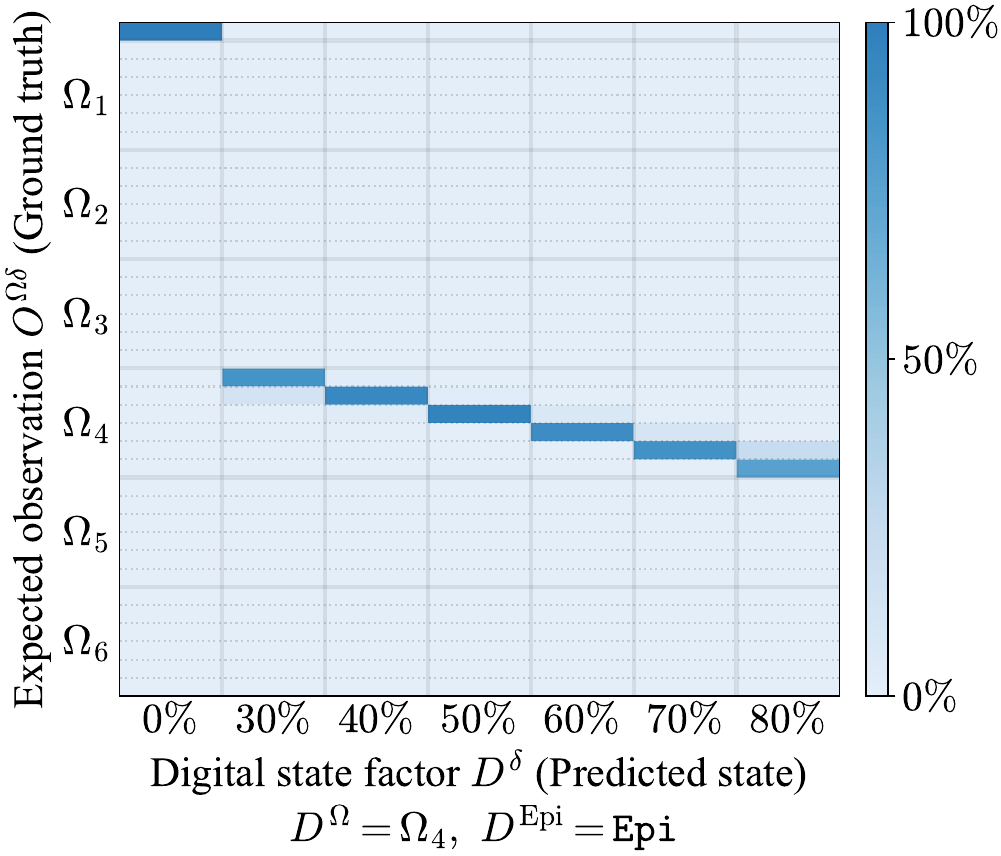}}\hspace{1cm}\subfloat[\label{fig:A_1_noepi}]{\includegraphics[width=.45\textwidth]{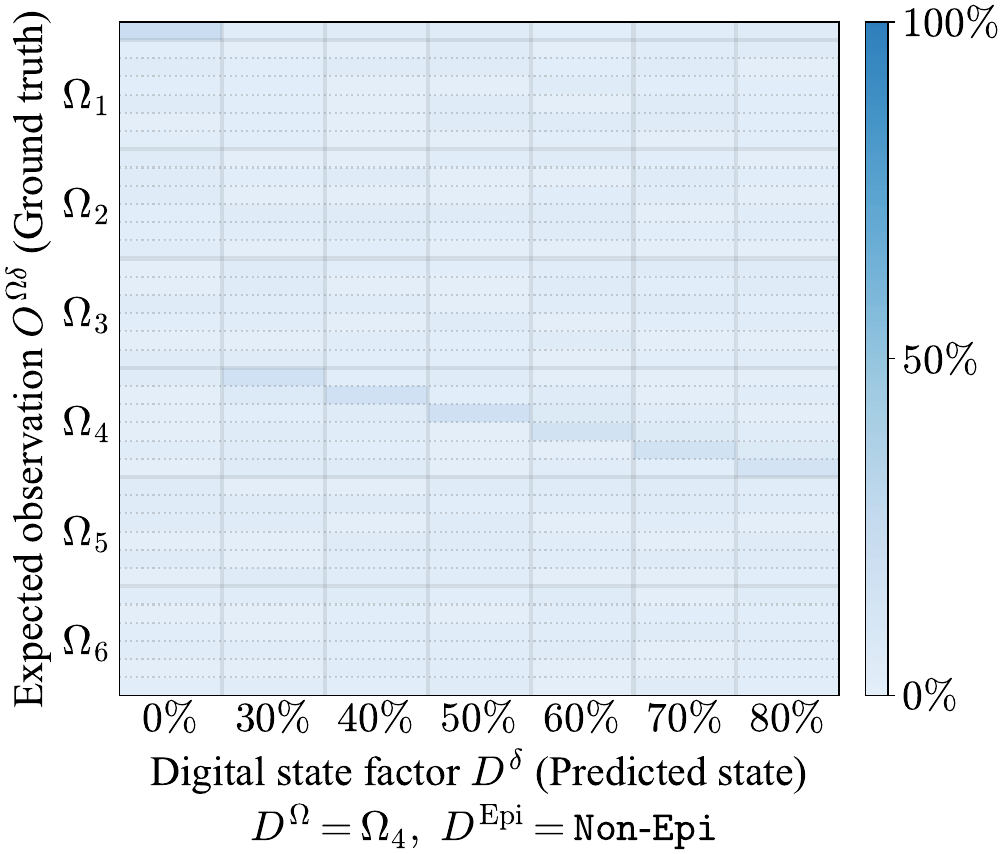}}\\
\vspace{-0.25cm}
\subfloat[\label{fig:A_2_epi}]{\includegraphics[width=.45\textwidth]{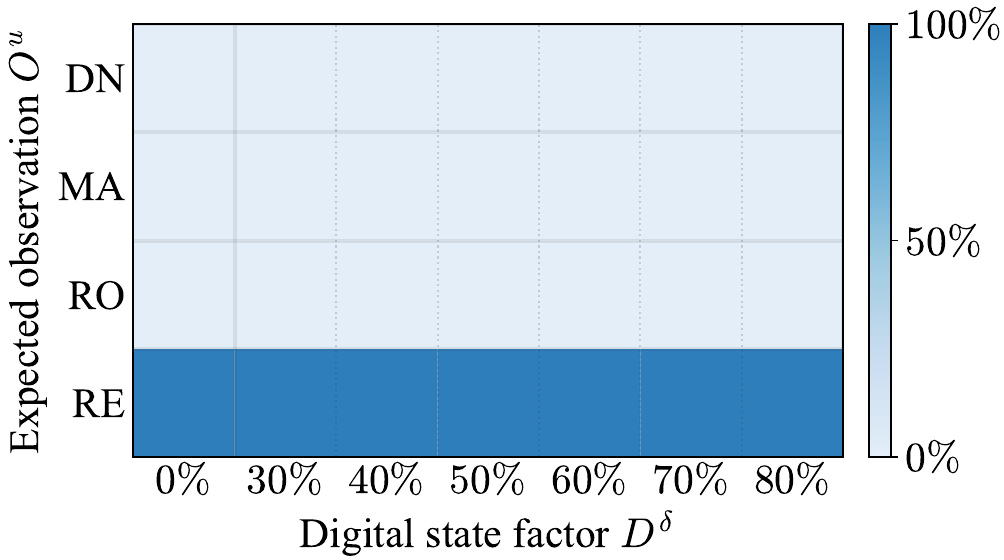}}\hspace{1cm}\subfloat[\label{fig:A_2_noepi}]{\includegraphics[width=.45\textwidth]{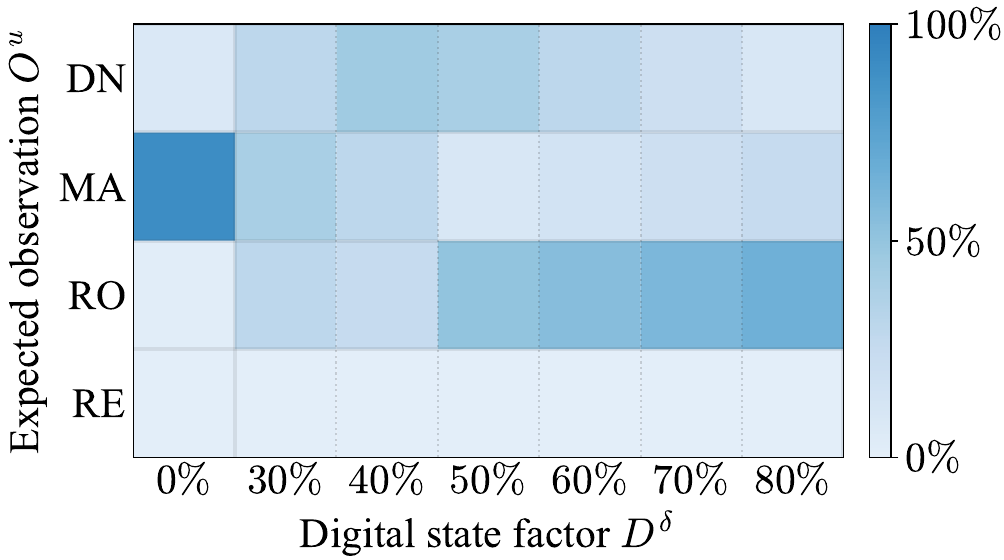}}
\caption{Visualization of the observation models: Panels (a) and (b) show slices of $\mathbf{A}^{\Omega\delta}$, corresponding to the sensory likelihoods (a) $p(O^{\Omega\delta}\mid D^\Omega=\Omega_4,D^\delta,D^\text{Epi}=\texttt{Epi})$ and (b) $p(O^{\Omega\delta}\mid D^\Omega=\Omega_4,D^\delta,D^\text{Epi}=\texttt{Non-Epi})$ for $\alpha=0.2$. Panels (c) and (d) show slices of $\mathbf{A}^{u}$, corresponding to the sensory likelihoods (c) $p(O^{u}\mid D^\Omega=\Omega_{1:6},D^\delta,D^\text{Epi}=\texttt{Epi})$ and (d) \mbox{$p(O^{u}\mid D^\Omega=\Omega_{1:6},D^\delta,D^\text{Epi}=\texttt{Non-Epi})$}.}\vspace{-0.25cm}
\end{figure}

The slice of $\mathbf{A}^{\Omega\delta}$ for the epistemic state, i.e., $p(O^{\Omega\delta}\mid D^\Omega,D^\delta,D^\text{Epi}=\texttt{Epi})$, is denoted by $\mathbf{A}^{\Omega\delta}_\text{Epi}\in\mathbb{R}^{\mid O^{\Omega\delta}\mid\times\mid D^\Omega\mid\times\mid D^\delta\mid}$. This observation model is derived from a confusion matrix that quantifies the offline (expected) performance of the DL models in identifying the digital state factors $D^\Omega$ and $D^\delta$. The confusion matrix is interpreted as a CPT, where rows correspond to ground-truth responses and columns to predicted outcomes. The offline evaluation has been performed using $4000$ noisy FOM solutions, achieving a classification accuracy of $91.39\%$. To mitigate the risk of inconsistencies due to zero-likelihood observations, i.e., evidence contradicting the confusion matrix, a small positive perturbation $10^{-5}$ is added to all entries of $\mathbf{A}^{\Omega\delta}_\text{Epi}$ prior to normalization. An exemplary slice of $\mathbf{A}^{\Omega\delta}_\text{Epi}$ associated with \mbox{$p(O^{\Omega\delta}\mid D^\Omega=\Omega_4,D^\delta,D^\text{Epi}=\texttt{Epi})$} is shown in \fig\ref{fig:A_1_epi}. 

While $\mathbf{A}^{\Omega\delta}_\text{Epi}$ serves as a relatively informative sensory likelihood, a higher-entropy likelihood is used to model the slice of $\mathbf{A}^{\Omega\delta}$ under the non-epistemic state, i.e., \mbox{$p(O^{\Omega\delta}\mid D^\Omega,D^\delta,D^\text{Epi}=\texttt{Non-Epi})$}, denoted by $\mathbf{A}^{\Omega\delta}_\text{Non-Epi}\in\mathbb{R}^{\mid O^{\Omega\delta}\mid\times\mid D^\Omega\mid\times\mid D^\delta\mid}$. This non-epistemic model is obtained via uniform random perturbation of $\mathbf{A}^{\Omega\delta}_\text{Epi}$ as the following linear combination:
\begin{equation}
\mathbf{A}^{\Omega\delta}_\text{Non-Epi} = (1-\alpha)\mathbf{A}^{\Omega\delta}_\text{Epi} + \alpha\mathbf{A}^{\Omega\delta}_\text{Entropic},
\end{equation}
which is then properly renormalized. Here, $\mathbf{A}^{\Omega\delta}_\text{Entropic}$ is a purely entropic observation model sampled from a uniform distribution over $[0,1]$, and $0\leq\alpha\leq1$ is a weighting coefficient controlling the degree of entropy introduced. Figure~\ref{fig:A_1_noepi} shows an exemplary slice corresponding to \mbox{$p(O^{\Omega\delta}\mid D^\Omega=\Omega_4,D^\delta,D^\text{Epi}=\texttt{Non-Epi})$} for $\alpha=0.2$. It is worth noting that modulating $\alpha$ can also be interpreted as a simple yet effective mechanism to account both for potential errors in, and for the decision-maker confidence about, the use of DL models to assimilate real-world data.

The slice of $\mathbf{A}^{u}$ for the epistemic state, i.e., encoding $p(O^{u}\mid D^\Omega,D^\delta,D^\text{Epi}=\texttt{Epi})$, is denoted as $\mathbf{A}_\text{Epi}^{u}\in\mathbb{R}^{\mid O^{u}\mid\times\mid D^\Omega\mid\times\mid D^\delta\mid}$. It is populated with Dirac delta distributions centered at the RE action for all possible combinations of $D^\Omega$ and $D^\delta$ (see also \fig\ref{fig:A_2_epi}). This design reflects the assumption that if the agent is in the state $D^\text{Epi}=\texttt{Epi}$, it knows with certainty that the previously taken action was the (epistemic) RE action, regardless of the values of the other digital state factors. From a data assimilation point of view, receiving $O^{u} = \text{RE}$ provides no informative cues for inferring $D^\Omega$ or $D^\delta$, but it deterministically sets $D^\text{Epi}=\texttt{Epi}$. In contrast, under the non-epistemic state, the corresponding observation model $\mathbf{A}_\text{Non-Epi}^{u}\in\mathbb{R}^{\mid O^{u}\mid\times\mid D^\Omega\mid\times\mid D^\delta\mid}$ is filled with entries that reflect a plausible causality for what the agent can infer about the previous action given $D^\Omega$ and $D^\delta$. This prior CPT (see also Fig.~\ref{fig:A_2_noepi}) is modeled consistently across the $D^\Omega$ factor, as follows: 
\begin{equation}
p(O^{u}\mid D^\Omega=\Omega_{1:6},D^\delta,D^\text{Epi}=\texttt{Non-Epi}) =
\begin{pmatrix}
0.08&0.3&0.45&0.4&0.3&0.2&0.1\\
0.9&0.4&0.3&0.1&0.15&0.2&0.25\\
0.02&0.3&0.25&0.5&0.55&0.6&0.65\\
0&0&0&0&0&0&0
\end{pmatrix}.
\end{equation}
For data assimilation, observing $O^{u} \neq \text{RE}$ has two implications: first, it deterministically sets $D^\text{Epi}=\texttt{Non-Epi}$; second, it introduces an inductive bias by leveraging the structure of $\mathbf{A}_\text{Non-Epi}^{u}$ to condition inference on the previous action, similar to the influence of the transition model.

\begin{figure}[!t]
\centering
    \subfloat[\label{fig:B_Omega}]{\includegraphics[width=.39\textwidth]{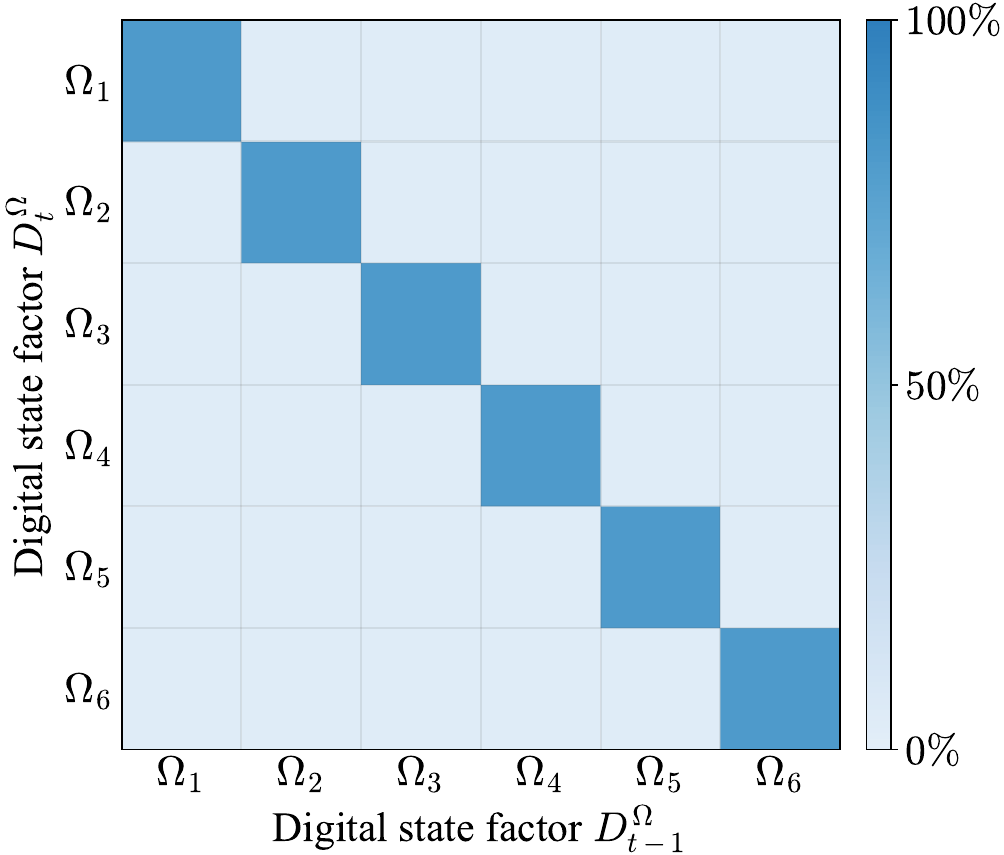}}\hspace{0.5cm}
\subfloat[\label{fig:B_delta_DN}]{\includegraphics[width=.39\textwidth]{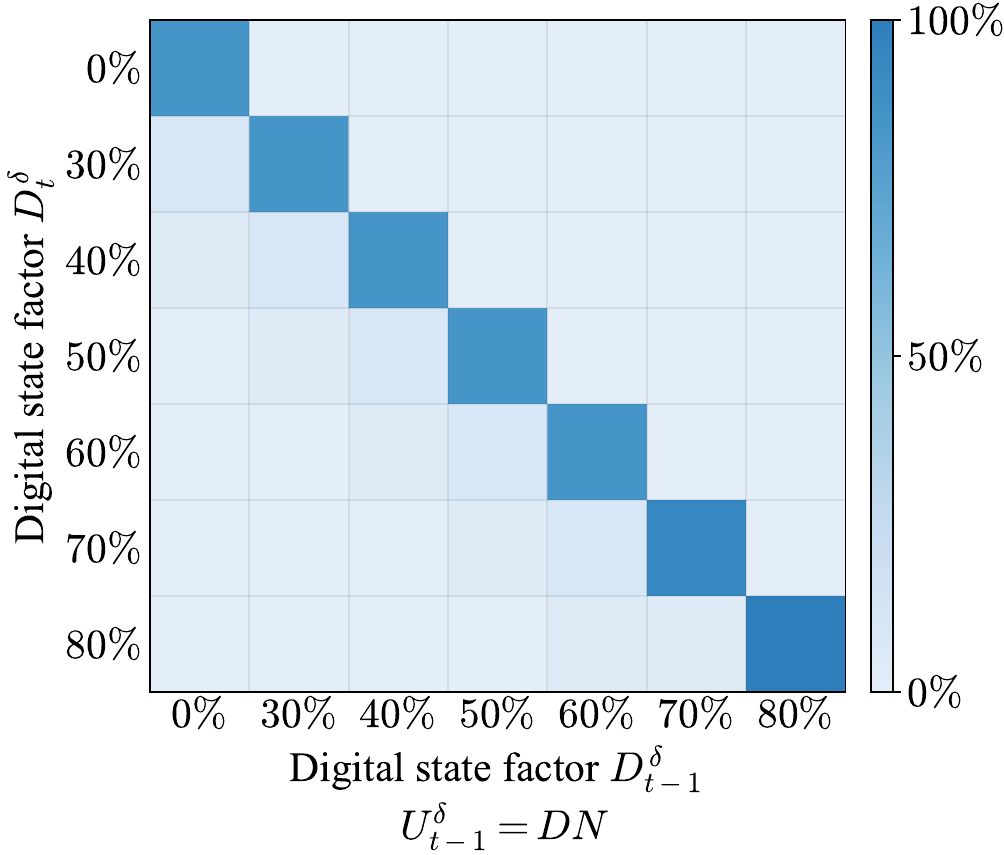}}\\\vspace{-0.25cm}
\subfloat[\label{fig:B_delta_RO}]{\includegraphics[width=.39\textwidth]{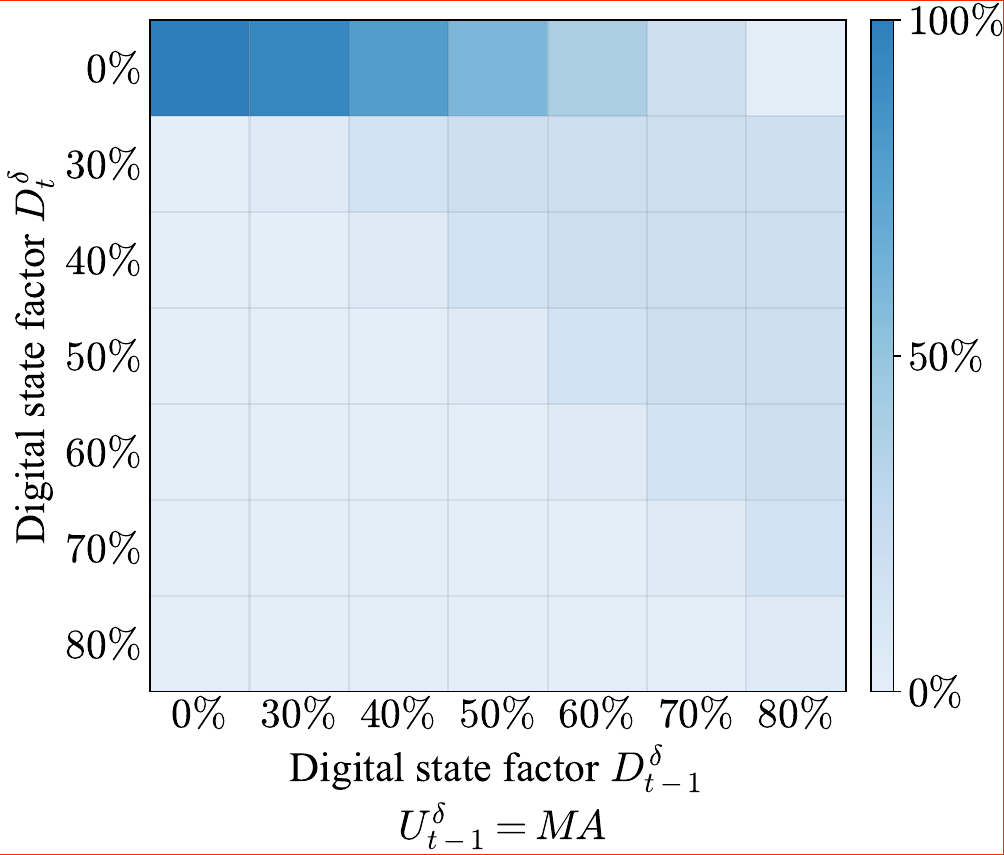}}\hspace{0.5cm}
\subfloat[\label{fig:B_delta_MA}]{\includegraphics[width=.39\textwidth]{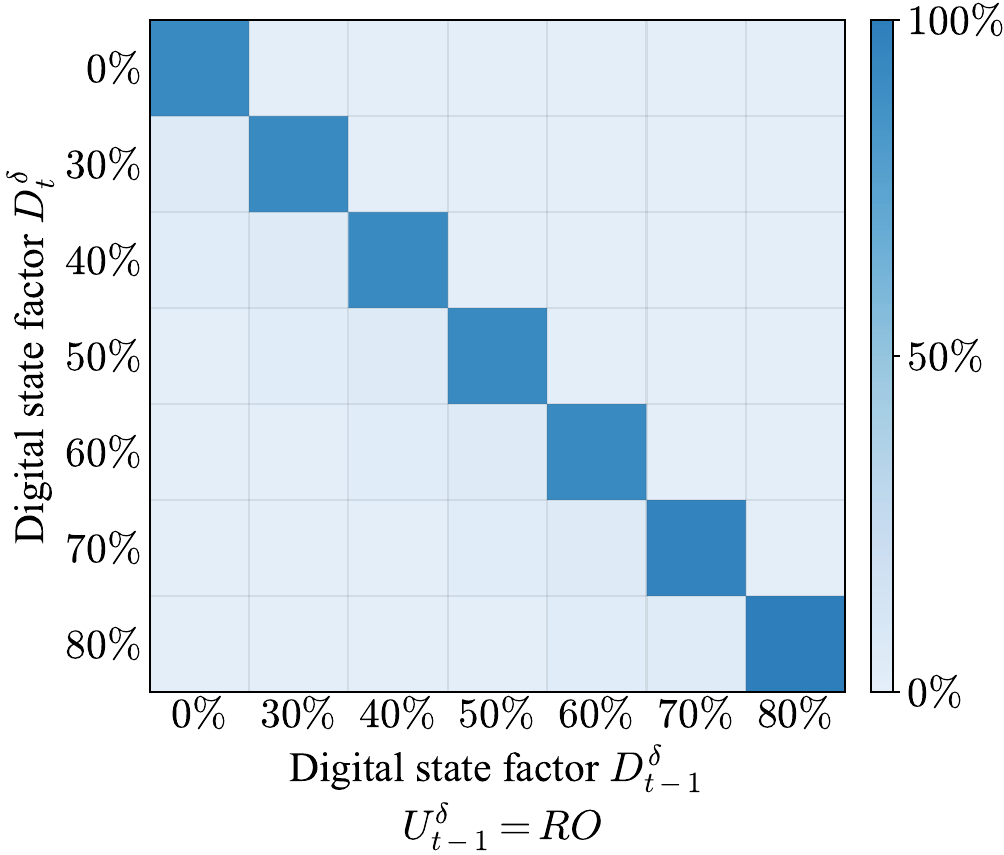}}\\\vspace{-0.25cm}
\subfloat[\label{fig:B_delta_RE}]{\includegraphics[width=.39\textwidth]{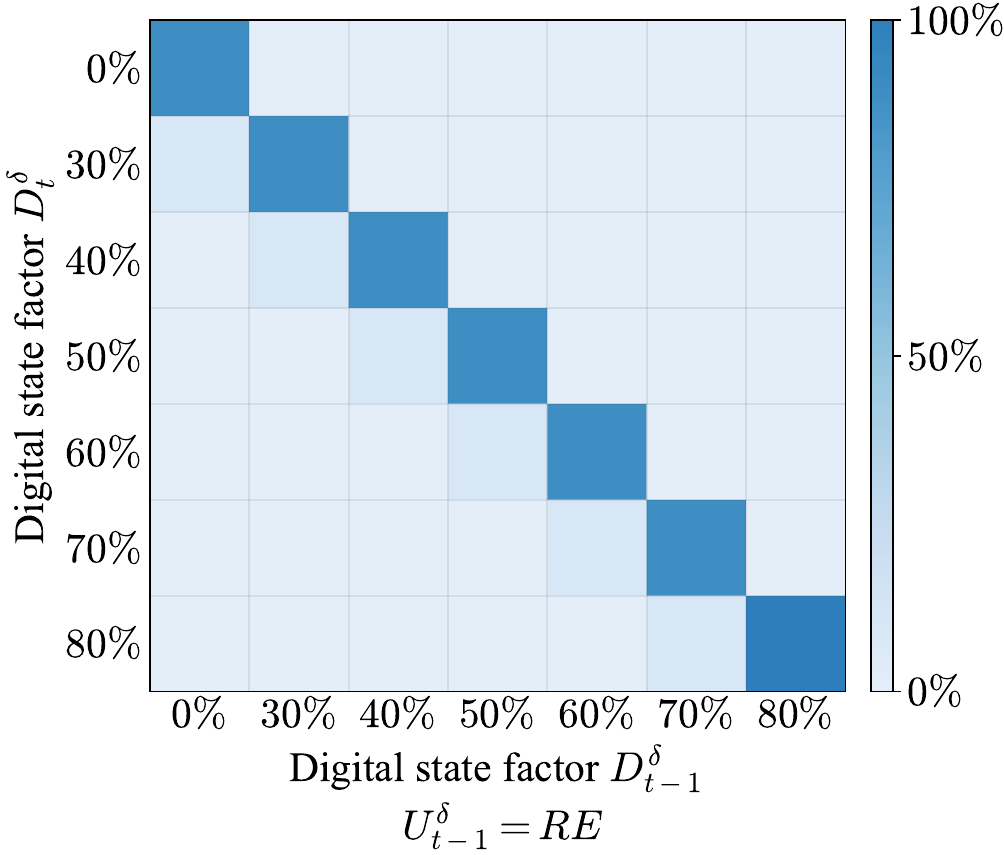}}\hspace{0.25cm}
\subfloat[\label{fig:B_epi_1}]{\includegraphics[width=.28\textwidth]{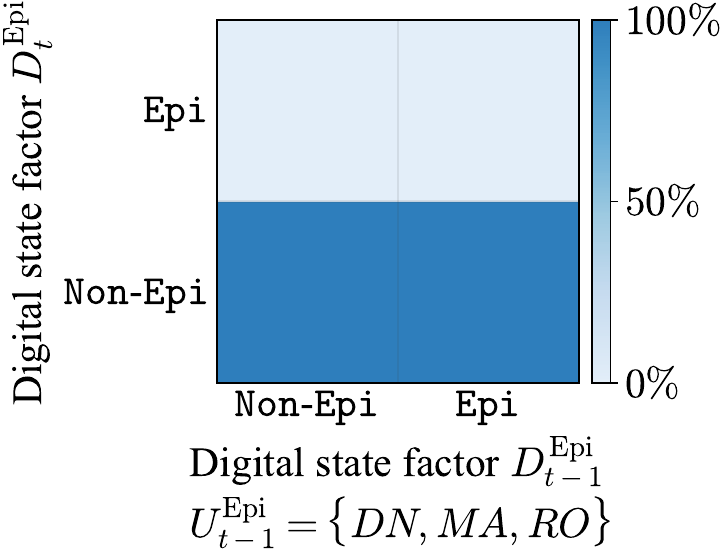}}\hspace{0.25cm}
\subfloat[\label{fig:B_epi_2}]{\includegraphics[width=.28\textwidth]{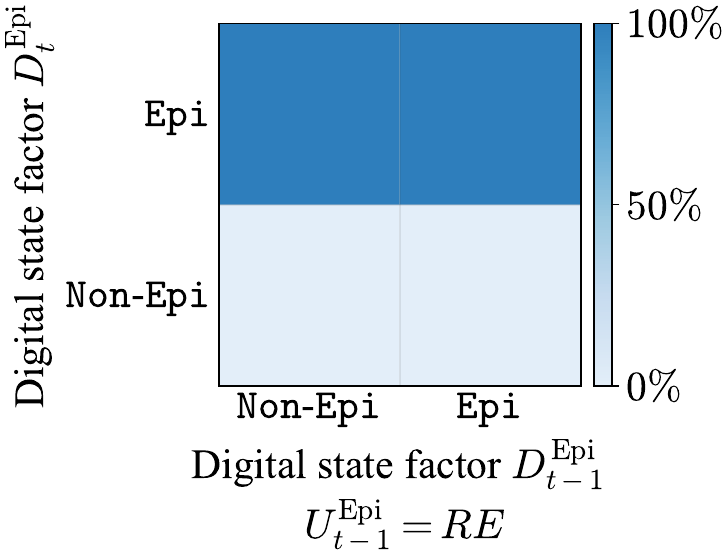}}
\caption{Visualization of the transition models: Panel (a) shows the uncontrollable $\mathbf{B}^{\Omega}$, corresponding to the transition likelihood $p(D_t^{\Omega}\mid D_{t-1}^{\Omega})$. Panels (b-e) show the action-specific slices of $\mathbf{B}^{\delta}$, corresponding to the transition likelihoods (b) $p(D_t^{\delta}\mid D_{t-1}^{\delta},U_{t-1}^{\delta}=\text{DN})$, (c) \mbox{$p(D_t^{\delta}\mid D_{t-1}^{\delta},U_{t-1}^{\delta}=\text{MA})$}, (d) $p(D_t^{\delta}\mid D_{t-1}^{\delta},U_{t-1}^{\delta}=\text{RO})$, and (e) $p(D_t^{\delta}\mid D_{t-1}^{\delta},U_{t-1}^{\delta}=\text{RE})$. Panels (f) and (g) shows the action-specific slices of $\mathbf{B}^\text{Epi}$, corresponding to the transition likelihoods (f) \mbox{$p(D_t^\text{Epi}\mid D_{t-1}^\text{Epi},U_{t-1}^\text{Epi}=\lbrace \text{DN,MA,RO}\rbrace)$} and (g) $p(D_t^\text{Epi}\mid D_{t-1}^\text{Epi},U_{t-1}^\text{Epi}=\text{RE})$.}\label{fig:transitions} \vspace{-0.25cm}
\end{figure}

The transition array $\mathbf{B}=\lbrace\mathbf{B}^{\Omega},\mathbf{B}^{\delta},\mathbf{B}^\text{Epi}\rbrace$ comprises three sub-arrays $\mathbf{B}^{f}\in\mathbb{R}^{\mid D^f\mid\times\mid D^f\mid\times\mid U^f\mid}$, each encoding the transition dynamics $p(D^{f}_t\mid D^{f}_{t-1},u_{t-1}^{f};\mathfrak{B}^f)$ of a specific digital state factor \mbox{$D^{f}\in\lbrace D^{\Omega},D^{\delta},D^\text{Epi}\rbrace$}, conditioned on its previous state and the corresponding control factor \mbox{$u^{f}\in\lbrace u^{\Omega},u^{\delta},u^\text{Epi}\rbrace$}. Starting with initial priors over the transition probabilities defined by $\mathfrak{B}^f$, these are iteratively refined by assimilating evidence from the system response to actions, as described in \sez\ref{sez:learning}. A graphical visualization of the initial transition models for each digital state and control factor is shown in \fig\ref{fig:transitions}. Note that these internal models do not replicate the ground-truth evolution, which remains unknown to the ADT. Moreover, assuming that digital state factors evolve independently, the control is factorized as $U=\lbrace U^{\Omega}=\varnothing,U^{\delta}=\overline{U},U^\text{Epi}=\overline{U}\rbrace$. This reflects that $D^\Omega$ is an uncontrollable factor, with a control dimensionality of 1, while $D^\delta$ and $D^\text{Epi}$ are both influenced by the same control variable $\overline{U}\in\mathcal{U}=\lbrace \text{DN},\text{MA},\text{RO},\text{RE}\rbrace$. The set of feasible policies $\Pi$ is constructed by combinatorially enumerating all possible sequences of actions from the action space $\mathcal{U}$ over the prediction horizon $t = t_c, \ldots, t_p$, resulting in a total of $4^{t_p-t_c}$ policies. 

The initial Dirichlet parameters $\mathfrak{b}^\Omega$ over the categorical distribution $\mathbf{\mathfrak{B}^\Omega}$ for the uncontrollable $\mathbf{B}^{\Omega}$ are selected to yield a $0.8$ probability that damage stays in the same subdomain $\Omega_m$, for $m=1,\ldots,6$. The remaining $0.2$ probability is evenly distributed across the other subdomains, reflecting a strong prior belief that damage is unlikely to move between different regions. 

For the action-conditioned $\mathbf{B}^{\delta}$, each action-specific slice encodes the probability of transitioning between discrete $\delta$ intervals. The diagonal entries represent the probability of remaining in the same damage state, while the lower-left and upper-right triangles denote the probabilities of deterioration and improvement, respectively. Under the DN action, the initial Dirichlet parameters $\mathfrak{b}^\delta$ for the categorical distribution $\mathbf{\mathfrak{B}^\delta}$ are configured to yield transition probabilities of $0.85$, $0.1$, and $0.05$ for degradation of zero, one, or two $\delta$ intervals, respectively. For the RO action, the corresponding probabilities are set to $0.92$, $0.05$, and $0.03$,  reflecting a slower rate of deterioration due to reduced structural load. The slice associated with the RE action is designed to reflect improved damage tracking. It assigns probabilities of $0.9$ and $0.1$ for degradation of zero and one $\delta$ intervals, capturing the higher confidence associated with epistemic control actions. In contrast, the MA action slice is designed to support transitions across up to six $\delta$ intervals, with probabilities $0.05$, $0.15$, $0.20$, $0.20$, $0.20$, and $0.20$, for improvements of zero to five intervals, respectively. To mitigate the risk of numerical inconsistencies caused by evidence that contradicts the assumed transition dynamics, a small perturbation of $10^{-3}$ is added to all entries of $\mathbf{B}^{\delta}$ prior to normalization.

The sub-array $\mathbf{B}^\text{Epi}$ serves as an epistemic switch, enabling deterministic transitions between the states $D^\text{Epi}=\texttt{Epi}$ and $D^\text{Epi}=\texttt{Non-Epi}$. This mechanism is implemented through Boolean matrices that enforce $D^\text{Epi}=\texttt{Non-Epi}$ -- regardless of its previous value -- whenever the ADT selects DN, MA, or RO actions. Conversely, selecting the RE action triggers a transition to the epistemic state $D^\text{Epi}=\texttt{Epi}$. This transition model is not subject to learning updates, as its structure is predefined and not expected to benefit from interaction with the generative process.

At each time step, the ADT selects a control action $u_t\in\mathcal{U}$ whose effects on the generative process are uncertain and may lead to unexpected outcomes. The costs associated with both the structural health state and the control actions are modeled as prior preferences via the array \mbox{$\mathbf{c}=\lbrace\mathbf{c}^{\Omega\delta},\mathbf{c}^u\rbrace$}. The components $\mathbf{c}^{\Omega\delta}\in\mathbb{R}^{\mid O^{\Omega\delta}\mid}$ and $\mathbf{c}^u\in\mathbb{R}^{\mid O^{u}\mid}$ assign relative log-probabilities to each outcome of the two observation modalities, respectively:
\begin{equation}
    \mathbf{c}^{\Omega\delta}\leftarrow\ln{\widetilde{p}(O^{\Omega\delta}_{t})} =
    \begin{cases}
        0 & \text{if $y = 0$}, \\
        -\exp(\delta) & \text{if } 30\% <\delta < 80\%,  \\
        -10 & \text{if $\delta = 80\%$},
    \end{cases}\ \
    \mathbf{c}^u\leftarrow\ln{\widetilde{p}(O^{u}_{t})} = 
    \begin{cases}
        +5.5 & \text{if $u_t$ = DN}, \\
        -5 & \text{if $u_t$ = MA}, \\
        +2.5 & \text{if $u_t$ = RO}, \\
        -0.5 & \text{if $u_t$ = RE}.
    \end{cases}
    \label{eq:priors}
\end{equation}
These log-probability vectors are passed through a Softmax function to produce valid probability distributions $\widetilde{p}(O^{\Omega\delta}_{t})$ and $\widetilde{p}(O^{u}_{t})$, which are then used to compute the expected utility term in the EFE. The structural health preferences penalize deterioration in proportion to the exponential of $\delta$, with a steep penalty for severely compromised states. The control action preferences reflect trade-offs between epistemic value of expected information gain and operational cost: DN and RO actions yield positive rewards but carry the risk of structural deterioration; RE similarly allows for deterioration, yet it is expected to reduce the entropy of the digital state posterior at the cost of a moderately negative reward. MA mitigates deterioration but carries a significantly negative reward due to its high cost. While these values are expressed in non-dimensional form, they represent indicative costs charged to the decision maker. Actual values may be derived from service and cost catalogs issued by governmental agencies or infrastructure operators. In particular, the health-related preference distribution $\widetilde{p}(O^{\Omega\delta}_{t})$ should reflect a prioritization analysis that accounts for both the likelihood and consequences of different damage scenarios -- such as loss of serviceability, increased accident risk, or structural failure -- as well as the risk tolerance of the decision-maker.

The array defining the initial state model $\mathbf{d}=\lbrace\mathbf{d}^\Omega,\mathbf{d}^\delta,\mathbf{d}^\text{Epi}\rbrace$ consists of three sub-arrays \mbox{$\mathbf{d}^f\in\mathbb{R}^{\mid D^f\mid}$}, each specifying the initial prior distribution $p(D^f_{t_c})$ over a digital state factor \linebreak\mbox{$D^{f}\in\lbrace D^{\Omega},D^{\delta},D^\text{Epi}\rbrace$}. Uniform probability distributions are adopted for $D^{\Omega}$ and $D^\text{Epi}$ to reflect initial uncertainty. In contrast, $D^{\delta}$ is initialized as a Dirac delta distribution centered at $0\%$, consistent with the assumption of undamaged structure when the ADT enters into operation.

The (unknown) ground-truth generative process evolves conditionally on the most recent control action. In particular, we assume that damage can develop in any predefined region, without propagating across different damageable subdomains. The evolution follows the degradation (or improvement) stochastic models described below. Under the DN, RO, and RE actions, structural health is assumed to degrade monotonically. For the DN action, the damage class $y$ is sampled from a categorical distribution \mbox{$y \sim \text{Cat}(\frac{1}{2}, \frac{1}{12}, \frac{1}{12}, \frac{1}{12}, \frac{1}{12}, \frac{1}{12}, \frac{1}{12})$}, which assigns half of the probability mass to the undamaged state $y=0$, and distributes the remaining half uniformly among the six damage classes \mbox{$y=1, \ldots,6$}. When damage first initiates, the magnitude $\delta$ is sampled uniformly within the range of the first damage interval \mbox{$\delta_t \mid  y_{t}\neq0,y_{t-1}=0 \sim \text{Uniform}(0.3\%, 0.35\%)$}. Subsequent damage progression is modeled by sampling $\delta$ increments from a truncated normal distribution centered at $1.5\%$ with a standard deviation of $1\%$, \mbox{$\delta_t - \delta_{t-1} \mid  y_{t-1}\neq0\sim \text{Normal}_{\geq 0}(1.5\%, 1\%)$}, with any increments below $0\%$ rounded up to $0\%$. For the RO action, a similar model is employed, but with a lower probability of damage initiation and slower deterioration. In this case, the damage class is sampled as $y \sim \text{Cat}(\frac{3}{4}, \frac{1}{24}, \frac{1}{24}, \frac{1}{24}, \frac{1}{24}, \frac{1}{24}, \frac{1}{24})$, and the damage magnitude evolves as \mbox{$\delta_t - \delta_{t-1} \mid  y_{t-1}\neq0\sim \text{Normal}_{\geq 0}(0.95\%, 0.5\%)$}. For the RE action, the generative process is the same as that under the DN or RO actions, respectively, depending on whether the system was previously in a restricted or unrestricted condition before engaging in information-seeking (epistemic) behavior. In contrast, the MA action is modeled as a healing process. If \mbox{$y=0$}, the system remains undamaged. If \mbox{$y\neq0$}, the damage magnitude decreases according to \mbox{$\delta_t - \delta_{t-1} \mid  y_t\neq0\sim\text{Normal}_{\leq 10\%}(-25\%, 15\%)$}, with any decrement below $10\%$ rounded up to $10\%$. The system is assumed to return to an undamaged condition ($y = 0$) if the resulting damage magnitude satisfies $\delta < 30\%$, reflecting a minimal detectable deterioration threshold.

\subsection{Results: Purely goal-directed behavior}
\label{sez:results_1}
In this section and the next, we present the results of several ADT simulations, each spanning $60$ time steps. At every time step, new observational data are generated based on the (unknown) ground-truth generative process. The ADT assimilates these data to infer the variational posterior $Q^*(D_{t_c})$ over the current digital state, and performs policy inference by computing the posterior $Q^*(\pi)$ over policies. Control actions are subsequently selected as the best-point estimate from the posterior $Q^*(U_{t_c})$ over control states, and the generative model is eventually learned by updating the variational posterior Dirichlet parameters $\widehat{\mathfrak{b}}^*$.

We adopt a policy horizon of $t_p-t_c=4$ and begin by analyzing a baseline scenario where the ADT operates under a purely goal-directed (pragmatic) behavior. This is achieved by retaining only the utility term associated with pragmatic value in the EFE formulation, excluding any contributions from information-seeking (epistemic) value and removing the epistemic RE action from the available action set. The entropy level in the observation model $\mathbf{A}^{\Omega\delta}_\text{Non-Epi}$ is set using $\alpha=0.5$. The inverse temperature parameter controlling the precision of policy selection is left to its default value  of $\gamma=16$. Furthermore, learning updates to the generative model are disabled in this baseline setting.

\begin{figure}[!t]
\centering
  \includegraphics[width=.78\linewidth]{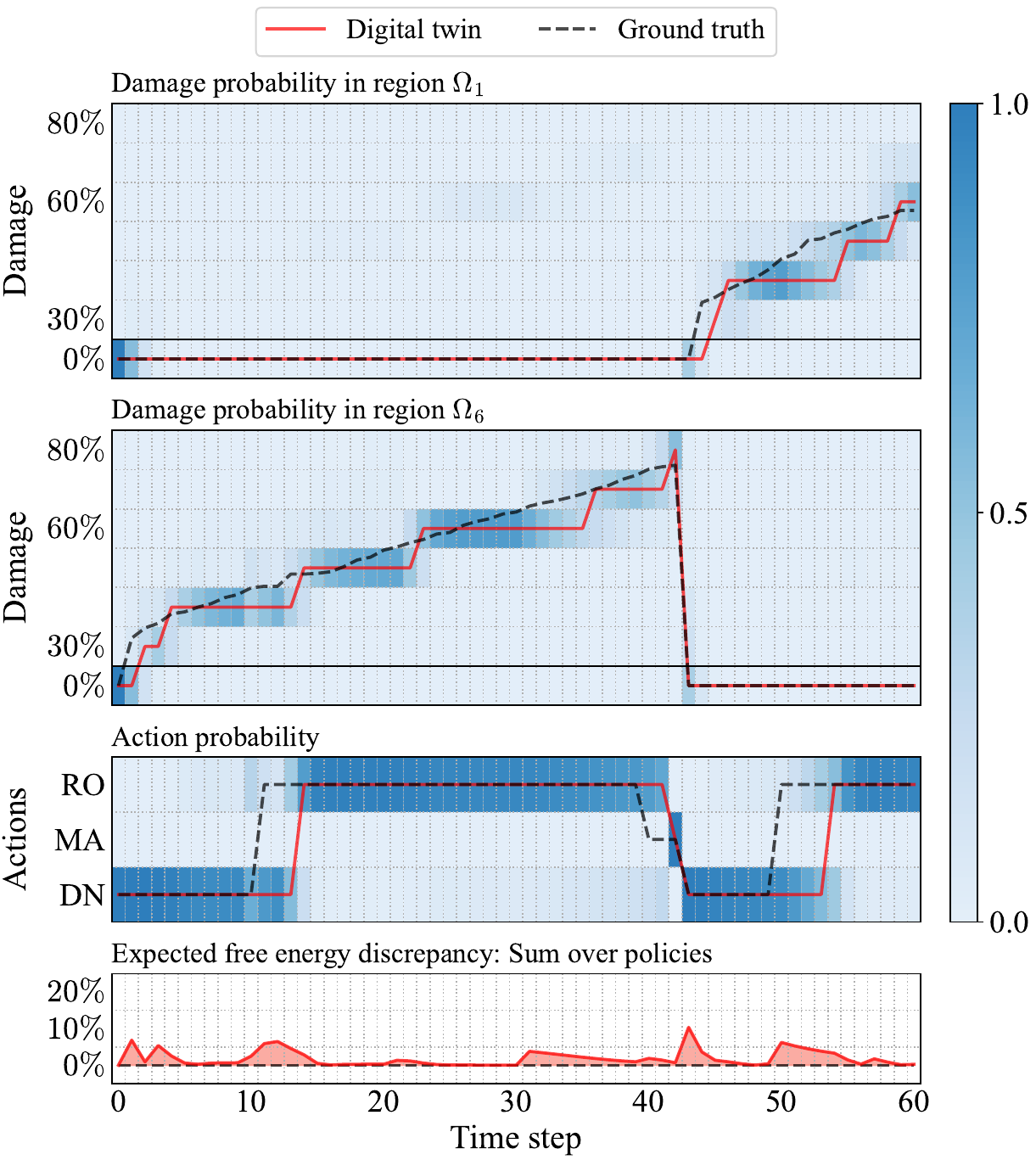}
  \caption{Active digital twin using purely goal-directed (pragmatic) behavior. Probabilistic and best-point estimates of: (top two panels) digital state evolution compared to the ground-truth physical state; (penultimate panel) control actions recommended by the digital twin versus the optimal action under the ground-truth generative process. In the top panels, background colors represent the belief distribution over the digital state at each time step. In the penultimate panel, background colors indicate the belief distribution over the control actions. The bottom panel quantifies simulation quality in terms of the percentage absolute discrepancy between the sum of the policy-specific expected free energies computed by the digital twin and those obtained under the ground truth.}
  \label{fig:history_noepi_nolearn_1}
  \vspace{-0.25cm}
\end{figure}

Figure~\ref{fig:history_noepi_nolearn_1} illustrates a representative ADT simulation. Results are reported in terms of both the ground-truth physical state and the corresponding ADT estimates obtained after assimilating observational data. The evolution of the digital state is shown only for regions that experience damage, although all damageable regions $\Omega_1,\ldots,\Omega_6$ are susceptible to degradation. Initially, damage develops in $\Omega_1$, and the posterior $Q^*(D_{t_c})$ reveals relatively high uncertainty, primarily due to the entropy in the observation model. Nevertheless, despite the severely corrupted observation model $\mathbf{A}^{\Omega\delta}_\text{Non-Epi}$, the ADT successfully follows the ground-truth evolution by leveraging prior information from the forward-time predictor $\mathbf{B}$. The corresponding sequence of control action estimates $Q^*(U_{t_c})$ is shown in the penultimate panel. The ADT initially recommends DN actions, aligned with the prior preferences over the two observation modalities encoded in $\mathbf{c}$, i.e., to maximize utility. Once a substantial probability mass in $Q^*(D_{t_c})$ is assigned to $D^\delta\geq45\%$, RO actions begin to be selected, enabling the ADT to continue monitoring degradation, which now evolves at a reduced rate. Eventually, an MA action is selected when the structural state becomes critically compromised, as indicated by a consistent probability mass over $D^\delta \geq 75\%$ in $Q^*(D_{t_c})$. A similar behavior is shown for the subsequent damage event in $\Omega_6$. 

For comparison, control actions under the ground-truth generative process are computed using a second AIF agent that mirrors the ADT architecture but has access to the true physical state. The ADT selects the appropriate control action from $Q^*(U_{t_c})$ with a delay of at most five time steps relative to the ground-truth-informed agent. This delay is mainly attributed to the continued use of a highly entropic observation model, which limits fast and accurate inference, along with the need to recursively update prior beliefs from earlier time steps. 
Note that including the RE action in the available action set would not affect the results in this case, as the ADT is driven solely by (pragmatic) utility maximization and does not engage in information-seeking (epistemic) behavior, i.e., it does not seek to reduce the entropy of $Q(D_{t_c:t_p})$ through exploratory actions. The bottom panel of the figure assesses simulation quality by tracking the evolution of the percentage absolute discrepancy between the sum of the policy-specific EFEs computed by the ADT and those obtained under the ground-truth-informed agent:
\begin{equation}
\Delta^\mathbf{G}=\left\lvert\frac{\sum_{\pi\in\Pi}(G^\pi - \widehat{G}^\pi)}{\sum_{\pi\in\Pi}\widehat{G}^\pi}\right\rvert\cdot100,
\end{equation}
where $\widehat{G}^\pi$ denotes the EFE associated with policy $\pi$ under the ground-truth-informed AIF agent.

\begin{figure}[!t]
\centering
  \includegraphics[width=.78\linewidth]{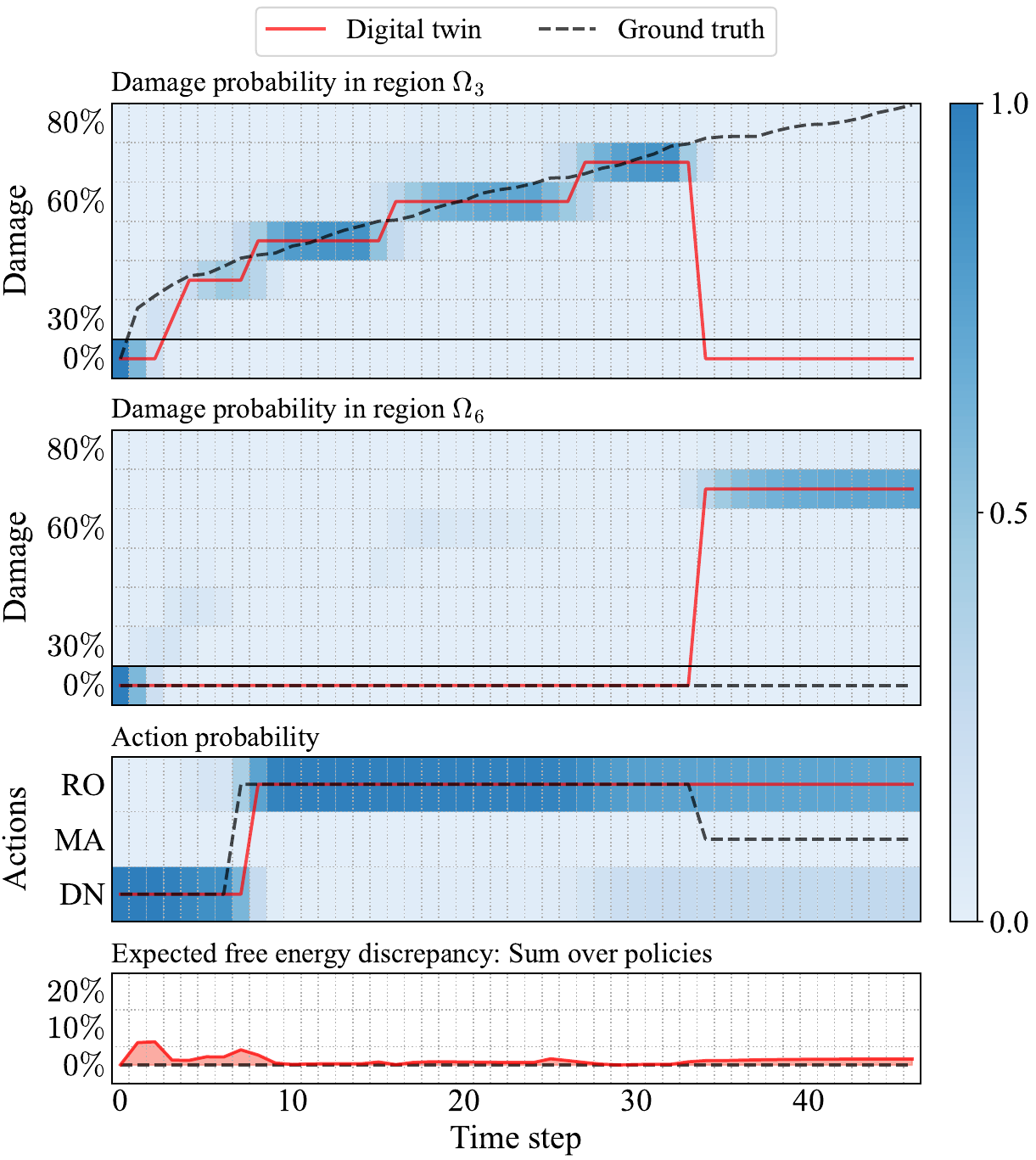}
  \caption{Failed active digital twin using purely goal-directed (pragmatic) behavior. Probabilistic and best-point estimates of: (top two panels) digital state evolution compared to the ground-truth physical state; (penultimate panel) control actions recommended by the digital twin versus the optimal action under the ground-truth generative process. In the top panels, background colors represent the belief distribution over the digital state at each time step. In the penultimate panel, background colors indicate the belief distribution over the control actions. The bottom panel quantifies simulation quality in terms of the percentage absolute discrepancy between the sum of the policy-specific expected free energies computed by the digital twin and those obtained under the ground truth.}
  \label{fig:history_noepi_nolearn_2}
  \vspace{-0.25cm}
\end{figure}

A second representative ADT simulation is shown in \fig\ref{fig:history_noepi_nolearn_2}, exemplifying the same purely goal-directed (pragmatic) behavior but with a different random seed. In this case, damage begins to develop in region $\Omega_3$, and the ADT initially behaves consistently with the previous results, tracking the generative process with relatively high-entropy estimates propagated forward in time. However, starting from time step $t=33$, the ADT begins to diverge from the ground truth, and the digital state posterior $Q^*(D_{t_c})$ progressively loses synchronization with the physical state. The probability mass in $Q^*(D_{t_c})$ gradually shifts from $D^\Omega=\Omega_3$ to $D^\Omega=\Omega_6$, where $D^\delta$ is consistently underestimated as lying within the range $[65\%,75\%]$, while the actual value is in the range $[75\%,80\%]$. As a result, the ADT fails to select an MA action for more than ten time steps, despite its necessity. The simulation eventually terminates at time step $t=46$, due to a digital failure at $t=47$, caused by $D^\delta>80\%$, and symbolizing structural collapse. The ADT inability to recover accurate tracking is attributed to the interplay between the poorly informative sensory likelihood and the recursive propagation of outdated prior beliefs, which degrade over time.

By running a cluster of $200$ simulations, each spanning $60$ time steps and initialized with a different random seed for both the observation model $\mathbf{A}^{\Omega\delta}_\text{Non-Epi}$ and the ground-truth generative process, the ADT operating under a purely goal-directed (pragmatic) behavior fails in $72$ out of $200$ cases. In this baseline setting, the failure rate is largely driven by the high entropy of the observation model $\mathbf{A}^{\Omega\delta}_\text{Non-Epi}$. In-simulation performance is assessed by measuring the accuracy of the maximum a-posteriori estimate of the $D^\Omega$ and $D^\delta$ digital state factors against the ground-truth generative process. The effect of the $\delta$ discretization is accounted for by computing accuracy with a $10\%$ tolerance (equal to the discretization step). Under these conditions, the ADT achieves a mean accuracy of $69\%$ with a $95\%$ confidence interval of $\pm2.6\%$. Although highly corrupted observations reflect challenging real-world conditions, the results presented in the following section show that equipping the ADT with both goal-directed and information-seeking (epistemic) components enables active exploration in response to critical uncertainty, resulting in a significant performance improvement over this purely pragmatic baseline.

The behavior described above can also be illustrated using a simplified scenario in which the ADT relies on two sensors, each providing partial observations to update its beliefs about the evolving damage state. If one sensor becomes faulty but the transition model closely approximates the actual dynamics of damage progression, the ADT may still track the system accurately, as the predictive power of the prior compensates for the degraded sensory evidence. However, in the more typical case where the transition model does not fully capture the actual system evolution, outdated priors dominate the inference process, and the compromised likelihood is unable to correct them. In such conditions, an information-seeking (epistemic) action should ideally be triggered to resolve ambiguity and restore confidence in the likelihood model -- for instance, by querying a redundant sensor, activating a dormant one, or scheduling a targeted diagnostic procedure.

It is interesting to note how the EFE discrepancy shown in the bottom panel of \fig\ref{fig:history_noepi_nolearn_2} does not indicate any critical issue. This is because the EFE is a subjective metric, reflecting how the ADT evaluates its own performance rather than measuring the correctness of the digital state estimates. Indeed, the $\Delta^\mathbf{G}$ indicator quantifies the misalignment in belief-driven action planning between the ADT and an idealized agent with access to the true physical state. As a result, $\Delta^\mathbf{G}$ remains low simply because the ADT is (mistakenly) confident in its estimated behavior, just as the ground-truth-informed agent is confident in its own. However, the resulting control actions differ. To address this limitation, one could instead employ objective performance indicators derived from the generative process. Examples include utility scores evaluated on realized system outcomes, or the survival time before reaching a critical condition.

\subsection{Results: Combining goal-directed and information-seeking behaviors}
\label{sez:results_2}

In this section, we present the results of ADT simulations combining goal-directed (pragmatic) and information-seeking (epistemic) behaviors. For this, we adopt the complete EFE formulation including both pragmatic and epistemic terms; furthermore, we include the epistemic RE action in the available actions. All other settings remain unchanged from the simulations presented earlier. 

\begin{figure}[!t]
\centering
  \includegraphics[width=.78\linewidth]{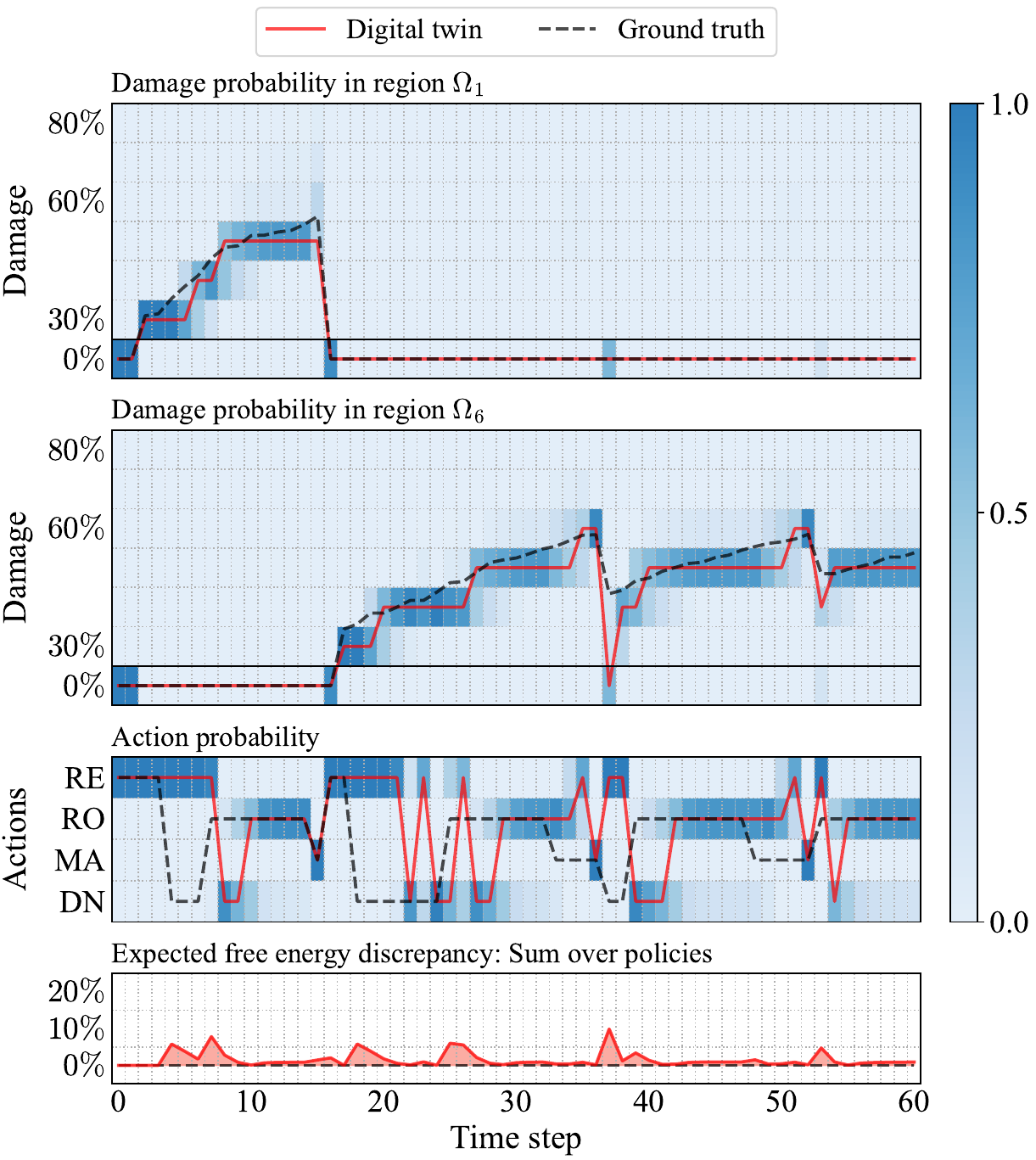}
  \caption{Active digital twin using a combination of goal-directed (pragmatic) and information-seeking (epistemic) behaviors. Probabilistic and best-point estimates of: (top two panels) digital state evolution compared to the ground-truth physical state; (penultimate panel) control actions recommended by the digital twin versus the optimal action under the ground-truth generative process. In the top panels, background colors represent the belief distribution over the digital state at each time step. In the penultimate panel, background colors indicate the belief distribution over the control actions. The bottom panel scores simulation quality in terms of the percentage absolute discrepancy between the sum of the policy-specific expected free energies computed by the digital twin and those obtained under the ground truth.}
  \label{fig:history_epi_nolearn}
  \vspace{-0.25cm}
\end{figure}

\begin{figure}[!t]
\centering
  \includegraphics[width=.78\linewidth]{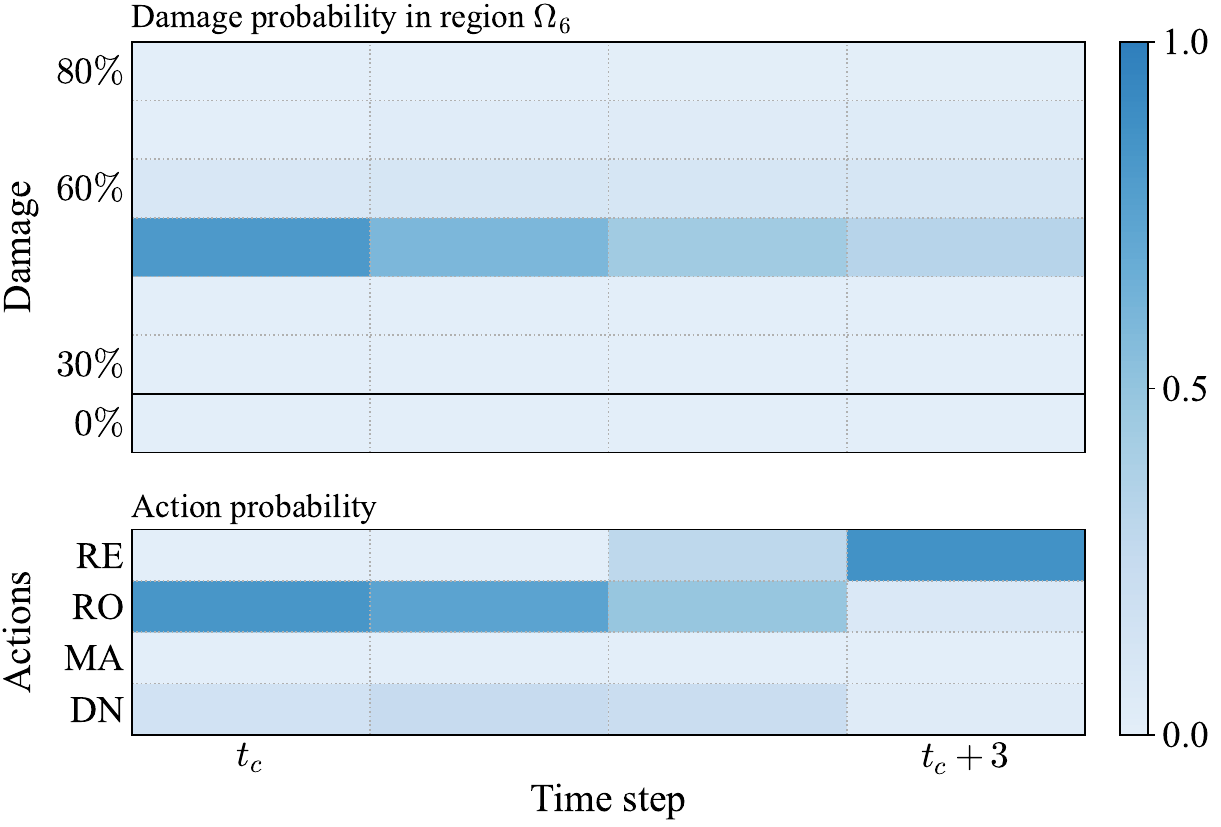}
  \caption{Active digital twin using a combination of goal-directed (pragmatic) and information-seeking (epistemic) behaviors. Posterior predictive densities beyond data assimilation over (future) digital states and control states, starting at $t_c=60$. In the top panel, background colors represent the belief distribution over the digital state at each time step. In the bottom panel, background colors indicate the belief distribution over the control actions.}
  \label{fig:prediction}
  \vspace{-0.25cm}
\end{figure}

Figure~\ref{fig:history_epi_nolearn} illustrates a representative simulation. The ADT initially exhibits information-seeking (epistemic) behavior, executing a sequence of RE actions to gather information about damage onset. Once the posterior $Q^*(D_{t_c})$ identifies evolving damage within $\Omega_1$ with relatively low uncertainty, the ADT shifts to DN actions aimed at (pragmatic) utility maximization. When a significant portion of $Q^*(D_{t_c})$ supports \mbox{$D^\delta\geq45\%$}, RO actions begin to emerge. An MA action is eventually selected when the risk of structural failure becomes substantial, i.e., for a significant probability mass over $D^\delta \geq 65\%$. A similar pattern is observed during the subsequent damage event in $\Omega_6$. In this case, sporadic RE actions are also triggered whenever the entropy of $Q^*(D_{t_c})$ increases, to prevent desynchronization from the physical state. These RE actions are interleaved with extended sequences of DN and RO decisions, depending on the evolving health state and the interaction between the sensory likelihood and the transition model. For instance, RE actions at $t=35$ and $t=51$ are deployed to disambiguate the digital state just before executing costly MA interventions. In contrast, the first MA action at $t=15$ is not preceded by RE behavior, as the ADT maintains a confident, low-entropy belief at that point. Note that the epistemic RE action carries a lower prior preference $\widetilde{p}(O^{u})$ than DN or RO actions and does not directly affect damage progression. As a result, it is employed only under epistemic-driven behavior, where its role is to support future goal-directed (pragmatic) decisions. Similarly, RE actions occurring immediately after MA interventions reflect the ADT effort to resolve uncertainty via active exploration of maintenance outcomes. In contrast, under the ground-truth generative process, RE actions are triggered exclusively to gather initial evidence about damage onset. Corrective inference is unnecessary in this setting due to perfect, uncertainty-free access to the physical state.

Figure~\ref{fig:prediction} shows the posterior predictive densities for the future digital states $Q^*(D_{t_c:t_p})$ and the corresponding control states $Q^*(U_{t_c:t_p})$, starting at $t_c=60$ and spanning four time steps. These predictions capture the expected progression of structural health, conditioned on the posterior over policies $Q^*(\pi)$, thereby supporting the planning of preventive interventions. When belief propagation leads to an overly flat digital state distribution, the likelihood of selecting an RE action increases, mitigating the risk of decisions based on unreliable or uncertain belief states.

By running a second cluster of $200$ simulations, each initialized with a different random seed, the ADT operating under combined goal-directed (pragmatic) and information-seeking (epistemic) behaviors exhibits zero failures. In-simulation performance also increases, with the maximum a-posteriori estimate of the $D^\Omega$ and $D^\delta$ digital state factors achieving a mean accuracy of $89\%$ with a $95\%$ confidence interval of $\pm0.5\%$. This result underscores the potential of fully equipped ADTs compared to the purely pragmatic baseline discussed in \sez\ref{sez:results_1}.

\begin{figure}[!t]
\centering
  \includegraphics[width=.78\linewidth]{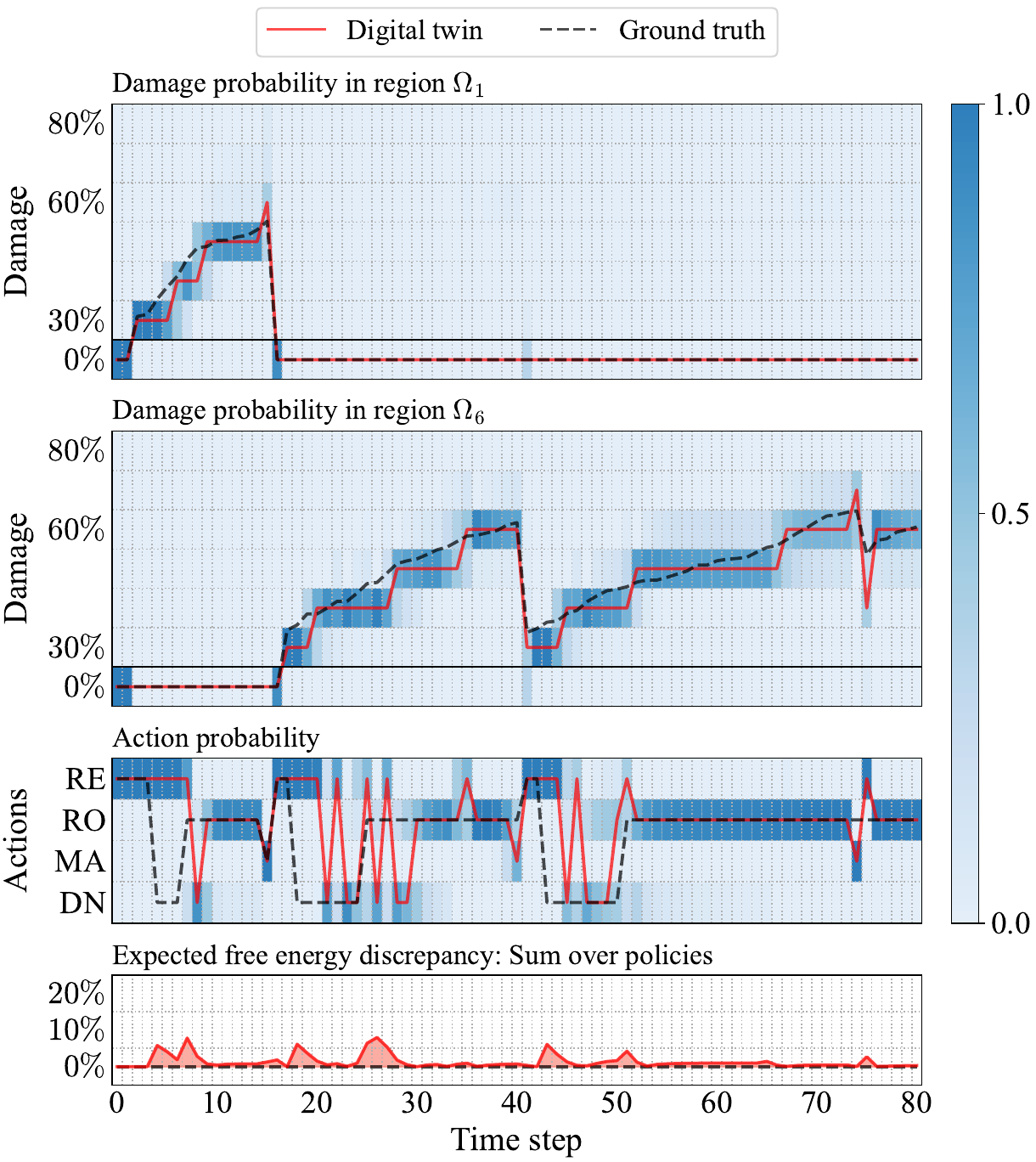}
  \caption{Active digital twin using a combination of goal-directed (pragmatic) and information-seeking (epistemic) behaviors and additionally incorporating learning updates to the generative model. Probabilistic and best-point estimates of: (top two panels) digital state evolution compared to the ground-truth physical state; (penultimate panel) control actions recommended by the digital twin versus the optimal action under the ground-truth generative process. In the top panels, background colors represent the belief distribution over the digital state at each time step. In the penultimate panel, background colors indicate the belief distribution over the control actions. The bottom panel quantifies simulation quality in terms of the percentage absolute discrepancy between the sum of the policy-specific expected free energies computed by the digital twin and those obtained under the ground truth.}
  \label{fig:history_epi_learn}
  \vspace{-0.25cm}
\end{figure}

The results of a complete ADT simulation, combining goal-directed (pragmatic) and information-seeking (epistemic) behaviors and additionally incorporating learning updates to the generative model, are shown in \fig\ref{fig:history_epi_learn} for the same initialization seed as in \fig\ref{fig:history_epi_nolearn}. This scenario spans $80$ time steps and introduces learning via updates to the transition model array $\mathbf{B}$, with a learning rate of $\eta=0.1$. Learning demonstrates beneficial in several aspects. First, it reduces the frequency of incorrect digital state inferences, thereby shortening the average response delay relative to the ground-truth agent. Second, as the transition dynamics become progressively tailored to the (unknown) generative process, the ADT gains confidence in its predictions, resulting in a reduced need for (corrective) RE actions. Third, the gradual reduction of uncertainty in the transition model enables the ADT to safely delay maintenance toward the end of the simulation. For example, the third maintenance action, previously triggered at $t=52$, is now postponed to $t=74$. Moreover, this intervention is no longer based on a maximum a-posteriori estimate of $D^\delta$ within the $[55\%,65\%]$ range, but instead within the higher $[65\%,75\%]$ range -- highlighting the potential for resource savings across the system operational lifespan. The runtime for this simulation is about $130~\textup{s}$, averaging $1.6~\textup{s}$ per time slice.

\subsection{Results: Robustness assessment}
\label{sez:results_3}

In this section, we assess the robustness of the ADT, fully equipped with goal-directed (pragmatic) and information-seeking (epistemic) behaviors, against incomplete data streams and uncertainty in the extent of the damaged region. We first consider two stress-testing scenarios: ($i$) progressively more incomplete data streams during the passage of a train, and ($ii$) progressively more frequent observation failures.

\begin{figure}[!t]
\begin{tikzpicture}[scale=.9, every node/.style={scale=1.}]
\node[draw=none,fill=none] at (0,0){\includegraphics[width=1\linewidth]{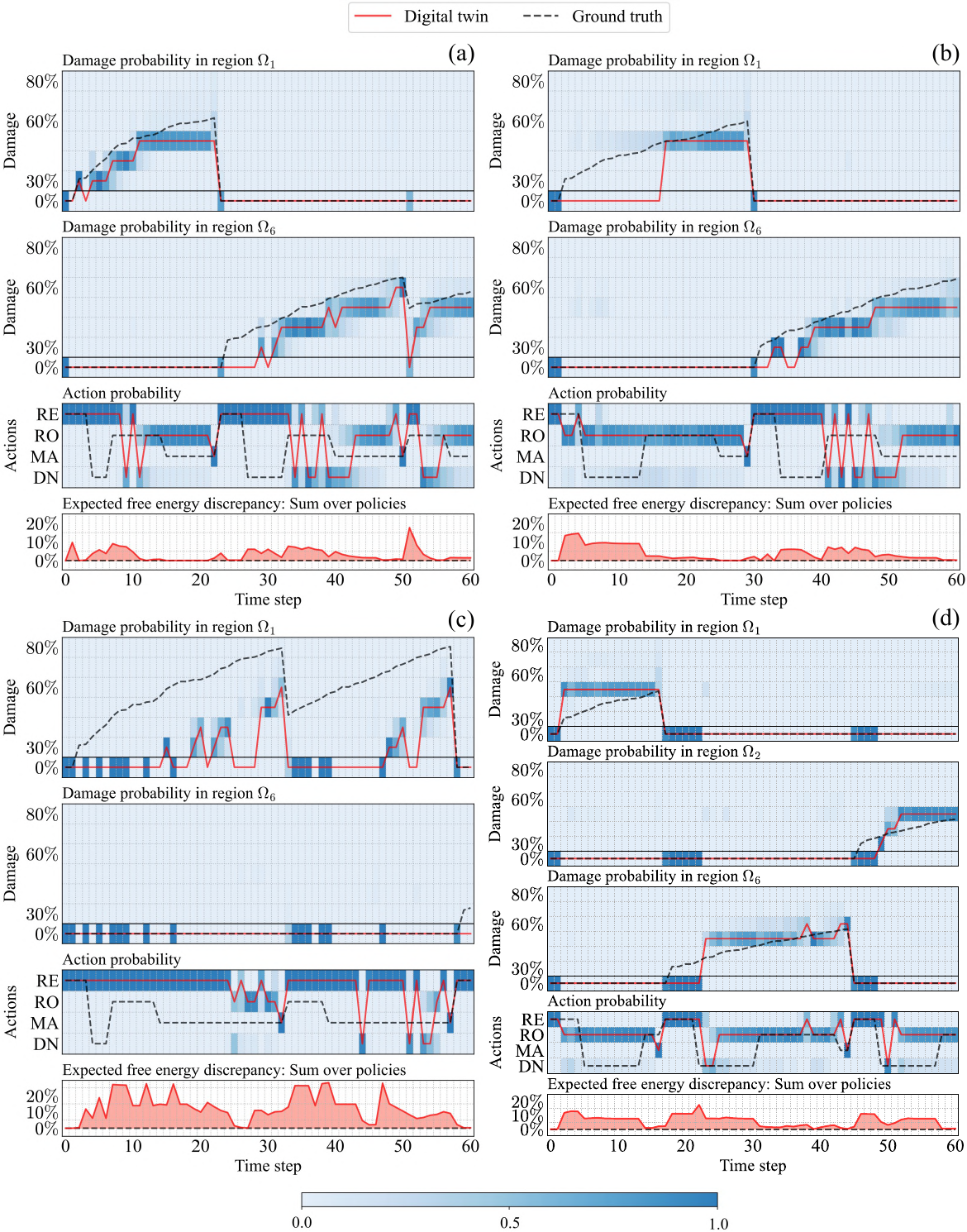}};
\node [rectvia] () at (-0.4,9.7) {};
\node [] () at (-0.4,9.6) {\small (a)};
\node [rectvia] () at (8,9.7) {};
\node [] () at (8,9.6) {\small (b)};
\node [rectvia] () at (-0.4,0) {};
\node [] () at (-0.4,-0.15) {\small (c)};
\node [rectvia] () at (8,0) {};
\node [] () at (8,-0.15) {\small (d)};
\end{tikzpicture}
\vspace{-0.6cm}
\caption{Active digital twin using a combination of goal-directed (pragmatic) and information-seeking (epistemic) behaviors. Performance under progressively more incomplete data streams during the passage of a train, affecting: (a) sensors $\lbrace \mathbf{u}_5,\mathbf{u}_{10}\rbrace$; (b) sensors $\lbrace \mathbf{u}_3,\mathbf{u}_5,\mathbf{u}_{10}\rbrace$; (c) sensors $\lbrace \mathbf{u}_1,\mathbf{u}_3,\mathbf{u}_5,\mathbf{u}_{10}\rbrace$; and (d) all sensors $\lbrace \mathbf{u}_1,\ldots,\mathbf{u}_{10}\rbrace$. Incomplete data streams correspond to the loss of one third of the normally sampled measurements, modeled by zeroing the signals over the central portion of the monitoring window $[0.5~\text{s},1~\text{s}]$. Probabilistic and best-point estimates of: (top panels) digital state evolution compared to the ground-truth physical state; (penultimate panels) control actions recommended by the digital twin versus the optimal action under the ground-truth generative process. In the top panels, background colors represent the belief distribution over the digital state at each time step. In the penultimate panels, background colors indicate the belief distribution over the control actions. The bottom panels scores simulation quality in terms of the percentage absolute discrepancy between the sum of the policy-specific expected free energies computed by the digital twin and those obtained under the ground truth.}
\vspace{-0.25cm}
  \label{fig:incomplete}
\end{figure}

Incomplete data streams during the passage of a train are modeled by zeroing vibration recordings in $\mathbf{U}$ over the central portion of the monitoring window $[0.5~\text{s},1~\text{s}]$. Figure~\ref{fig:incomplete} shows representative cases where the corruption affects an increasing number of sensors, up to all sensors. As information loss increases, the ADT becomes more prone to desynchronize from the ground truth, with the risk of failure due to the delayed selection of a necessary MA action. To counteract this, the ADT autonomously increases the probability of selecting corrective RE actions, promoting physical--digital realignment. Interestingly, corrupting only a subset of sensors proves more detrimental than disabling all sensors over the same time window. This effect is due to the disruption of cross-channel convolution patterns in the DL models used to construct the observation modality $O^{\Omega\delta}$. Partial corruption generates inconsistent patterns not encountered during training, whereas removing all channels leads to information loss without injecting misleading correlations. In this latter case, the ADT still exploits the remaining signal segments, although with reduced tracking accuracy and an increased need for RE actions.

\begin{figure}[!t]
\begin{tikzpicture}[scale=.9, every node/.style={scale=1.}]
\node[draw=none,fill=none] at (0,0){\includegraphics[width=1\linewidth]{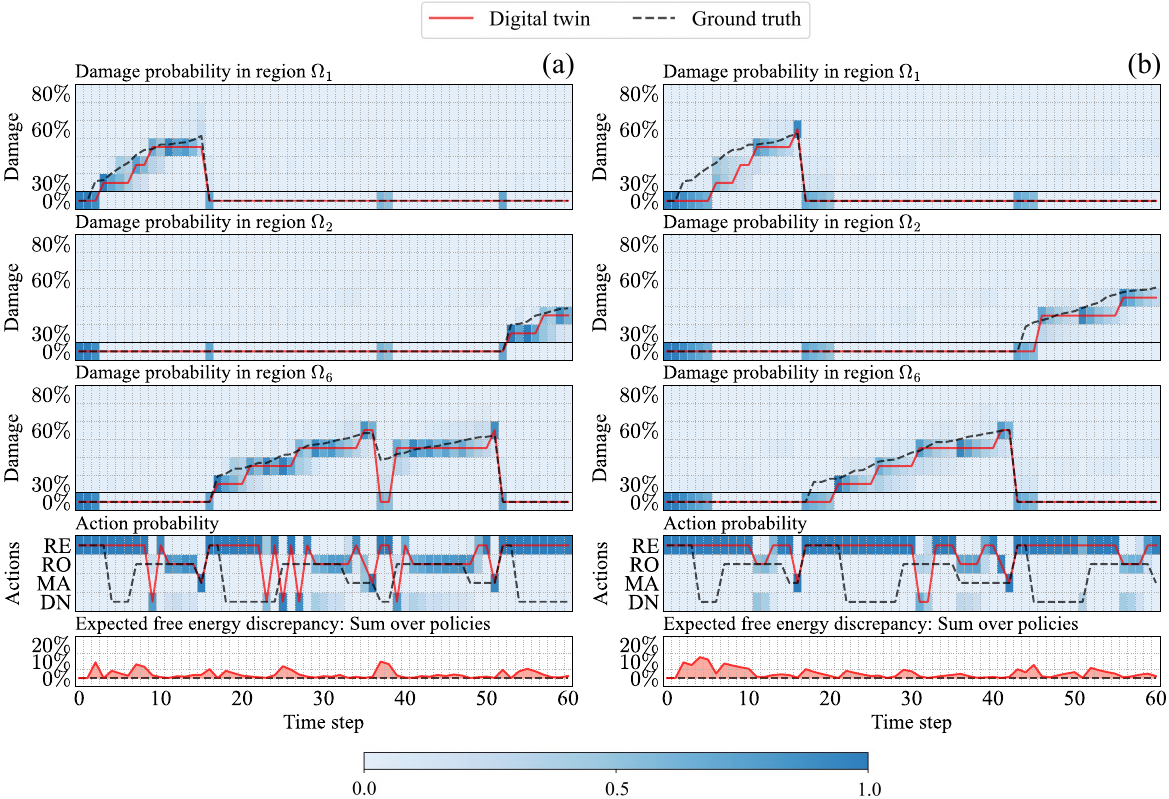}};
\node [rectvia] () at (-0.4,4.8) {};
\node [] () at (-0.4,4.7) {\small (a)};
\node [rectvia] () at (8,4.8) {};
\node [] () at (8,4.7) {\small (b)};
\end{tikzpicture}
\vspace{-0.6cm}
\caption{Active digital twin using a combination of goal-directed (pragmatic) and information-seeking (epistemic) behaviors. Performance under progressively more frequent observation failures, with the active digital twin receiving observational data: (a) every two simulation steps; and (b) every five simulation steps. When a measurement fails, observational evidence for the first modality $O^{\Omega\delta}$ is unavailable to the active digital twin. Probabilistic and best-point estimates of: (top three panels) digital state evolution compared to the ground-truth physical state; (penultimate panels) control actions recommended by the digital twin versus the optimal action under the ground-truth generative process. In the top panels, background colors represent the belief distribution over the digital state at each time step. In the penultimate panels, background colors indicate the belief distribution over the control actions. The bottom panels scores simulation quality in terms of the percentage absolute discrepancy between the sum of the policy-specific expected free energies computed by the digital twin and those obtained under the ground truth.}
  \label{fig:incomplete2}
  \vspace{-0.25cm}
\end{figure}

From a slightly different perspective, observation failures are instead modeled by making observational evidence for $O^{\Omega\delta}$ unavailable to the ADT at selected time steps. Figure~\ref{fig:incomplete2} presents two representative simulations in which measurements are received only every two or five simulation steps. In both cases, the ADT remains synchronized with the ground truth by leveraging the predictive capabilities of the transition model. The main effect is a periodic widening and sharpening of the belief over the digital state, reflecting the observation frequency. This leads to a stronger tendency to select information-gathering RE actions rather than the more conservative RO action, despite the latter having a higher prior preference  $\widetilde{p}(O^{u})$.

\begin{figure}[!t]
\centering
  \includegraphics[width=.78\linewidth]{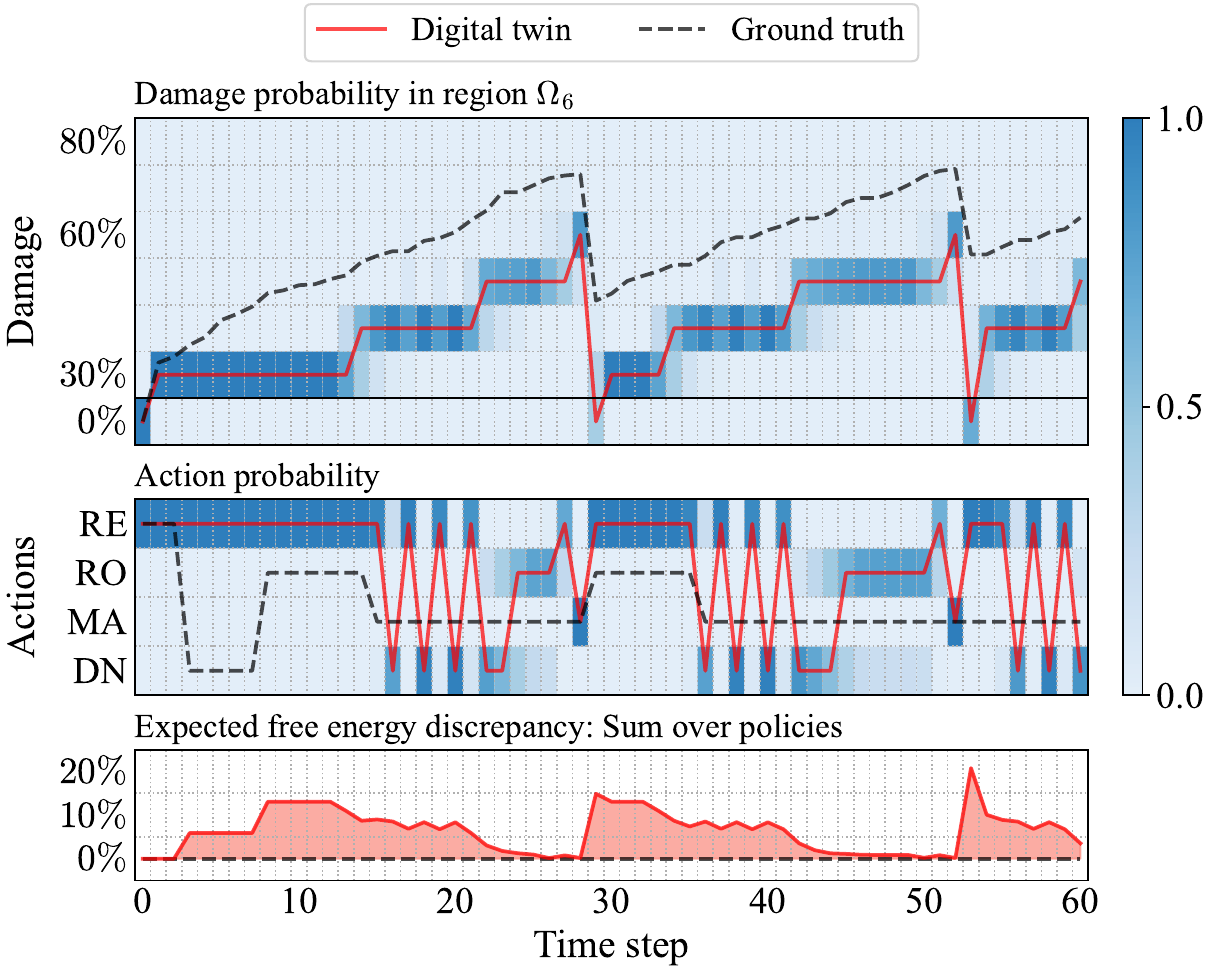}
  \caption{Active digital twin using a combination of goal-directed (pragmatic) and information-seeking (epistemic) behaviors. Performance for a damaged region of size reduced by $50\%$ compared to the training regime. Probabilistic and best-point estimates of: (top panel) digital state evolution compared to the ground-truth physical state; (middle panel) control actions recommended by the digital twin versus the optimal action under the ground-truth generative process. In the top panel, background colors represent the belief distribution over the digital state at each time step. In the middle panel, background colors indicate the belief distribution over the control actions. The bottom panel scores simulation quality in terms of the percentage absolute discrepancy between the sum of the policy-specific expected free energies computed by the digital twin and those obtained under the ground truth.}
  \label{fig:reduced}
  \vspace{-0.25cm}
\end{figure}

Finally, robustness to modeling inaccuracies is assessed by introducing uncertainty in the size of the damaged region. As in previous cases, test instances are derived from FOM solutions corrupted with additive Gaussian noise. However, here the extent of the damaged subdomains $\Omega_1,\ldots,\Omega_6$ is reduced by $25\%$ and $50\%$ in the testing data only; the offline training dataset remains unchanged.

Figure~\ref{fig:reduced} shows a representative case for a $50\%$ reduction in the extent of $\Omega_6$. The ADT correctly identifies the damaged region but systematically underestimates its magnitude. The resulting behavior is consistent with what seen in \sez\ref{sez:results_2}: an initial sequence of epistemic RE actions is followed by pragmatic DN actions, interleaved with RE actions to compensate for increased $Q^*(D_{t_c})$ entropy. Subsequently, RO actions emerge to slow degradation, and a MA action is selected when failure risk becomes substantial, anticipated by a preventive RE action.

While the persistent underestimation may degrade tracking performance, it does not necessarily imply desynchronization. We quantify this risk through a cluster of $200$ simulations with different random seeds for a $50\%$ size reduction. The mean accuracy of the maximum a-posteriori estimates of $D^\Omega$ and $D^\delta$ decreases to $40\%$ ($95\%$ confidence interval of $\pm3.2\%$), yet only a single failure occurs. For a $25\%$ size reduction, the mean accuracy increases to $80.5\%$ (95\% confidence interval $\pm 0.8\%$), with no observed failures. Despite the degradation in tracking accuracy, these results demonstrate robustness to misspecification of damage extent, in addition to previously considered uncertainties such as measurement noise, varying operational conditions, and damage severity.

\section{Discussion}
\label{sez:discussion}

We have shown that framing a digital twin as an AIF agent transforms it from a passive observer into an autonomous decision-making entity. The considered case study, among the few AIF applications in engineering, demonstrates how pragmatic and epistemic behaviors emerge naturally from expected free energy minimization, without explicit programming. Beyond optimizing structural health and maintenance costs, the ADT maintains synchronization with the evolving structural condition by autonomously deciding when to acquire additional information. Actions are selected not only to respond to the current belief state, but also to shape future observations and reduce anticipated uncertainty. The ADT also proves robust to incomplete or unreliable observations and to uncertainty in the extent of the damaged region, adapting its policy to mitigate desynchronization risks. 

These capabilities stem directly from the advantages of AIF. Utility maximization is not the sole driver of policy selection; rather, it complements information-gain maximization within a unified functional objective that balances exploration and exploitation. Unlike reinforcement learning, which typically does not consider information-gain maximization, and relies on reward functions shaped through extensive trial-and-error over large datasets, AIF offers greater modeling and learning flexibility by encoding preferences and beliefs as probability distributions within a generative model. This feature is particularly valuable in complex, nonstationary, and nonlinear environments, where robustness and adaptivity are critical. Moreover, explicitly incorporating hypotheses and assumptions into the generative model enhances interpretability, boosting explainable and trustworthy decision-making~\cite{albarracin2023designing}. However, we note that reinforcement learning can be viewed as a special case of AIF~\cite{Friston02102015}. Since rewards or utility functions can be expressed as log-probabilities encoding prior preferences, reinforcement learning can be reformulated as inference over policies that generate preferred outcomes.

{\bf Limitations:} The effectiveness of the ADT framework depends on the design of the generative model, including the choice of the digital state space and the specification of the observation and transition models, which remain problem-dependent. In the presented application, one observation modality relies on deep learning models trained offline on synthetic data. Although this supports generalization to realistic scenarios, performance is sensitive to modeling assumptions and to the representativeness of the training dataset. Capturing more complex damage patterns or geometries would require more detailed numerical models and larger datasets, increasing computational demands. These aspects are also linked to sensor deployment and the associated value of information~\cite{andtioris2021102072,malings201845}, highlighting the importance of optimal sensor placement~\cite{Capellari}.

{\bf Policy recommendations:} From an operational perspective, digital twins should incorporate decision-making and information-seeking capabilities, rather than being limited to state estimation and prediction. Moreover, probabilistic representations such as PGMs are essential to explicitly manage uncertainty and enable adaptive behavior. While ADTs enjoy these features, their behavior is not fixed but depends on the objectives encoded in the generative model. Prior preferences can be tuned to reflect specific safety or operational requirements, allowing stakeholders to explore how varying levels of risk sensitivity influence decisions. Active digital twin-based systems can thus support the transition from periodic or reactive maintenance to predictive strategies, where inspection and monitoring are scheduled autonomously based on quantified uncertainty and expected utility.

\section{Conclusions and outlook}
\label{sez:conclusion}

{\bf Summary of contributions:} This paper introduces active digital twins based on the active inference paradigm. At the core of active digital twins lies a self-updating generative model that interacts with a partially observable dynamical environment, extending the abstraction of physical--digital systems proposed by Kapteyn et al.~\cite{pgm_wilcox_dt}. By leveraging the variational free energy minimization process that drives active inference agents~\cite{parr2022active,friston2010free}, we have enabled active digital twins capable of adaptively monitoring, interacting with, and learning from uncertain and dynamic environments. Of particular interest is their ability to autonomously balance goal-directed (pragmatic) and information-seeking (epistemic) behaviors. Within this dual objective, decision-support from the active digital twin becomes an integral part of the inference process, allowing uncertainty about hidden states to be resolved as a means to maximize utility, in the spirit of intelligent automation~\cite{san2026evolution}.

Results have been presented for a case study on structural health monitoring and predictive maintenance of a railway bridge. The application has focused on the construction of a generative model enabling bidirectional perception--action interaction. Simulations of active digital twins have been carried out using active inference agents with progressively richer behaviors: purely goal-directed; combined goal-directed and information-seeking; with or without learning updates to the generative model. The results demonstrate that active digital twins autonomously balance the joint optimization of structural health and maintenance costs objectives with the need to acquire information in a principled manner, as dictated by free energy minimization. In particular, active exploration has proved essential for maintaining synchronization between the digital and physical states, while incorporating learning updates has improved inference accuracy, reduced the need for corrective epistemic actions, and enabled the safe postponement of costly interventions.

{\bf Opportunities for future research:} Beyond the structural health monitoring application presented here, the proposed framework offers a generalizable methodology applicable across a wide range of domains. Active digital twins are envisioned as key enablers of autonomous agents in the development of smart structures and systems~\cite{Kon-Well}, as well as in fields such as medicine and neuroscience~\cite{amunts2024coming}. Future extensions can leverage the epistemic value associated with expected information gain not only over digital states but also over model parameters, as anticipated in \sez\ref{sez:active_behav}, enabling the generative model underlying active digital twins to be learned online from controlled experience. This capability will support complex behaviors combining goal-directed, information-gathering, and curiosity-driven components~\cite{friston2017active}, fostering adaptation and continual self-learning from real-world data. Online learning may also be complemented by offline updates via Bayesian model reduction~\cite{friston2018bayesian}, enabling an optimal trade-off between the complexity and accuracy of the generative model.

\vspace{6pt} 
\noindent{\bf Data Accessibility:} The implementation code used for the experiments presented in \sez\ref{sez:results} is available in the public repository \texttt{activeDT}~\cite{repo}. The code implements the proposed active digital twin framework and can be used to simulate and generate the plots for digital state estimation, future prediction, and policy inference, as reported in this paper. The observational data used to run the experiments, along with the deep learning models trained according to the implementation details provided in the Appendix of~\cite{Torzoni_DT}, are also available in the same repository. The \texttt{Matlab} library for finite element simulation and reduced-order modeling of partial differential equations employed to generate these data is available in the repository \texttt{Redbkit}~\cite{Redbkit}.

\vspace{6pt} 
\noindent{\bf Competing Interests:} The authors declare that they have no known competing financial interests or personal relationships that could have appeared to influence the work reported in this paper.

\vspace{6pt} 
\noindent{\bf Acknowledgments:} The authors thank Dr.~Luca Rosafalco (Politecnico di Milano) and Eng.~Giacomo Mondello (Politecnico di Milano) for the insights and contributions during our discussions.

\vspace{6pt} 
\noindent{\bf Funding:} This work is supported by the ERC advanced grant IMMENSE (Grant Agreement 101140720), funded by the European Union. Views and opinions expressed are however those of the authors only and do not necessarily reflect those of the European Union or the European Research Council Executive Agency. Neither the European Union nor the granting authority can be held responsible for them. 
Author AM also acknowledges the financial support from the FIS starting grant DREAM (Grant Agreement FIS00003154), funded by the Italian Science Fund (FIS) - Ministero dell’Università e della Ricerca.  
Authors DM, FD, and GP acknowledge financial support from the ERC consolidator grant ThinkAhead (Grant Agreement 820213), funded by the European Union.


\medskip

\bibliographystyle{elsarticle-num}
\biboptions{sort&compress}

\appendix
\section{Expected free energy derivations}
\label{sez:efe_eq}
In this Appendix, we provide the complete derivation of the expected free energy expressions, as adapted to the adaptive digital twin framework from~\cite{pymdp}:
\\

\paragraph{Expected free energy}
\begin{equation}
\begin{split}
G_t^\pi&=\mathbb{E}_{Q(O_t,D_t\mid\pi)}[\ln{Q(D_{t}\mid\pi)}-\ln{\widetilde{p}(O_t,D_t\mid\pi)}]\\
&=\mathbb{E}_{Q(O_t,D_t\mid\pi)}[\ln{Q(D_{t}\mid\pi)}-\ln{\widetilde{p}(O_t,D_t\mid\pi)} +\underbrace{\ln{Q(D_{t}\mid O_{t},\pi)} -\ln{Q(D_{t}\mid O_{t},\pi)}}_{=0}]\\
&=\mathbb{E}_{Q(O_t,D_t\mid\pi)}[\ln{Q(D_{t}\mid\pi)}-\ln{Q(D_{t}\mid O_{t},\pi)}-\ln{\widetilde{p}(O_{t})} \\
&\hspace{50pt}- \ln{p(D_{t}\mid O_{t},\pi)} +\ln{Q(D_{t}\mid O_{t},\pi)}]\\
&=-\mathbb{E}_{Q(O_t,D_t\mid\pi)}[\ln{Q(D_{t}\mid O_{t},\pi)}-\ln{Q(D_{t}\mid\pi)}]-\mathbb{E}_{Q(O_t,D_t\mid\pi)}[\ln{\widetilde{p}(O_{t})}] \\
&\hspace{50pt}+\mathbb{E}_{Q(O_t,D_t\mid\pi)}[\ln{Q(D_{t}\mid O_{t},\pi)} - \ln{p(D_{t}\mid O_{t},\pi)}]\\
&= -\underbrace{\mathbb{E}_{Q(O_t\mid\pi)}[\text{D}_\text{KL}[Q(D_{t}\mid O_{t},\pi)\mid\mid Q(D_{t}\mid\pi) ] ]}_{\text{Epistemic value (information gain)}} - \underbrace{\mathbb{E}_{Q(O_t\mid\pi)}[\ln{\widetilde{p}(O_{t})}]}_{\text{Pragmatic value (utility)}}\\
&\hspace{50pt}+\underbrace{\mathbb{E}_{Q(O_t\mid\pi)}[\text{D}_\text{KL}[Q(D_{t}\mid O_{t},\pi)\mid\mid p(D_{t}\mid O_{t},\pi) ] ]}_{\text{Expected variational approximation error ($\geq0$)}}.
\end{split}
\end{equation}
\\

\paragraph{Expected free energy with model parameters}
\begin{equation}
\begin{split}
G_t^\pi&=\mathbb{E}_{Q(O_t,D_t,\boldsymbol{\phi}\mid\pi)}[\ln{Q(D_{t},\boldsymbol{\phi}\mid\pi)}-\ln{\widetilde{p}(O_t,D_t,\boldsymbol{\phi}\mid\pi)}]\\
&=\mathbb{E}_{Q(O_t,D_t,\boldsymbol{\phi}\mid\pi)}[\ln{Q(D_{t},\boldsymbol{\phi}\mid\pi)}-\ln{\widetilde{p}(O_t,D_t,\boldsymbol{\phi}\mid\pi)}\\
&\hspace{50pt}+\underbrace{\ln{Q(D_{t},\boldsymbol{\phi}\mid O_{t},\pi)} -\ln{Q(D_{t},\boldsymbol{\phi}\mid O_{t},\pi)}}_{=0}]\\
&=\mathbb{E}_{Q(O_t,D_t,\boldsymbol{\phi}\mid\pi)}[\ln{Q(D_{t}\mid\pi)}+\ln{Q(\boldsymbol{\phi}\mid\pi)} -\ln{\widetilde{p}(O_t)}-\ln{p(D_t,\boldsymbol{\phi}\mid O_t,\pi)}\\&\hspace{50pt}+\ln{Q(D_{t},\boldsymbol{\phi}\mid O_{t},\pi)} -\ln{Q(D_{t},\boldsymbol{\phi}\mid O_{t},\pi)}]\\
&=\mathbb{E}_{Q(O_t,D_t,\boldsymbol{\phi}\mid\pi)}[\ln{Q(D_{t}\mid\pi)}-\ln{Q(D_{t}\mid O_{t},\pi)}+\ln{Q(\boldsymbol{\phi}\mid\pi)}-\ln{Q(\boldsymbol{\phi}\mid O_{t},\pi)}\\&\hspace{50pt}-\ln{\widetilde{p}(O_t)}-\ln{p(D_t,\boldsymbol{\phi}\mid O_t,\pi)}+\ln{Q(D_{t},\boldsymbol{\phi}\mid O_{t},\pi)}]\\
&=-\mathbb{E}_{Q(O_t,D_t,\boldsymbol{\phi}\mid\pi)}[\ln{Q(D_{t}\mid O_{t},\pi)}-\ln{Q(D_{t}\mid\pi)}]\\&\hspace{50pt}-\mathbb{E}_{Q(O_t,D_t,\boldsymbol{\phi}\mid\pi)}[\ln{Q(\boldsymbol{\phi}\mid O_{t},\pi)}-\ln{Q(\boldsymbol{\phi}\mid\pi)}]-\mathbb{E}_{Q(O_t,D_t,\boldsymbol{\phi}\mid\pi)}[\ln{\widetilde{p}(O_t)}]\\&\hspace{50pt}+\mathbb{E}_{Q(O_t,D_t,\boldsymbol{\phi}\mid\pi)}[\ln{Q(D_{t},\boldsymbol{\phi}\mid O_{t},\pi)}-\ln{p(D_t,\boldsymbol{\phi}\mid O_t,\pi)}]\\
&=-\underbrace{\mathbb{E}_{Q(O_t\mid\pi)}[\text{D}_\text{KL}[Q(D_{t}\mid O_{t},\pi)\mid\mid Q(D_{t}\mid\pi)]]}_{\text{Epistemic value (digital state information gain)}}-\underbrace{\mathbb{E}_{Q(O_t\mid\pi)}[\text{D}_\text{KL}[Q(\boldsymbol{\phi}\mid O_{t},\pi)\mid\mid Q(\boldsymbol{\phi}\mid\pi)]]}_{\text{Epistemic value (model parameters information gain)}}\\&\hspace{50pt}-\underbrace{\mathbb{E}_{Q(O_t\mid\pi)}[\ln{\widetilde{p}(O_t)}]}_{\text{Pragmatic value (utility)}}+\underbrace{\mathbb{E}_{Q(O_t\mid\pi)}[\text{D}_\text{KL}[Q(D_{t},\boldsymbol{\phi}\mid O_{t},\pi)\mid\mid p(D_t,\boldsymbol{\phi}\mid O_t,\pi)]]}_{\text{Expected variational approximation error ($\geq0$)}}.
\end{split}
\end{equation}
\end{document}